\pdfoutput=1

\documentclass[preprint2]{aastex}

\shorttitle{Variability survey in the CoRoT SRa01 field}
\shortauthors{Klagyivik et al.}

\begin{document}

\title{Variability survey in the CoRoT SRa01 field:\\
Implications of eclipsing binary distribution on cluster formation in NGC 2264}

\author{P.~Klagyivik\altaffilmark{1,2}, Sz.~Csizmadia\altaffilmark{2}, T.~Pasternacki\altaffilmark{2},
T.~Fruth\altaffilmark{2}, A.~Erikson\altaffilmark{2}, J.~Cabrera\altaffilmark{2},
R.~Chini\altaffilmark{3,4}, P.~Eigm\"uller\altaffilmark{2}, P.~Kabath\altaffilmark{5}, S.~Kirste\altaffilmark{2},
R.~Lemke\altaffilmark{3}, M.~Murphy\altaffilmark{6}, H.~Rauer\altaffilmark{2,7}, and R.~Titz-Weider\altaffilmark{2}}

\altaffiltext{1}{Konkoly Observatory, Research Centre for Astronomy and Earth Sciences, Hungarian Academy of Sciences, H-1121 Budapest XII, Konkoly Thege \'ut 15-17., Hungary}
\altaffiltext{2}{Institut f\"ur Planetenforschung, Deutsches Zentrum f\"ur Luft- und Raumfahrt, Rutherfordstra\ss e 2, 12489 Berlin, Germany}
\altaffiltext{3}{Astronomisches Institut, Ruhr-Universit\"at Bochum, 44780 Bochum, Germany}
\altaffiltext{4}{Instituto de Astronom\'ia, Universidad Cat´olica del Norte, Antofagasta, Chile}
\altaffiltext{5}{European Southern Observatory, Alonso de C{\'o}rdova 3107, Vitacura, Casilla 19001, Santiago Chile}
\altaffiltext{6}{Depto. F\'isica, Universidad Cat\'olica del Norte, PO 1280, Antofagasta, Chile}
\altaffiltext{7}{Technische Universit\"at Berlin, Zentrum f\"ur Astronomie und Astrophysik, Hardenbergstra\ss e 36, 10623 Berlin, Germany}

\begin{abstract}
Time-series photometry of the CoRoT field SRa01 was carried out with the
Berlin Exoplanet Search Telescope II (BEST II) in 2008/2009.
A total of 1,161 variable stars were detected,
of which 241 were previously known and 920 are newly found.
Several new, variable young stellar objects have been discovered.
The study of the spatial distribution of eclipsing binaries revealed
the higher relative frequency of Algols toward the center of the
young open cluster NGC 2264.
In general Algol frequency obeys an isotropic distribution
of their angular momentum vectors, except inside the cluster, where
a specific orientation of the inclinations is the case.
We suggest that we see the orbital plane of the binaries almost edge-on.
\end{abstract}


\section{Introduction}

NGC 2264 is among the best studied star formation regions. The NGC-designation
simultaneously denotes a young, 3 Myr old open cluster and an extended
HII-region with some small molecular knots where star formation is still in
progress \citep{mayne08}. Hundreds of pre-main sequence stars have been
identified in this cluster \citep[e.g.][]{makidon04,ramirez04,furesz06}.
Signs of current formation of low-mass companion objects (stellar or planetary objects) can be
found in this cluster, too \citep[e.g.][]{hamilton05}.
The cluster is well studied at a broad wavelength range from X-rays \citep[e.g.][]{dahm07}, through visual and infrared
wavelengths \citep{ramirez04,furesz06}, to radio observations \citep[e.g.][]{trejo08}.

The cluster contains about 500 well-identified
cluster members \citep{ramirez04,furesz06}, but the true number of cluster members
reaches $\sim2000$ \citep[][and references therein]{cody13}.

A very special and unique simultaneous time-series photometric study was carried
out in near-infrared by Spitzer, in X-ray by Chandra and in optical by MOST and
CoRoT space satellites during the winter 2010/2011. CoRoT also observed this
cluster in 2007/2008 for 30 days obtaining
another set of ultraprecise time-series photometry \citep{cody13}. We used the 25 cm
ground-based BEST II (Berlin Exoplanet Search Telescope) in 2008/2009 to support the CoRoT observations in
two ways: to keep the validity of the ephemeris of several eclipsing binary
stars and planetary candidates for future follow-up observations and to help the
determination of possible variable contamination sources inside the CoRoT
photometric mask. Since CoRoT's photometric mask is large (typically $80 \times 20$
arcseconds) due to a dispersion element in the optical path, CoRoT often
measures the total flux from several light sources.
Therefore, the origin of the light variation is sometimes not
the CoRoT-target but one of the contaminating objects
or the variation of the main CoRoT-target is
diluted by another star. Our observations help to decide which one is the
variable source if the amplitude of such variation is above our detection limit.
Similar observations for the same purpose were reported in \citet{karo07},
\citet{kaba07, kaba08, kaba09a, kaba09b}, \citet{pasternacki11} and \citet{fruth12}.
Since our observations on NGC 2264 cover a whole season, time-series photometric
data can also be used to investigate the photometric optical behaviour of the stars
between the epochs of the two CoRoT studies.

In this paper we present our study on the variable stars in NGC 2264.
Section \ref{data} presents the description of the telescope used and the observations.
We report our flux measurements on the known variable stars, we confirm or reject the variability
of several previously suspected variable stars as well as we report 920 new
variable stars in and around NGC 2264 (Section \ref{variable_stars}).

In Section \ref{yso} we present a set of light curves of an interesting
type of pre-main sequence stars. One similar light curve shape was found by
\citet{rodrigez-ledesma12}, and here we add a selection of new examples to this
rare variable star class.

In Section~\ref{models} we present the fitted models of 11 selected
eclipsing binaries, while in Section~\ref{others} we discuss a set of interesting variable stars in our data set.

In Section~\ref{eclipsing} we shall demonstrate that the observable frequency of Algol-type eclipsing
binaries is extremely high inside the cluster area compared to the field.
We point out that there is no other possible explanation for this phenomenon but a
well-determined, special orientation of the orbital planes of the binaries in
the cluster, giving an additional, new constraint for the star
formation theories.

Eventually the summary and conclusions of this paper can be found in Section~\ref{summary}.

\section{Data acquisition}
\label{data}

The observations were performed with the BEST II telescope located
at the Observatorio Cerro Armazones, Chile. The system consists of
a Takahashi 25 cm Baker-Ritchey-Chr\'etien telescope equipped with
a 4k $\times$ 4k Finger Lakes CCD. The corresponding field of view is
$1.7^\circ\times 1.7^\circ$ with an angular resolution of $1\farcs5$ pixel$^{-1}$.
In order to maximize the photon yield no filter was used.

BEST II observed the CoRoT target field SRa01 during a total of
50 nights between 2008 December 06 and 2009 February 10. In order
to cover the larger FOV of CoRoT completely, SRa01 was split into
two BEST II target fields that were observed alternately.
These subfields are denoted SRa01a and SRa01b in the following.

The acquired observations were processed using the BEST automated
photometric pipeline as described in \citet{kaba09a}, \citet{raue10} and \citet{fruth12}.
The resulting datasets consist of 90,065 light curves of stellar objects in the CoRoT SRa01 field.

\section{Variable stars}
\label{variable_stars}

Since the CoRoT satellite observed this field in 2007/2008
and 2010/2011, our photometric data obtained in 2008 and 2009
are an excellent supplement for characterizing both short and
long term variability, thus extending the CoRoT observations.

\subsection{Detection}

For detecting variable stars, we apply the efficient method
described by \citet{fruth12}. It is based on the widely-used variability
index $J$ \citep{stet96,zhan03} and a multiharmonic period search
\citep{sccz96}, but additionally involves an automatic treatment
of systematic variability. All light curves with $J > 0$ ($86\%$)
were fitted with seven harmonics and ranked using the modified
Analysis-of-Variance statistic $q$ \citep[see][]{fruth12}. A cut-off
limit was set to $q > 8$ based on empirical experiments, yielding
1,805 variable star candidates (728 in SRa01a, 1,077 in SRa01b).

\begin{figure*}
\centering
\includegraphics[width=0.24\textwidth]{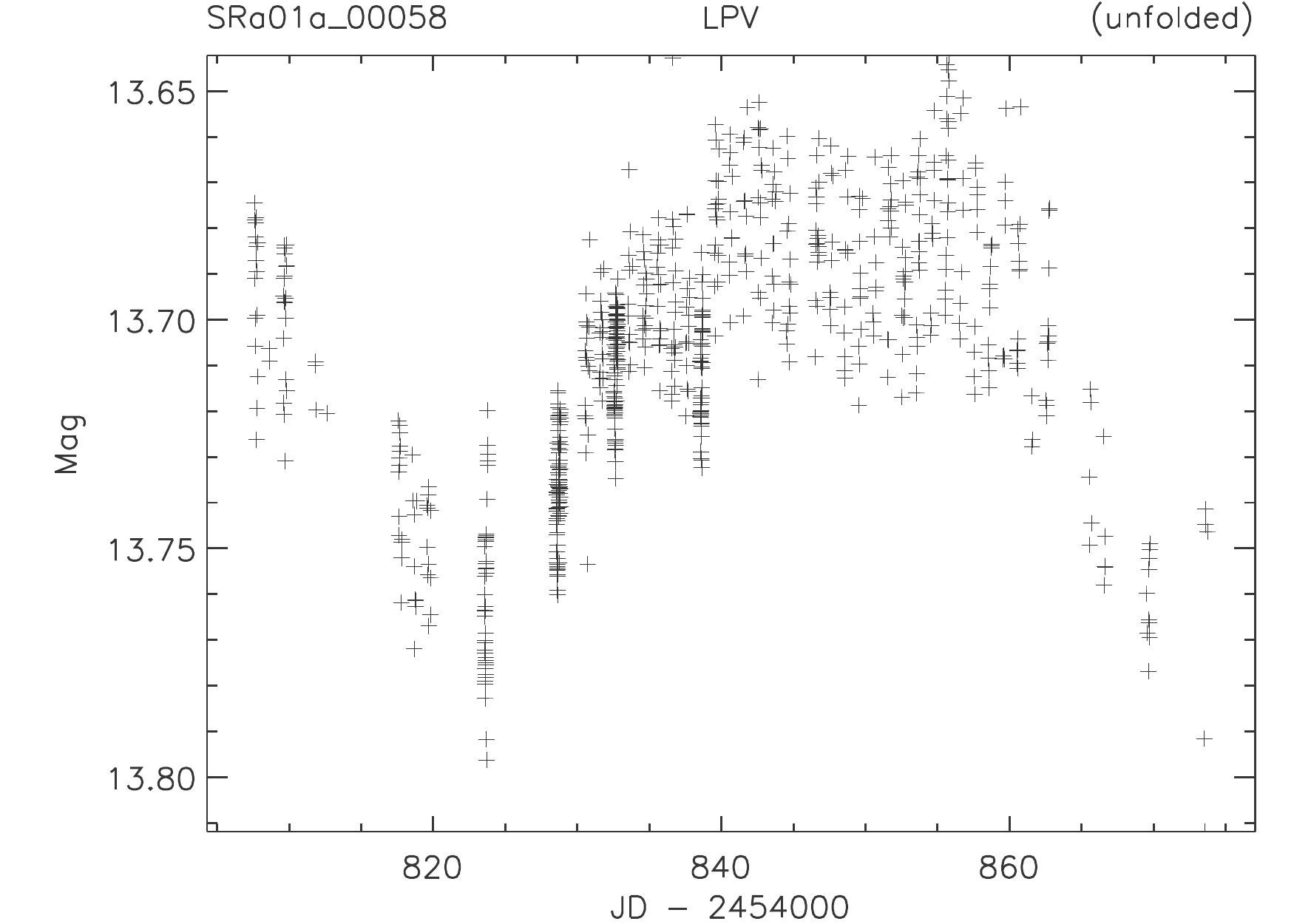}
\includegraphics[width=0.24\textwidth]{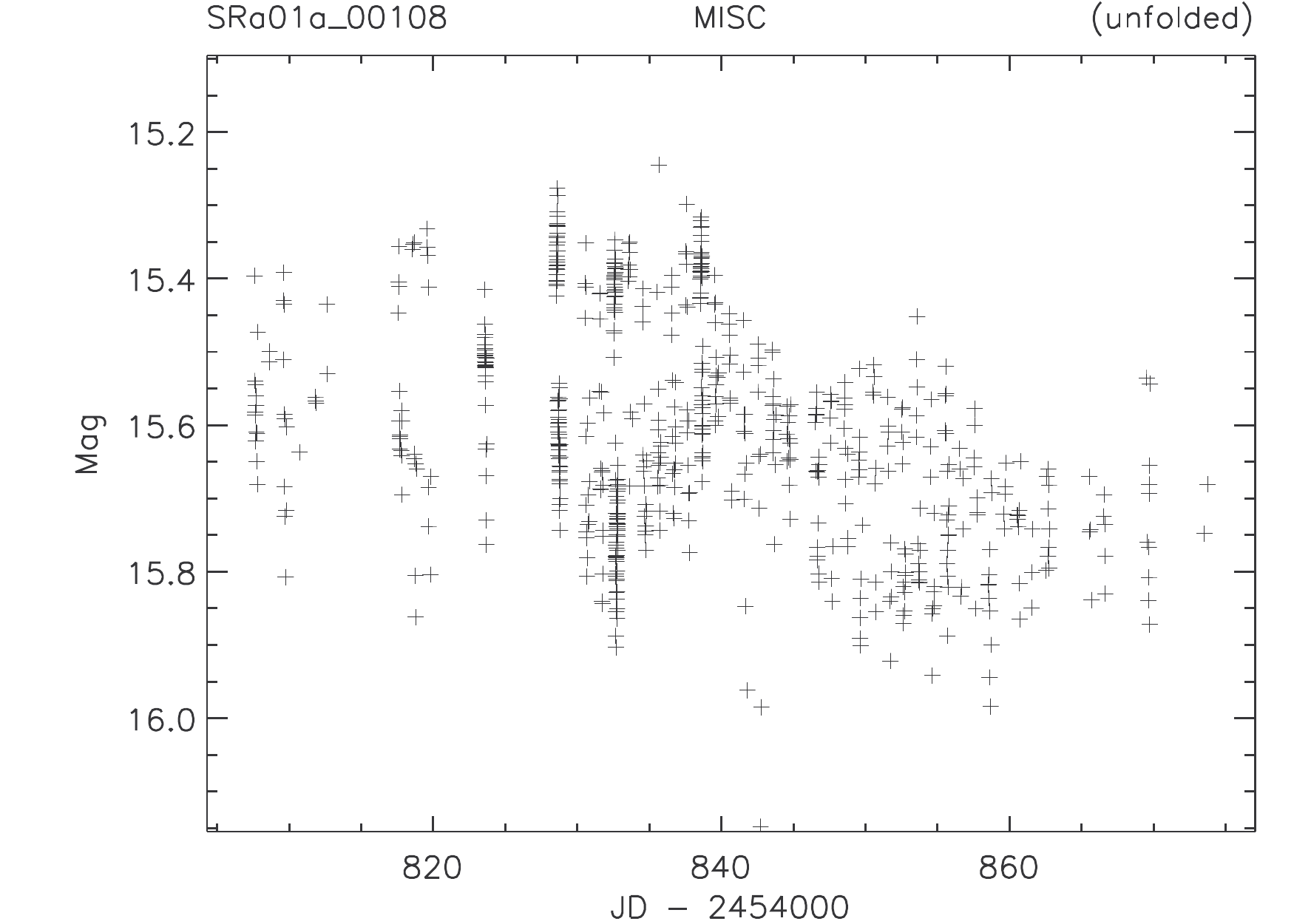}
\includegraphics[width=0.24\textwidth]{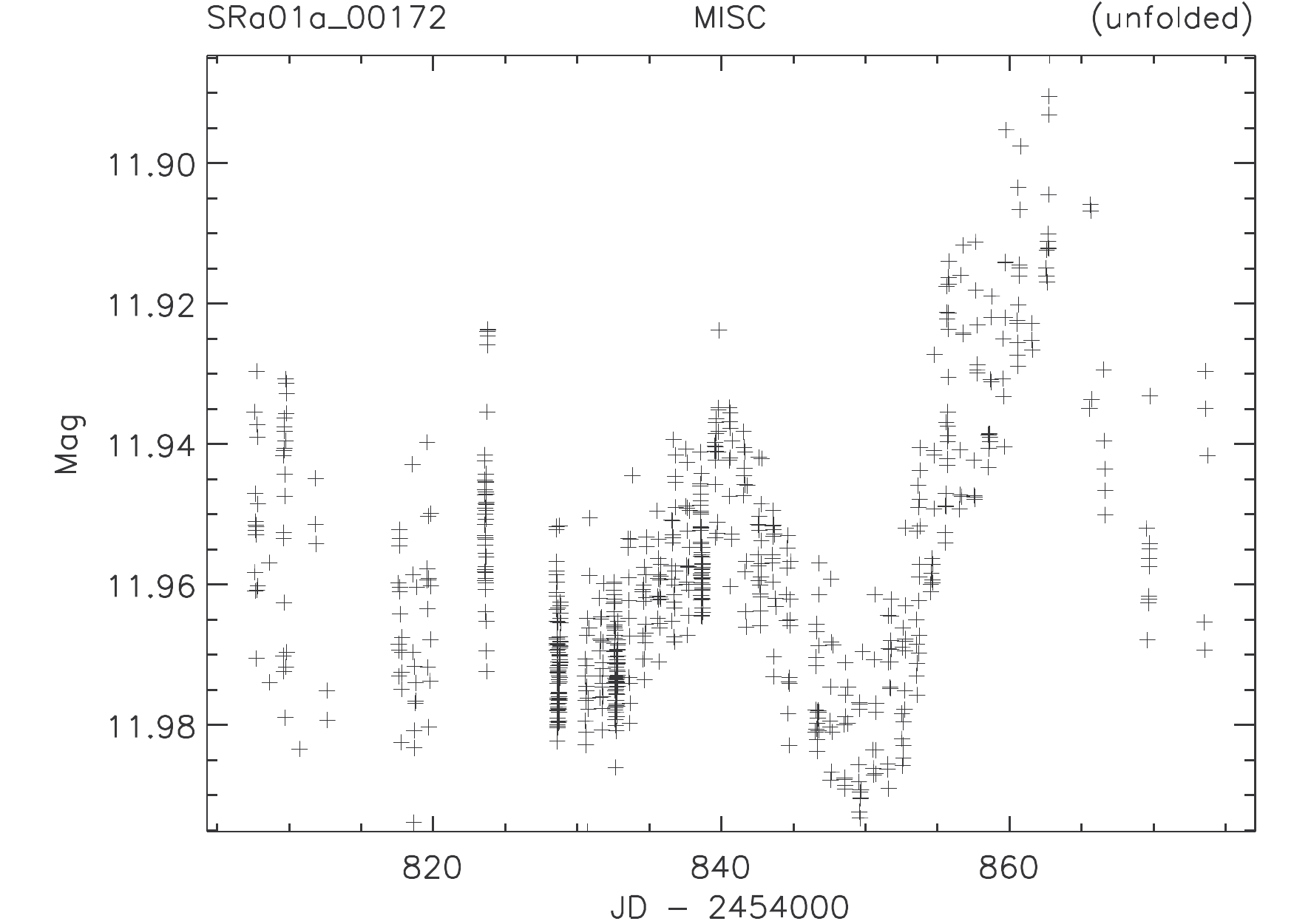}
\includegraphics[width=0.24\textwidth]{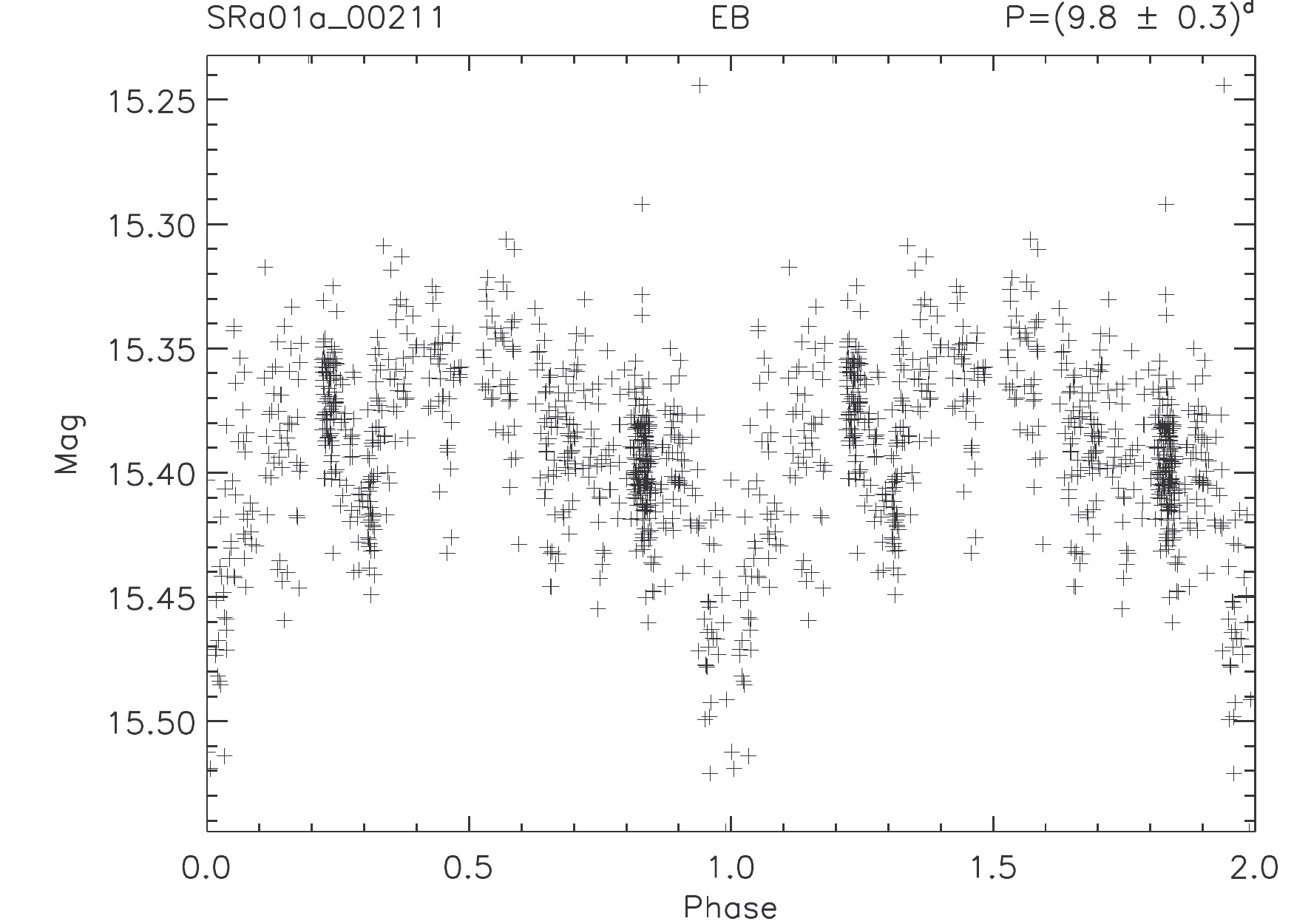}
\includegraphics[width=0.24\textwidth]{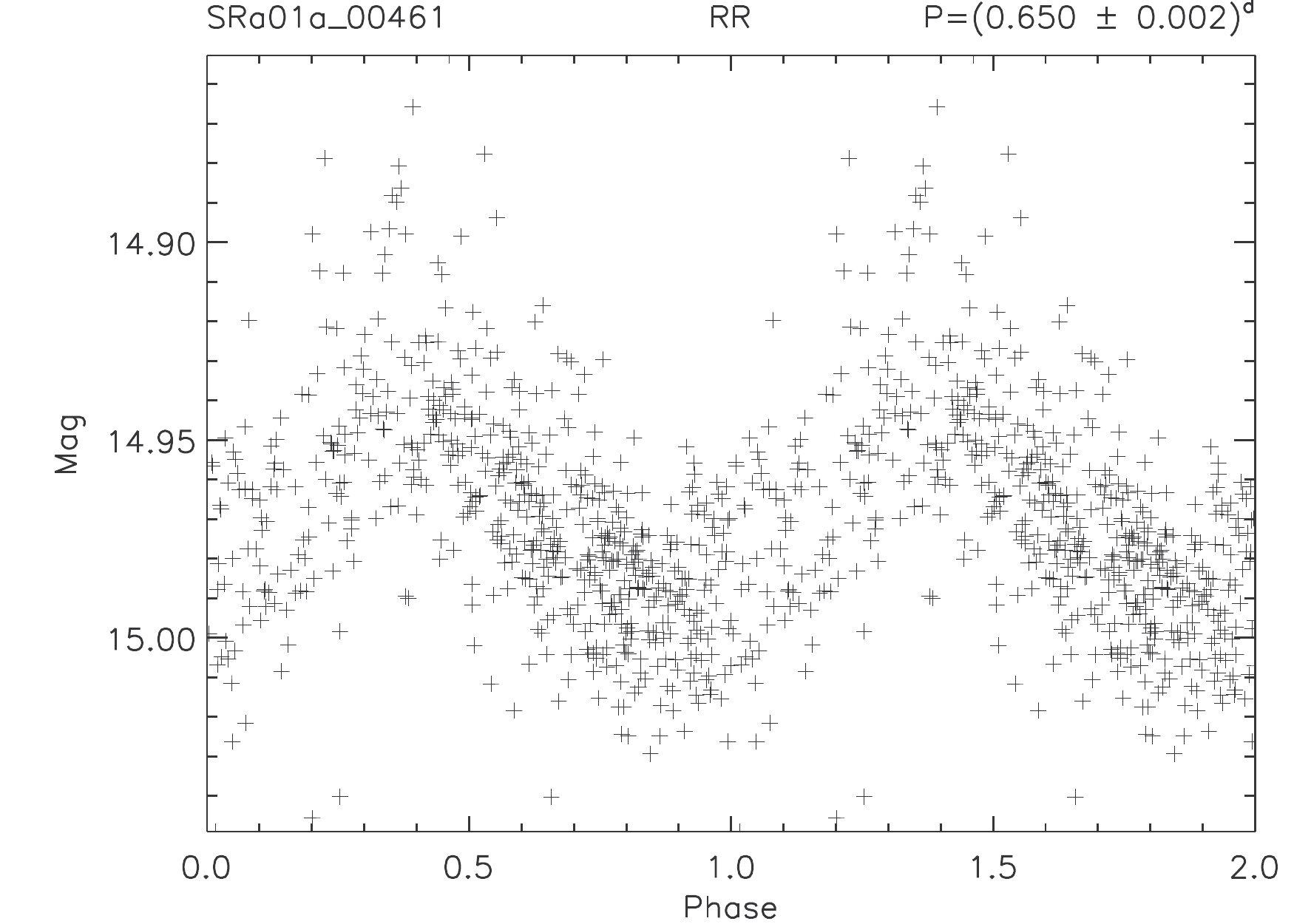}
\includegraphics[width=0.24\textwidth]{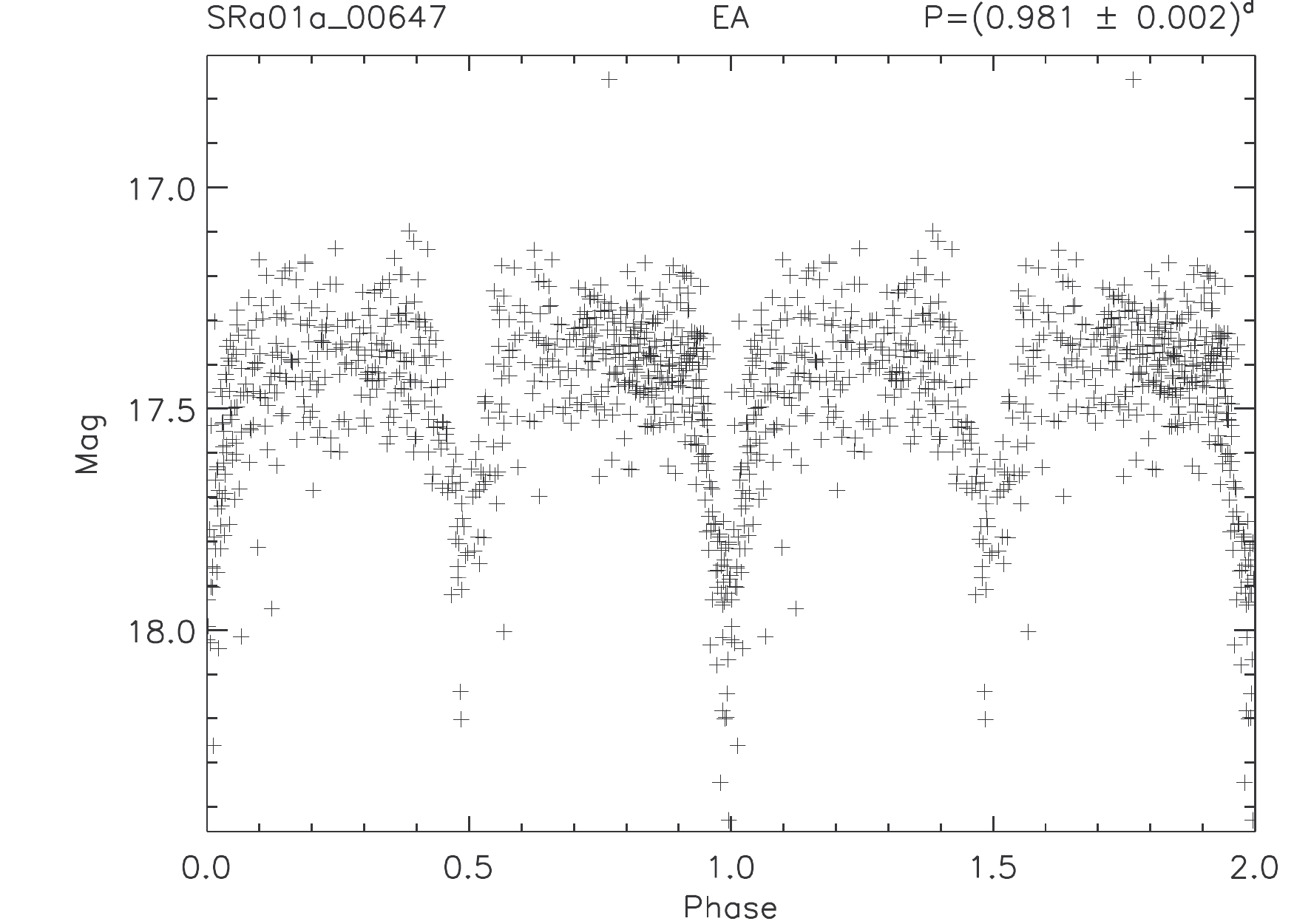}
\includegraphics[width=0.24\textwidth]{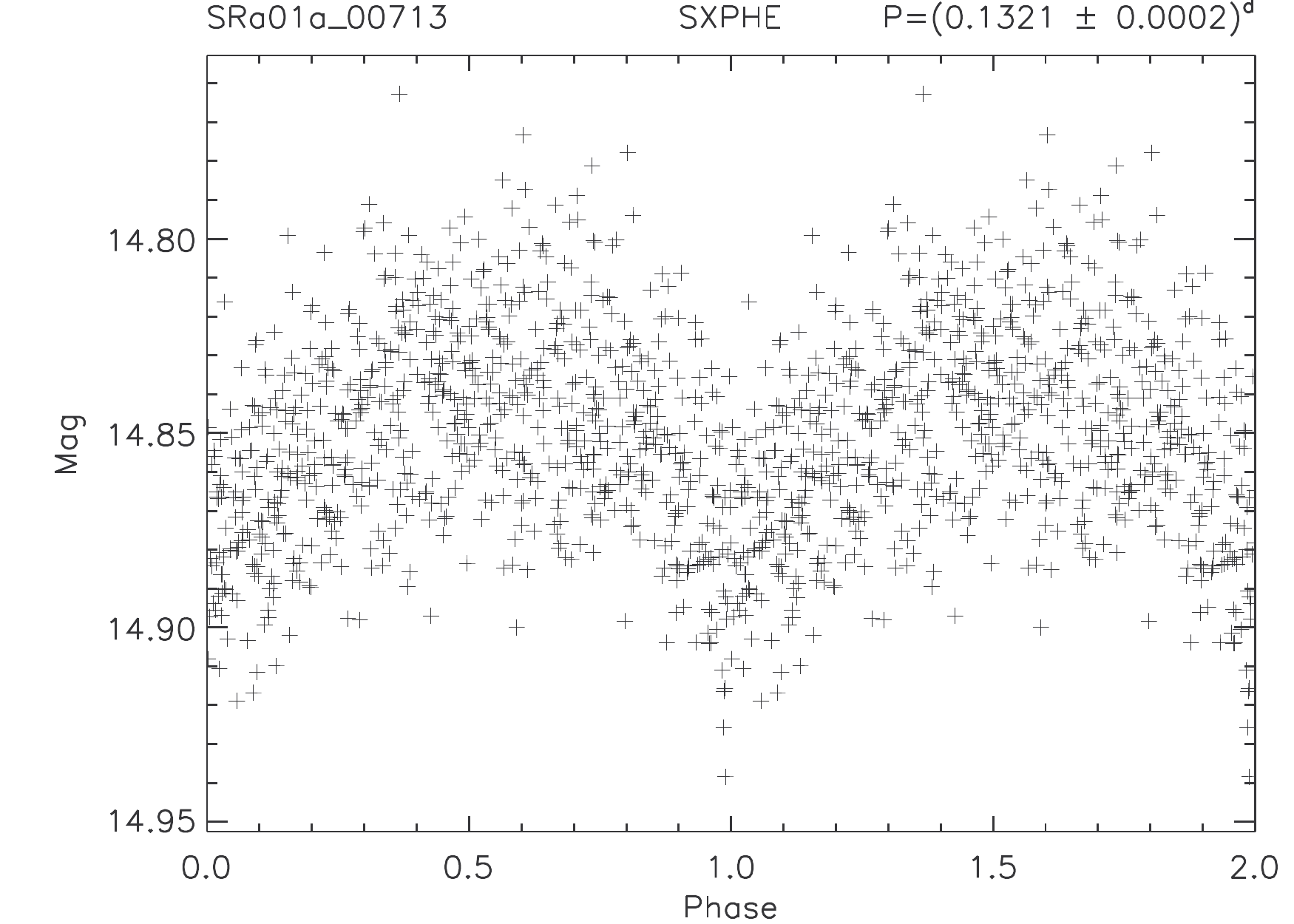}
\includegraphics[width=0.24\textwidth]{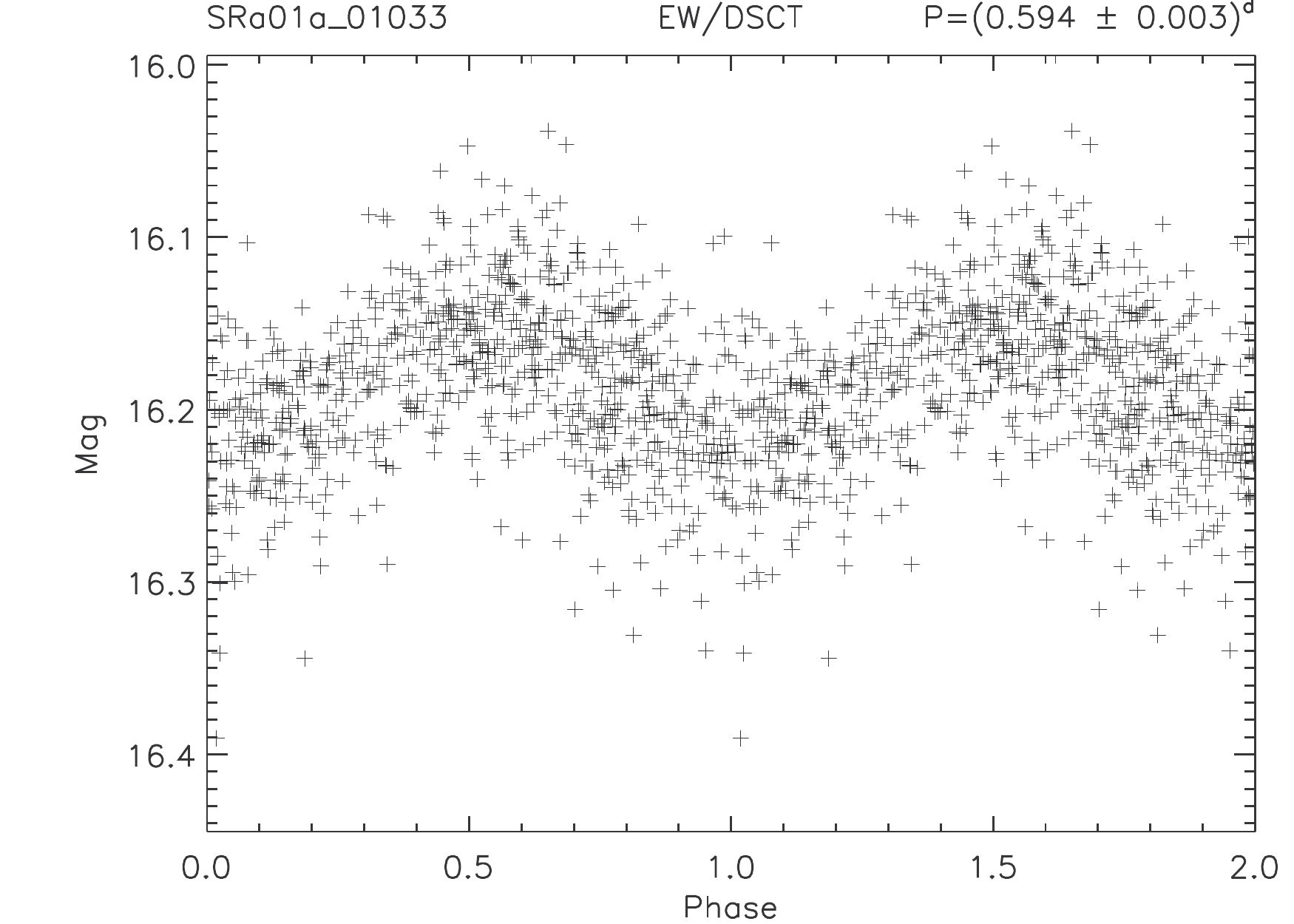}
\label{fig:lc_sample}
\caption{The light curves of the first eight variable stars in Table~\ref{tab:varcat}. All the light curves
are available in the electronic edition of the journal.}
\end{figure*}

All candidates were inspected visually and classified on an
individual basis. We detected a total of 1,161 variable stars,
of which 241 were previously known and 920 are new discoveries.
The variable star catalogue and the observed light curves are presented here in Table \ref{tab:varcat}
and Figure \ref{fig:lc_sample}, respectively, for guidance regarding its form and content. Table \ref{tab:varcat}
and Figure \ref{fig:lc_sample} are published in its entirety in the electronic edition of {\it Astrophysical Journal}.

\begin{deluxetable}{lccccccccccccc}
\rotate
\tablecolumns{11}
\tablewidth{0pc}
\tabletypesize{\scriptsize} 
\tablecaption{\scriptsize Catalog of variable stars detected in {\it CoRoT} field SRa01, sorted by internal BEST II identifiers. \label{tab:varcat}}
\tablehead{
\colhead{BESTID} & \colhead{flag} & \colhead{2MASSID} & \colhead{$\alpha(J2000.0)$} & \colhead{$\delta(J2000.0)$} & 
\colhead{R$_B$ [mag]} &\colhead{J [mag]} &\colhead{H [mag]} &\colhead{K [mag]} & 
\colhead{T$_0$ [rHJD]} & \colhead{P [d]} & \colhead{A [mag]} &  \colhead{Type} & \colhead{Other names}
}
\startdata
SRa01a\_00058  &    &06372843+1002552  &$06^h37^m28.4^s$  &$ 10^\circ02'55.8"$  &13.70  &  9.47 &  8.22 &  7.77 & \nodata  &       \nodata             &         \nodata      &LPV  &\\
SRa01a\_00108  &    &06364462+0931232  &$06^h36^m44.7^s$  &$ 09^\circ31'24.4"$  &15.62  & 15.07 & 14.41 & 14.09 & \nodata  &       \nodata             &         \nodata      &MISC  &\\
SRa01a\_00172  &    &06365888+0941207  &$06^h36^m58.9^s$  &$ 09^\circ41'20.9"$  &11.96  &  8.18 &  7.06 &  6.62 & \nodata  &       \nodata             &         \nodata      &MISC  &\\
SRa01a\_00211  &    &06363124+0921174  &$06^h36^m31.2^s$  &$ 09^\circ21'17.4"$  &15.39  & 12.57 & 11.77 & 11.50 &  10.742  &            $9.8 \pm 0.3$  &          $0.12 \pm 0.04$  &EB  &\\
SRa01a\_00461  &    &06362054+0912226  &$06^h36^m20.6^s$  &$ 09^\circ12'22.4"$  &14.97  & 13.80 & 13.62 & 13.51 &   8.150  &        $0.650 \pm 0.002$  &          $0.06 \pm 0.04$  &RR  &\\
SRa01a\_00647  &    &06363213+0919546  &$06^h36^m32.1^s$  &$ 09^\circ19'54.7"$  &17.42  & 15.76 & 15.18 & 15.00 &   8.248  &        $0.981 \pm 0.002$  &            $0.6 \pm 0.2$  &EA  &\\
SRa01a\_00713  &    &06380069+1022493  &$06^h38^m00.7^s$  &$ 10^\circ22'49.3"$  &14.85  & 13.45 & 12.82 & 12.67 &   7.706  &      $0.1321 \pm 0.0002$  &          $0.04 \pm 0.03$  &SXPHE  &\\
SRa01a\_01033  &    &06373374+1002111  &$06^h37^m33.7^s$  &$ 10^\circ02'11.3"$  &16.19  & 14.39 & 13.80 & 13.59 &   7.968  &        $0.594 \pm 0.003$  &          $0.08 \pm 0.05$  &EW/DSCT  &\\
SRa01a\_01086  &    &06373720+1004170  &$06^h37^m37.2^s$  &$ 10^\circ04'17.6"$  &12.24  &  7.95 &  6.76 &  6.24 & \nodata  &       \nodata             &         \nodata      &MISC  &\\
SRa01a\_01414  &    &06381538+1029312  &$06^h38^m15.4^s$  &$ 10^\circ29'31.4"$  &14.58  & 11.41 & 10.86 & 10.57 & \nodata  &       \nodata             &         \nodata      &MISC  &\\
SRa01a\_01515  &    &06365626+0932389  &$06^h36^m56.3^s$  &$ 09^\circ32'39.0"$  &16.12  & 15.47 & 15.28 & 15.38 &   7.625  &        $0.579 \pm 0.003$  &          $0.07 \pm 0.05$  &ELL  &\\
SRa01a\_01522  &    &06380652+1022424  &$06^h38^m06.5^s$  &$ 10^\circ22'43.1"$  &12.81  &  8.46 &  7.27 &  6.76 & \nodata  &       \nodata             &         \nodata      &LPV  &\\
SRa01a\_01556  &    &06365672+0932430  &$06^h36^m56.7^s$  &$ 09^\circ32'43.1"$  &16.07  & 14.32 & 13.77 & 13.55 &   8.945  &          $1.38 \pm 0.02$  &          $0.12 \pm 0.05$  &ELL  &\\
SRa01a\_01612  &    &06372722+0954167  &$06^h37^m27.2^s$  &$ 09^\circ54'16.8"$  &14.14  & 12.48 & 12.22 & 12.09 &   8.096  &        $1.921 \pm 0.007$  &          $0.08 \pm 0.02$  &EA  &\\
SRa01a\_01754  &    &06372176+0949348  &$06^h37^m21.8^s$  &$ 09^\circ49'35.1"$  &13.91  & 10.33 &  9.41 &  9.01 & \nodata  &       \nodata             &         \nodata      &EA  &\\
\enddata
\tablecomments{Previously known objects are flagged with $k$. The flag $c$ denotes stars affected by crowding.
Their IDs from VSX or GCVS can be found in the last column.
$R_B$ is the brightness in BEST II photometric system.
$J$, $H$ and $K$ are the brightness in 2MASS photometric system.
The Epoch $T_0$ is given in reduced Julian date [rHJD] in respect to $T=2,454,800.0$.
It denotes the first minimum in the light curve.
$P$ is the period and $A$ is the amplitude of the variability.
This table is published in its entirety in the electronic edition
of the {\it Astrophysical Journal}. A portion is shown here for guidance regarding its form and content.}
\end{deluxetable}

\subsection{Classification}
\label{classification}

The classification of the newly discovered periodic variable
stars was based on the shape, period and
amplitude of their variability according to a simplified scheme based on
the General Catalogue of Variable Stars \citep[GCVS, ][]{samu09}. Pulsating variables were sorted into
Delta Scuti (DSCT), $\gamma$ Doradus (GDOR), RR Lyrae
(RR) and Cepheid variable (DCEP) types.
Eclipsing binary star systems were classified as Algol
type (EA), Beta Lyrae type (EB), or W Ursae Majoris
type (EW).
Stars having sinusoidal-like light curves are classified as
ellipsoidal variables (ELL), whereas light
curves that exhibit typical features of starspots
are marked as spotted stars (SP) or BY Draconis-type stars (BY).
Non-periodic variables are classified
as miscellaneous (MISC). Most of these stars are likely
non-periodic pre-main sequence stars.
Stars that vary on time scales longer than the observational baseline
are classified as long periodic (LP). These stars need further observations
to determine their accurate period or to demonstrate their non-periodic behaviour.
In the case of questionable light curves we marked as
mixed types (EW/DSCT, ELL/SP).
Since NGC 2264 is a young open cluster, there are many periodic pre-main sequence stars
which were marked as young stellar objects (YSO).

The accuracy of our classification is expected to be around $70\%$ \citep{fruth12}.
An overview of all classifications is given in Table \ref{classification_table}.

\begin{table}[t]
\begin{center}
\caption{Variable stars in and around the CoRoT SRa01 field. The number of newly detected variables
are in brackets.}
\label{classification_table}
\begin{tabular}{lccc}
\hline
\hline
\noalign{\smallskip}
Type & \multicolumn{2}{c}{Subfield} & All\\
\noalign{\smallskip}
\hline
\noalign{\smallskip}
 & $a$ & $b$\\
\noalign{\smallskip}
\cline{2-3}
\noalign{\smallskip}
\noalign{\smallskip}
\multicolumn{4}{l}{Intrinsic variables}\\
\noalign{\smallskip}
DSCT    &  28  (28) & 44 (43) & 72 (71)\\
GDOR    &  38  (30) & 17 (17) & 55 (47)\\
DCEP    &  10   (9) &  8  (7) & 18 (16)\\
RR      &  20  (18) & 10 (10) & 30 (27)\\
SXPHE   &   2   (2) &  1  (1) &  3  (3)\\
\noalign{\smallskip}
\hline
\noalign{\smallskip}
\noalign{\smallskip}
\multicolumn{4}{l}{Extrinsic variables}\\
\noalign{\smallskip}
EA      &  71  (67) & 43 (41) & 114 (108)\\
EW      &  18  (18) & 24 (24) &  42  (42)\\
EB      &  11   (7) &  8  (8) &  19  (15)\\
E       &   1   (1) & 11 (11) &  12  (12)\\
ELL     & 158 (134) & 28 (28) & 186 (162)\\
ELL/SP  &  10   (8) & 32 (32) &  42  (40)\\
BY      &   3   (0) &    -    &   3   (0)\\
YSO     & 155  (89) & 61 (61) & 216 (150)\\
LP      &  13  (10) & 10  (9) &  23  (19)\\
EW/DSCT &   9   (9) & 19 (19) &  28  (28)\\
MISC    & 229 (113) & 69 (66) & 298 (179)\\
\noalign{\smallskip}
\hline
\noalign{\smallskip}
All     & 776 (543) & 385 (377) & 1161 (920)\\
\noalign{\smallskip}
\hline
\hline
\end{tabular}
\end{center}
\end{table}

\subsection{Known variables}
\label{known_var}

The stars observed with BEST II are crosschecked with the Variable Star Index (VSX)\footnote{http://www.aavso.org/vsx} of
the American Association of Variable Star Observers and with the GCVS
\citep{samu09}. Within the observed target field, in total 495 previously known
variable stars could be found according to these catalogues, 469 in the subfield 'a' and
26 in the subfield 'b'. In the subfield 'a' we found 233 of the 469 as variable star
while in 'b' only 8 of the 26. All of the previously reported variables
without any detected variability in our data are pre-main sequence flare stars.

The variable star catalogue (Table \ref{tab:varcat}) contains only the stars that show
clear variability in our data. A review of all known
variables one-by-one is beyond the limits of this paper.

Previously known variables are marked with a 'k' flag in Table \ref{tab:varcat} and
the names in other catalogue(s) are also given.

\subsection{Newly detected variables}
\label{new_var}

920 out of 1161 are newly detected variables. 543 of them are located in subfield 'a' and 377
in subfield 'b'. The most populated types among the newly detected variables
are of short period pulsators (DSCT and GDOR), eclipsing binaries (EA and EW),
stars with sinusoidal light curve and low amplitude (ELL), young stellar objects (YSO)
and stars that show irregular variations (MISC). The stars classified as
MISC as well as the YSOs are concentrating toward the center of the cluster.

The parameters of all new and previously known variable stars are listed in Table \ref{tab:varcat}.

\section{Pre-main sequence stars}
\label{yso}

Characteristic of a young star forming region, NGC 2264 contains many stars
that show rotational modulation-like as well as irregular light variations.

\begin{figure*}[tb]
\centering
\includegraphics[width=0.24\textwidth]{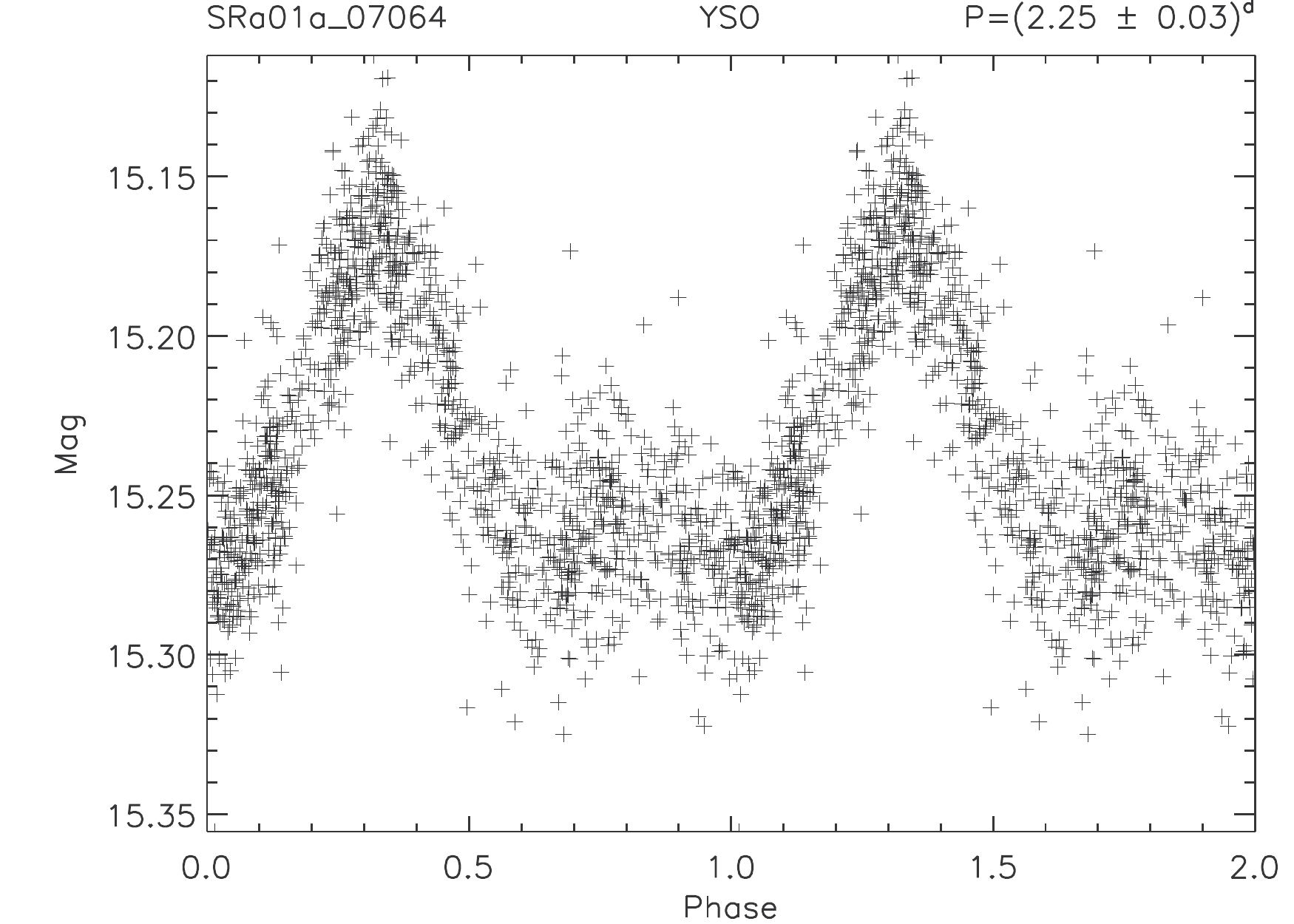}
\includegraphics[width=0.24\textwidth]{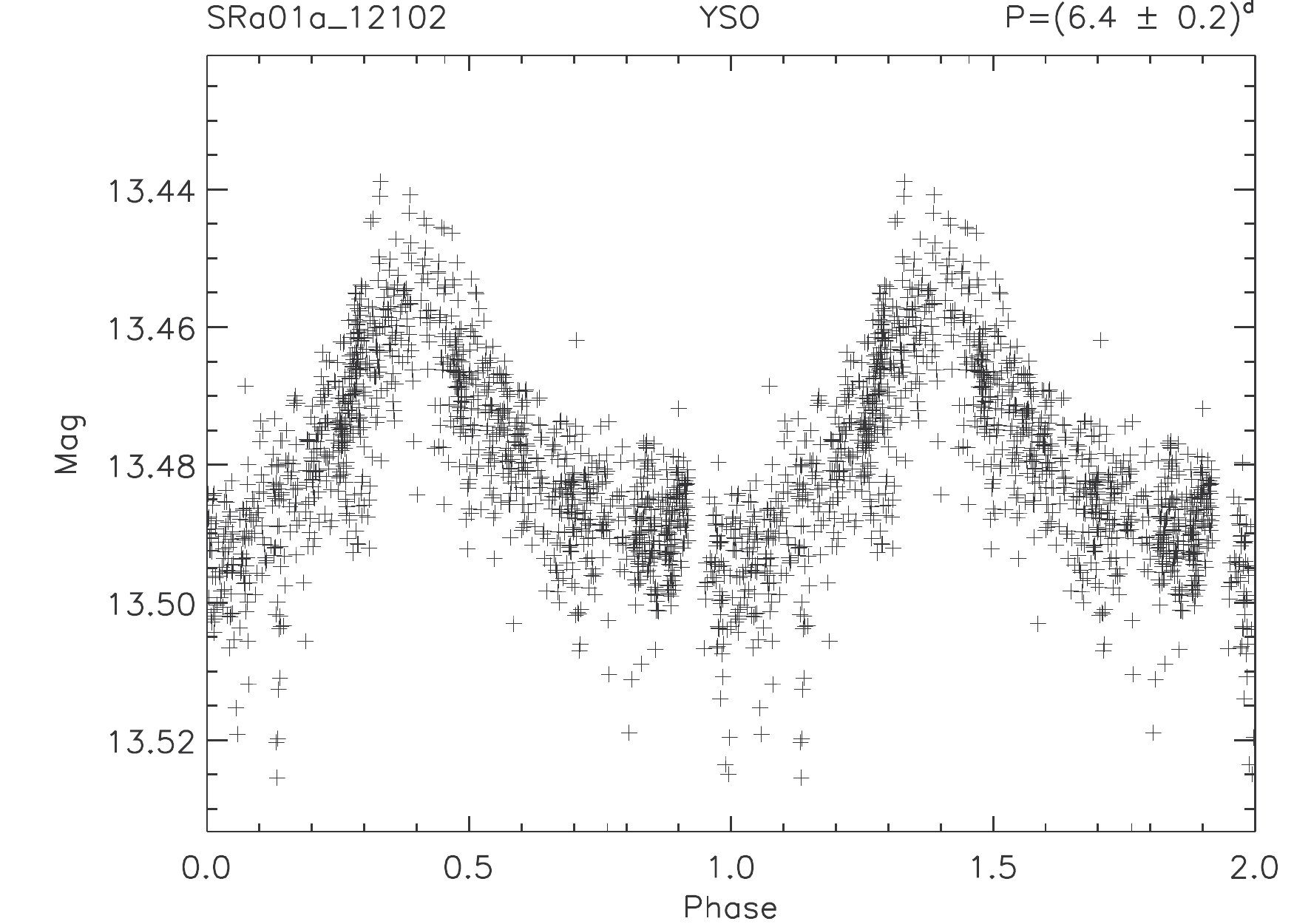}
\includegraphics[width=0.24\textwidth]{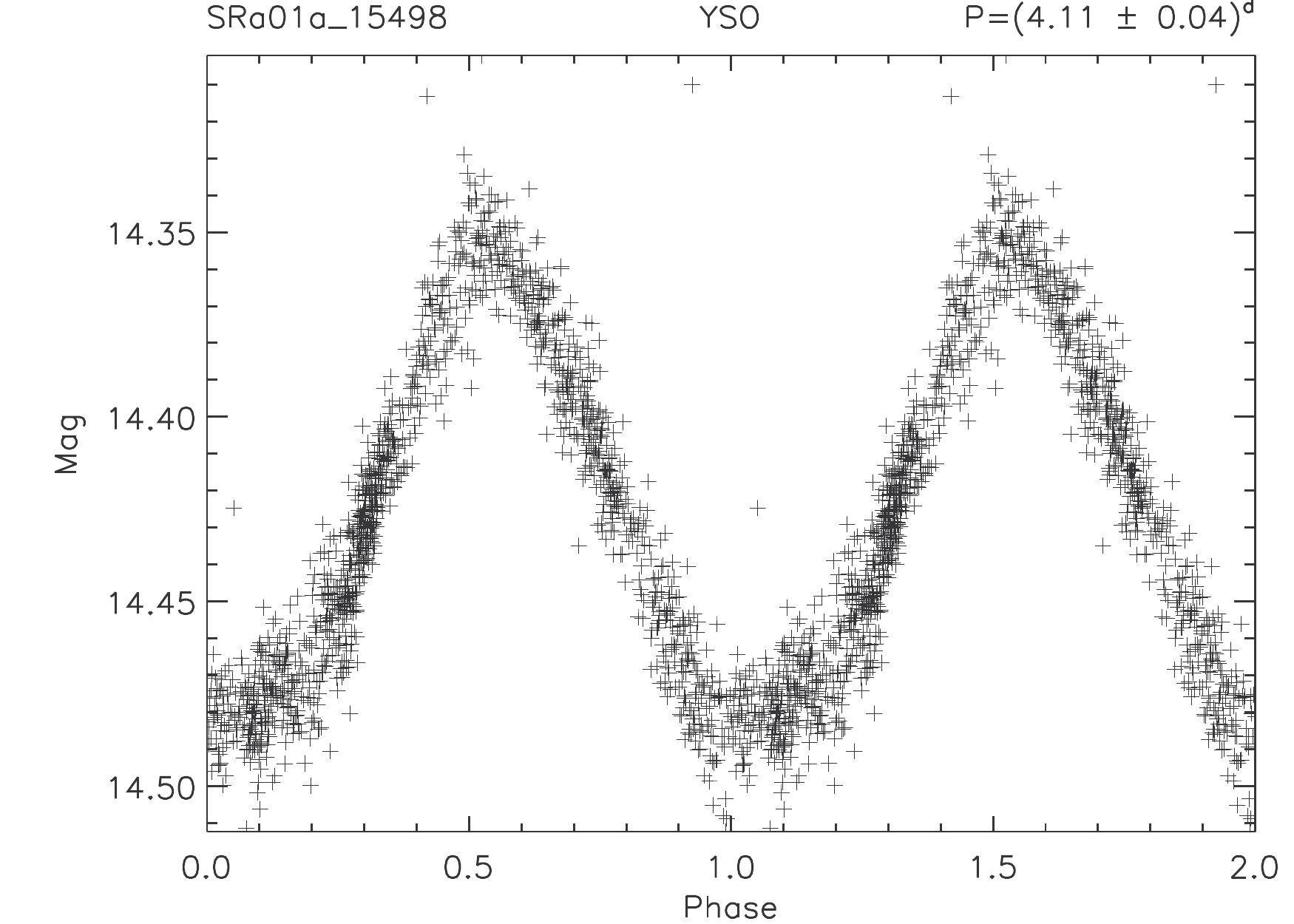}
\includegraphics[width=0.24\textwidth]{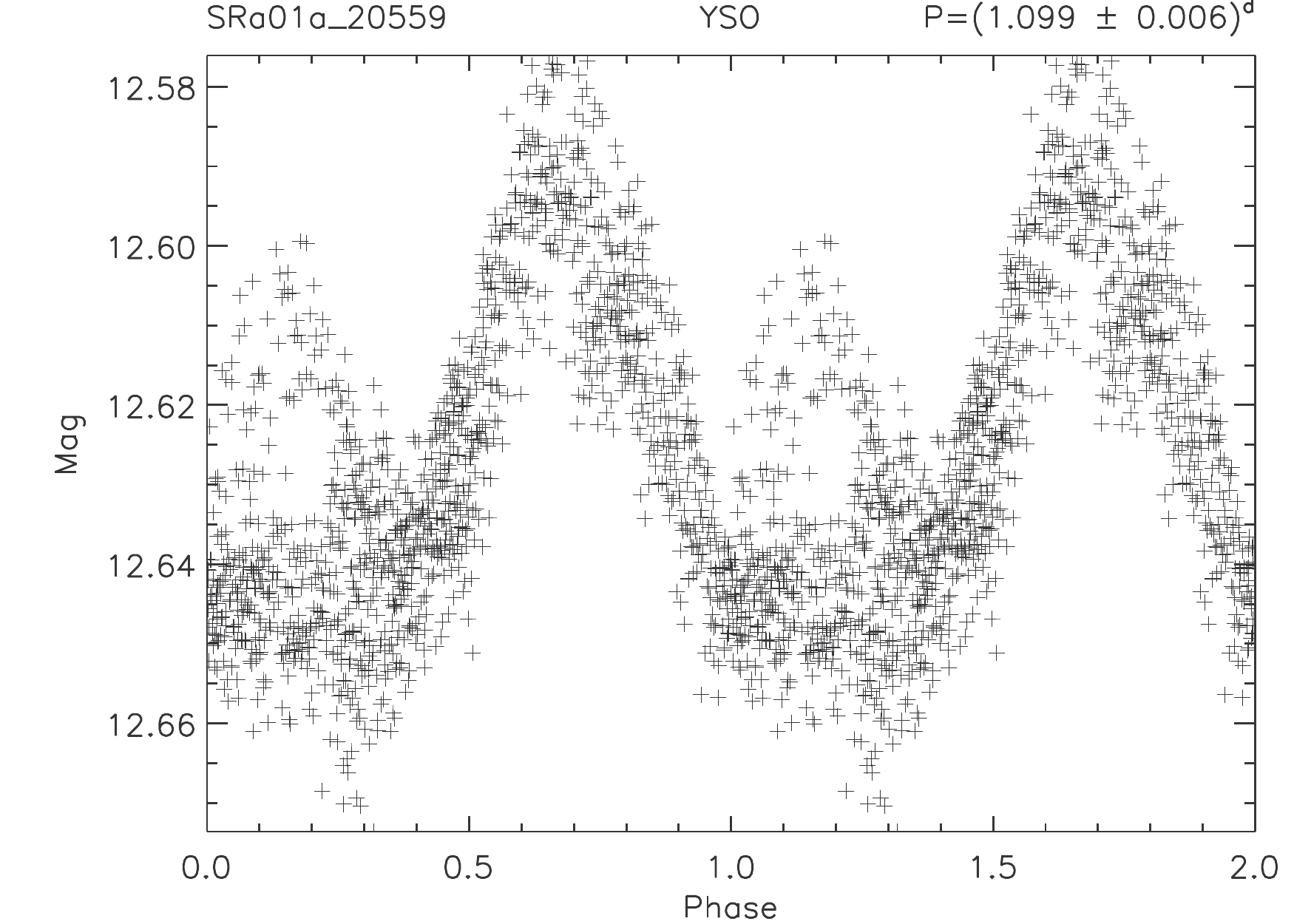}
\includegraphics[width=0.24\textwidth]{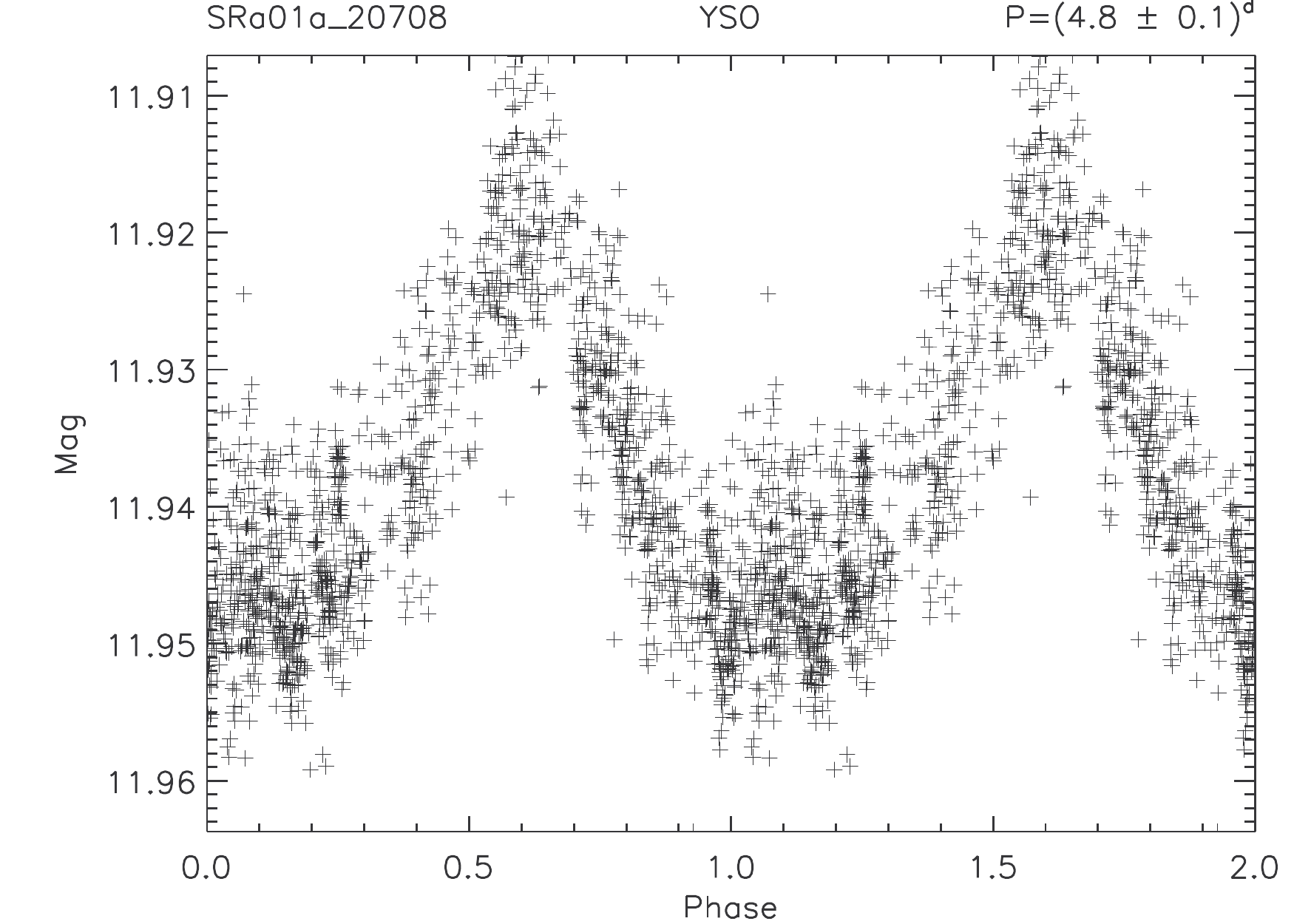}
\includegraphics[width=0.24\textwidth]{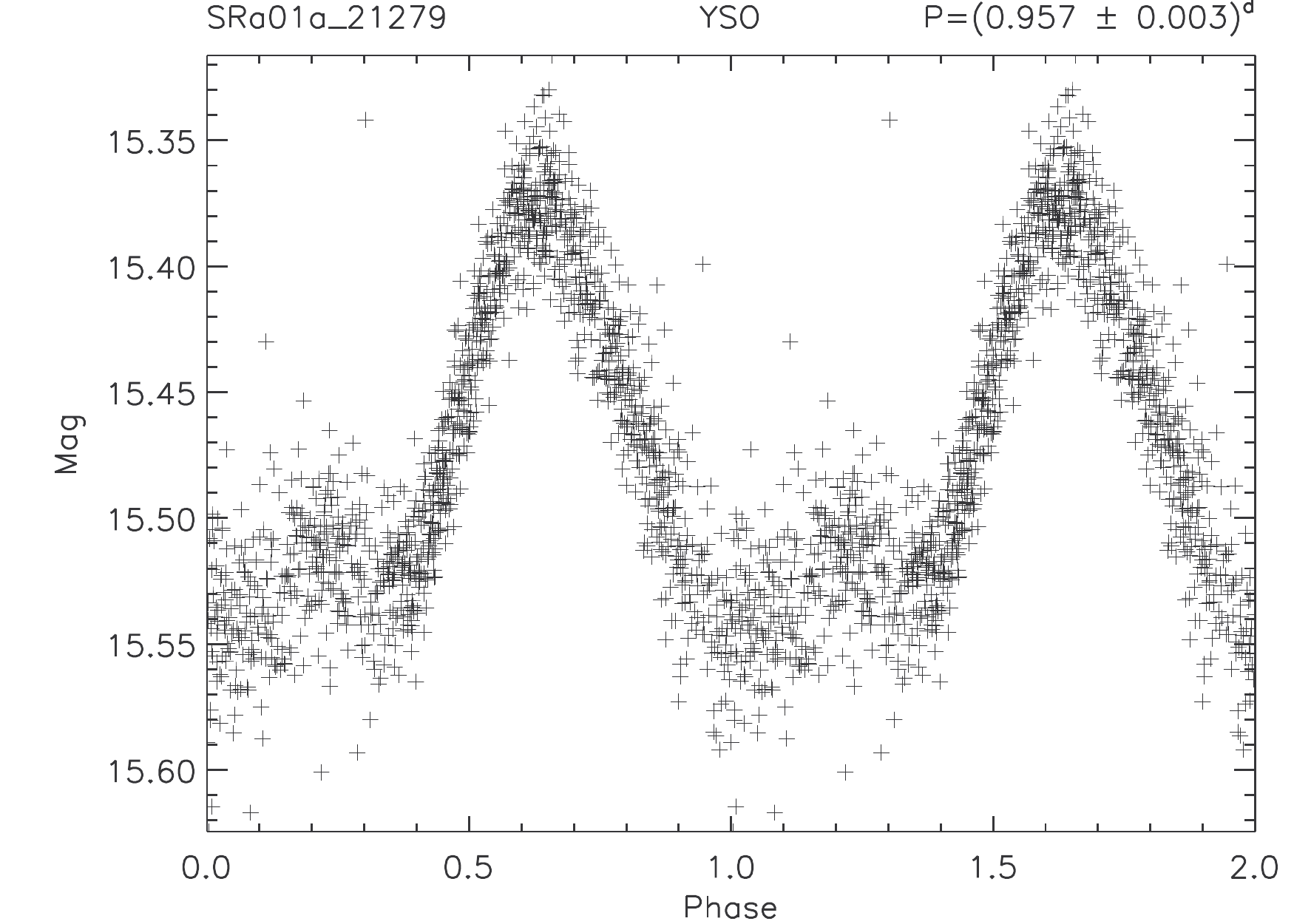}
\includegraphics[width=0.24\textwidth]{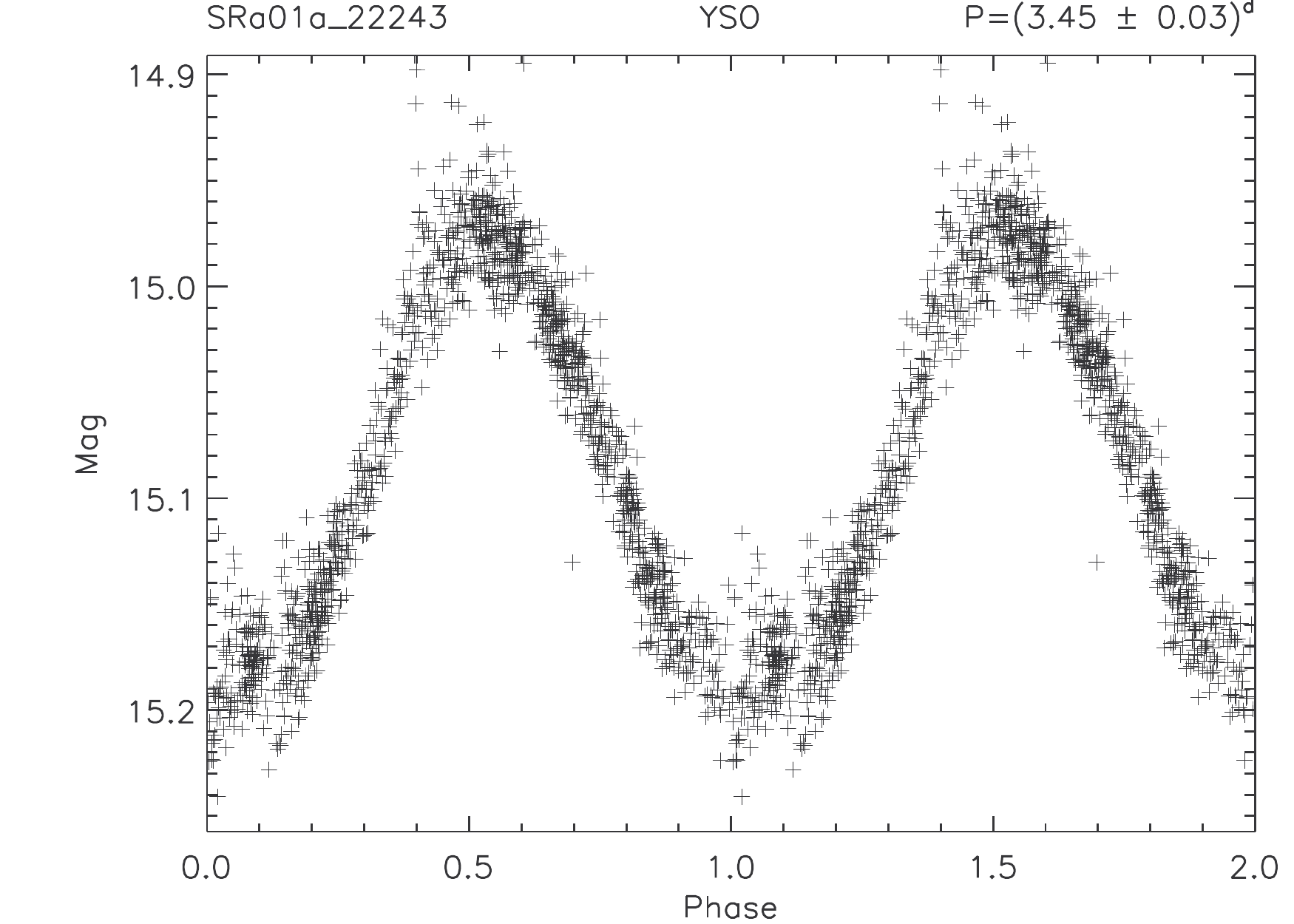}
\includegraphics[width=0.24\textwidth]{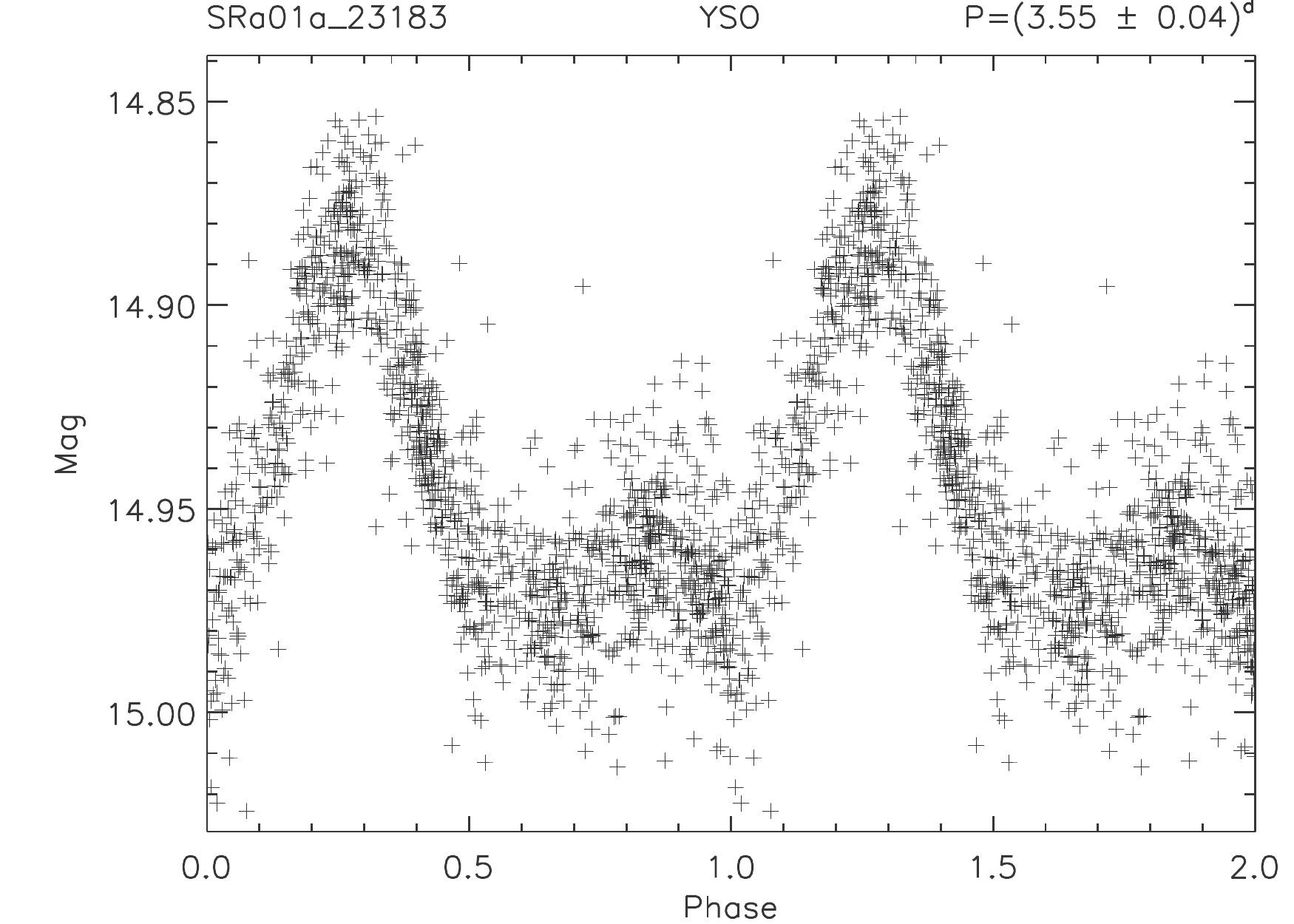}
\includegraphics[width=0.24\textwidth]{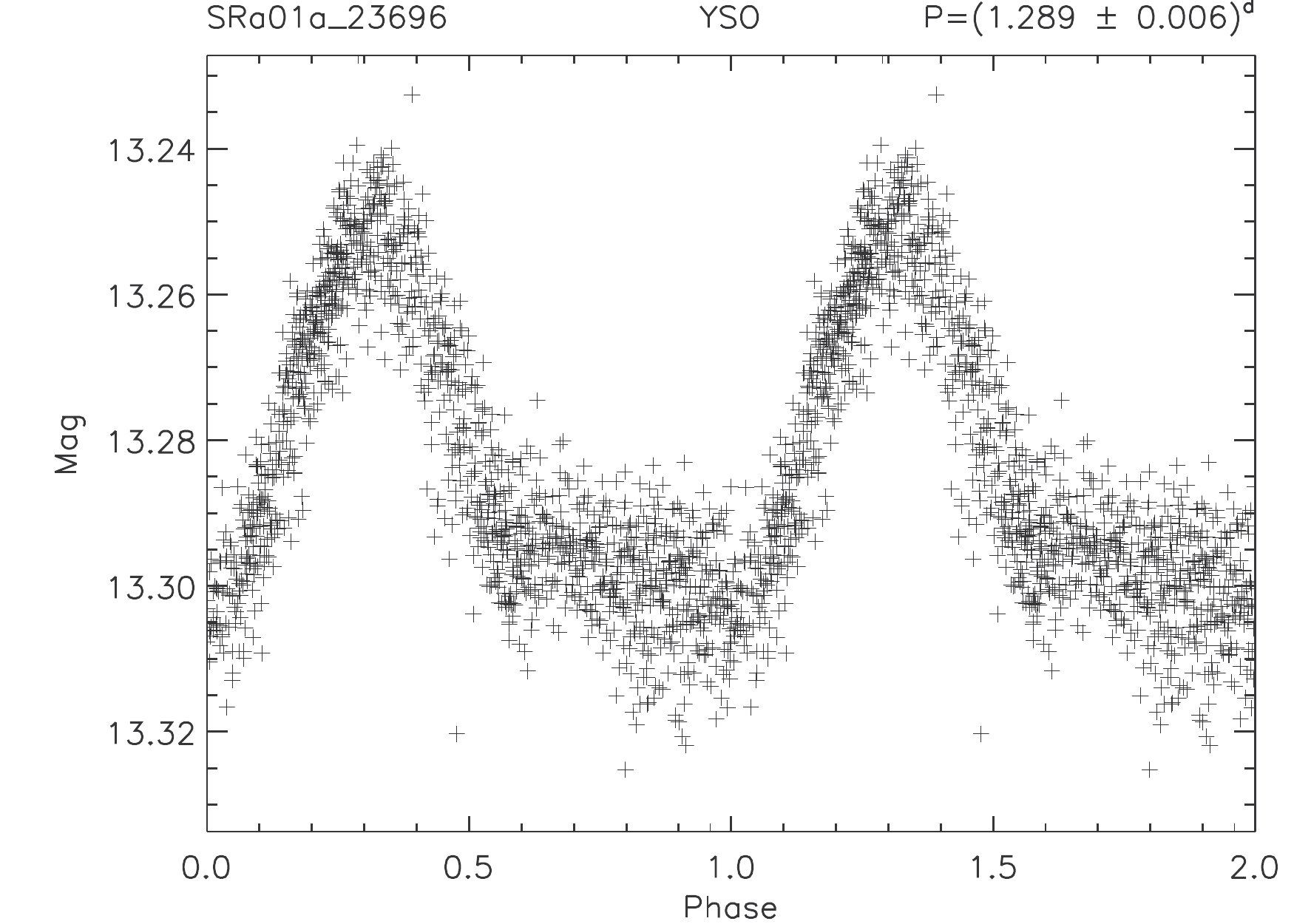}
\includegraphics[width=0.24\textwidth]{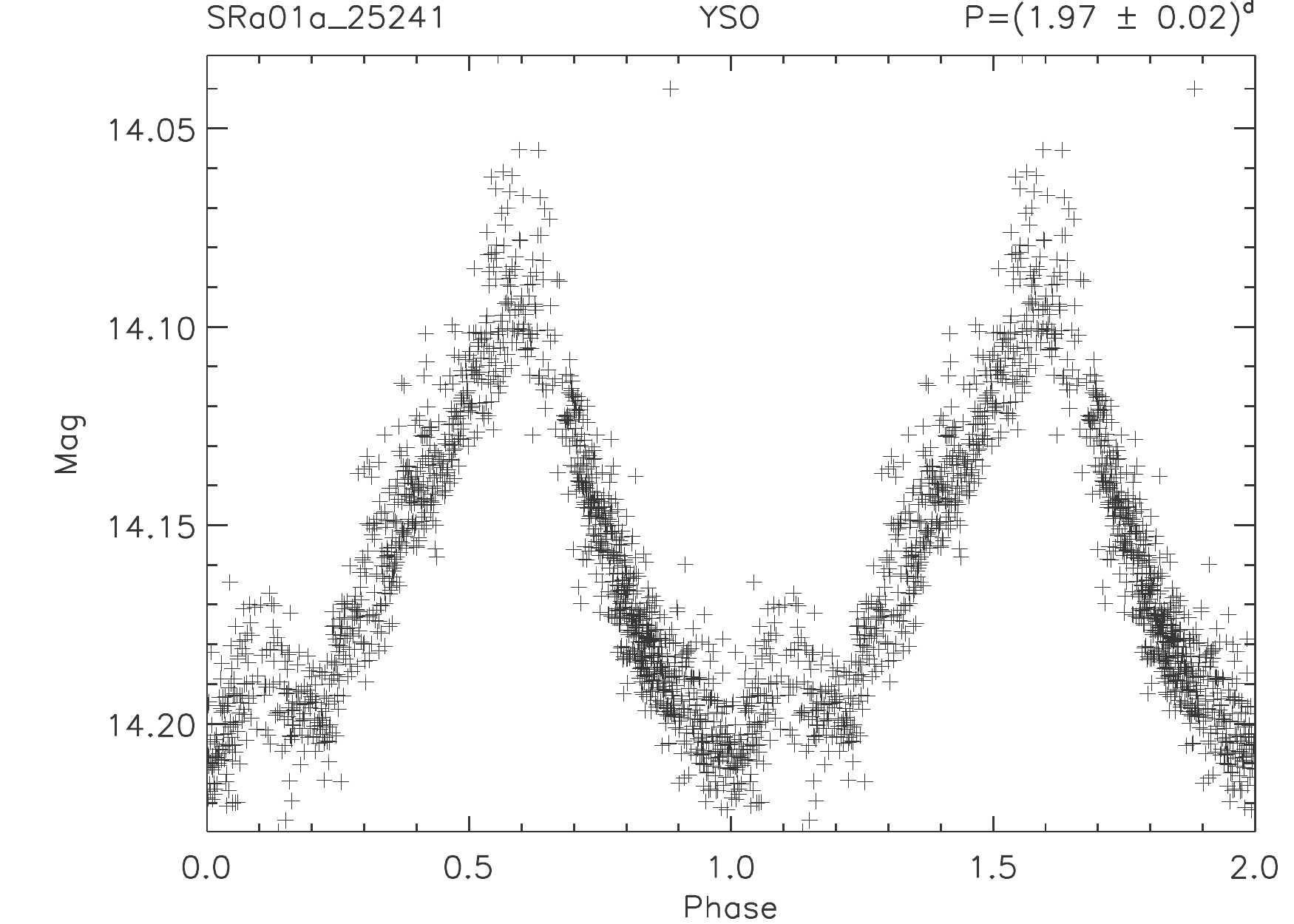}
\includegraphics[width=0.24\textwidth]{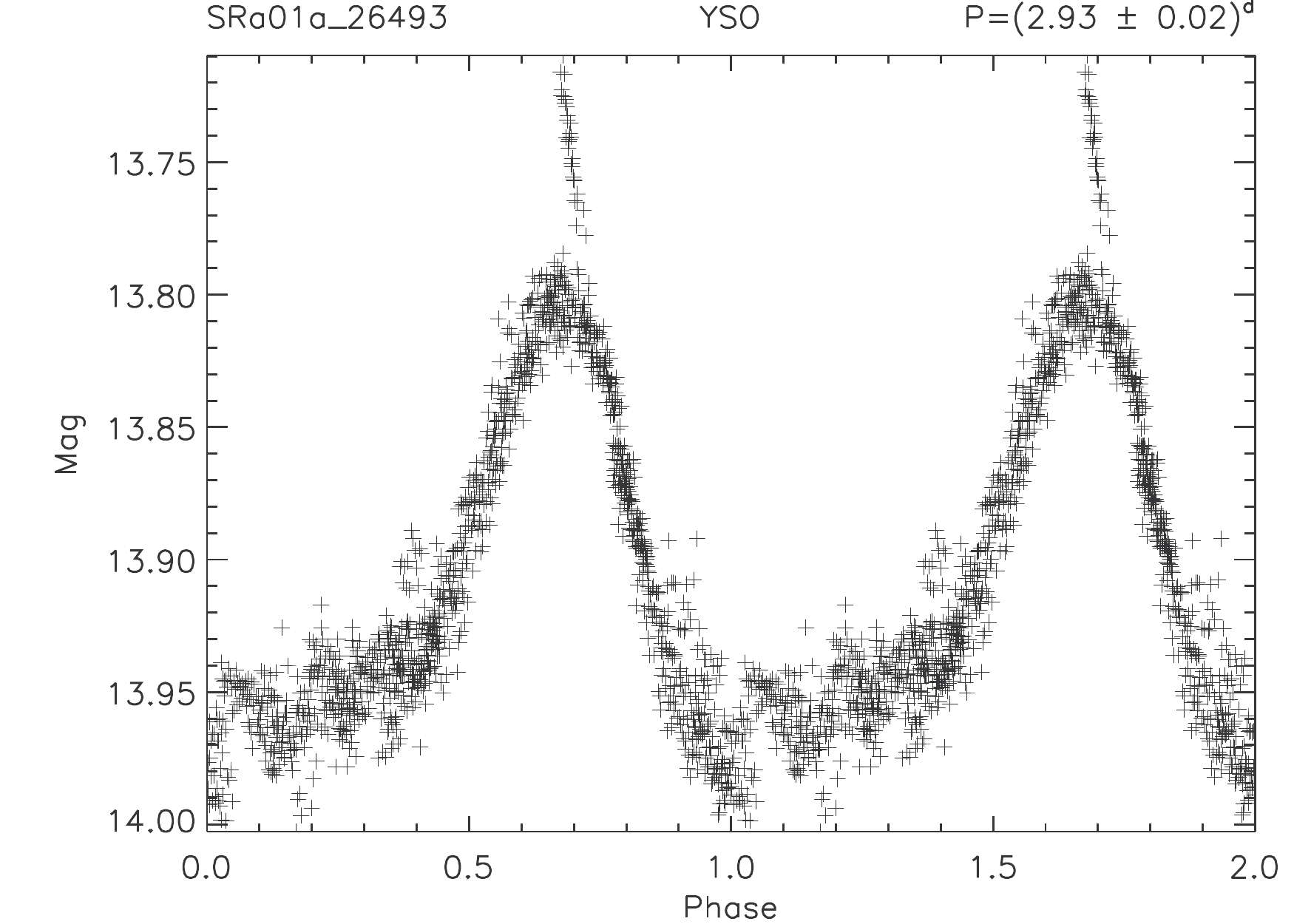}
\includegraphics[width=0.24\textwidth]{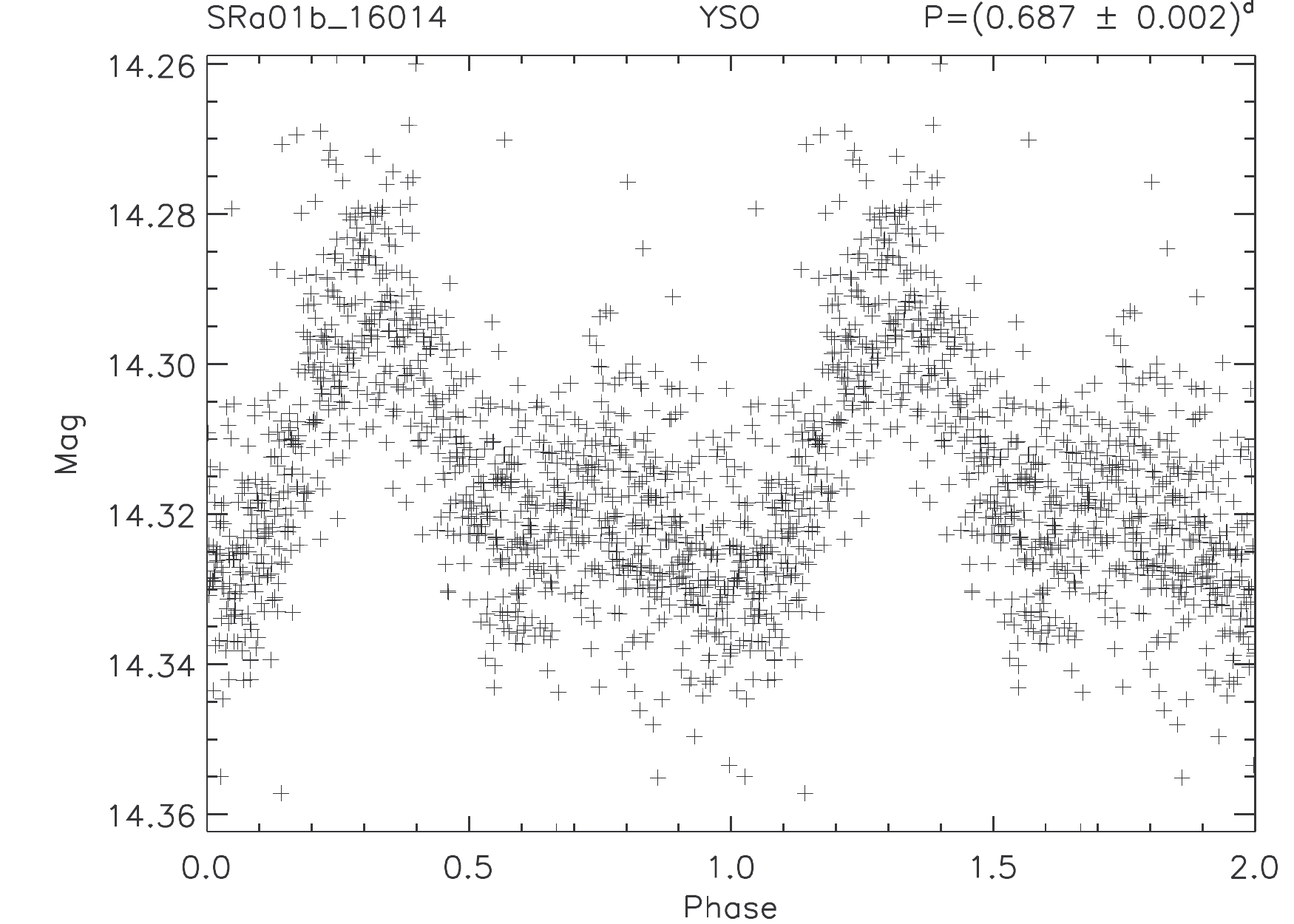}
\caption{Light curves of young stellar objects (YSOs) with rotational modulation.}
\label{fig:yso}
\end{figure*}

Fig. \ref{fig:yso} displays a selected set of unusual light curves.
All show similar features e.g. a sharp bump followed by a constant brightness region.
This phenomenon was stable during the 67 days between the first and last night of our observations.
Only SRa01a\_20559 (top right panel of Fig. \ref{fig:yso}) is an exception, where a second bump
appeared in the constant region. This is consistent with a
rotational modulation with a hot spot either on the star or on
an accretion disk around it. In most cases the bump lasts half of the period which is
explained if we are viewing the systems edge-on.
\citet{rodrigez-ledesma12} found a similar variation in one star in the Orion Nebula Cluster, but
the amplitude of this variation was much larger and the period was much longer than in NGC 2264.
We are not aware of any similar light curves in the literature which may be caused by a particular
spatial distribution of the inclinations in other clusters (c.f. Sect. \ref{EA_excess}).

A detailed analysis of the pre-main sequence stars is out of the scope of this paper.

\section{Notes on selected eclipsing binaries}
\label{models}

\begin{figure*}
\centering
\includegraphics[width=0.32\textwidth]{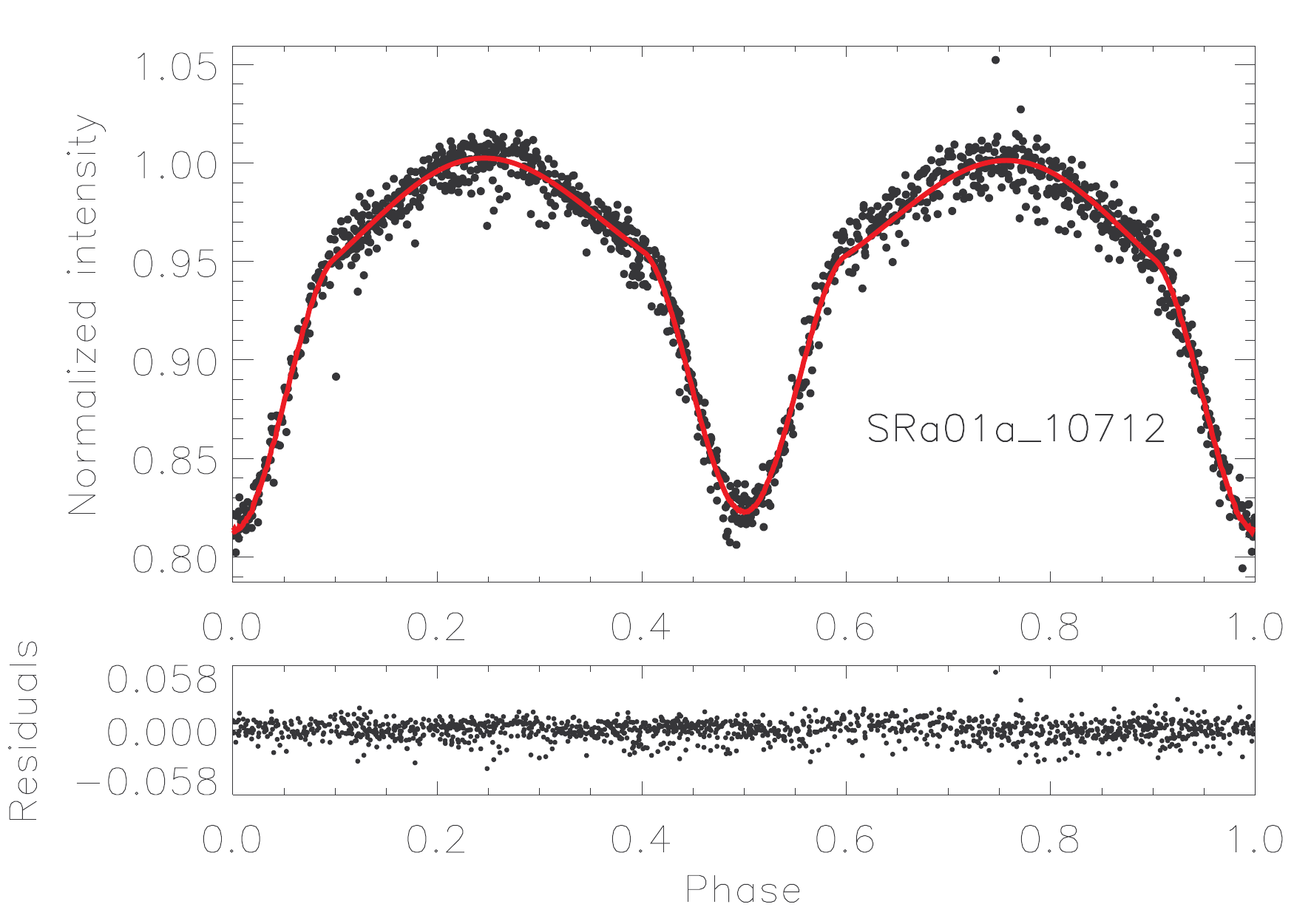}
\includegraphics[width=0.32\textwidth]{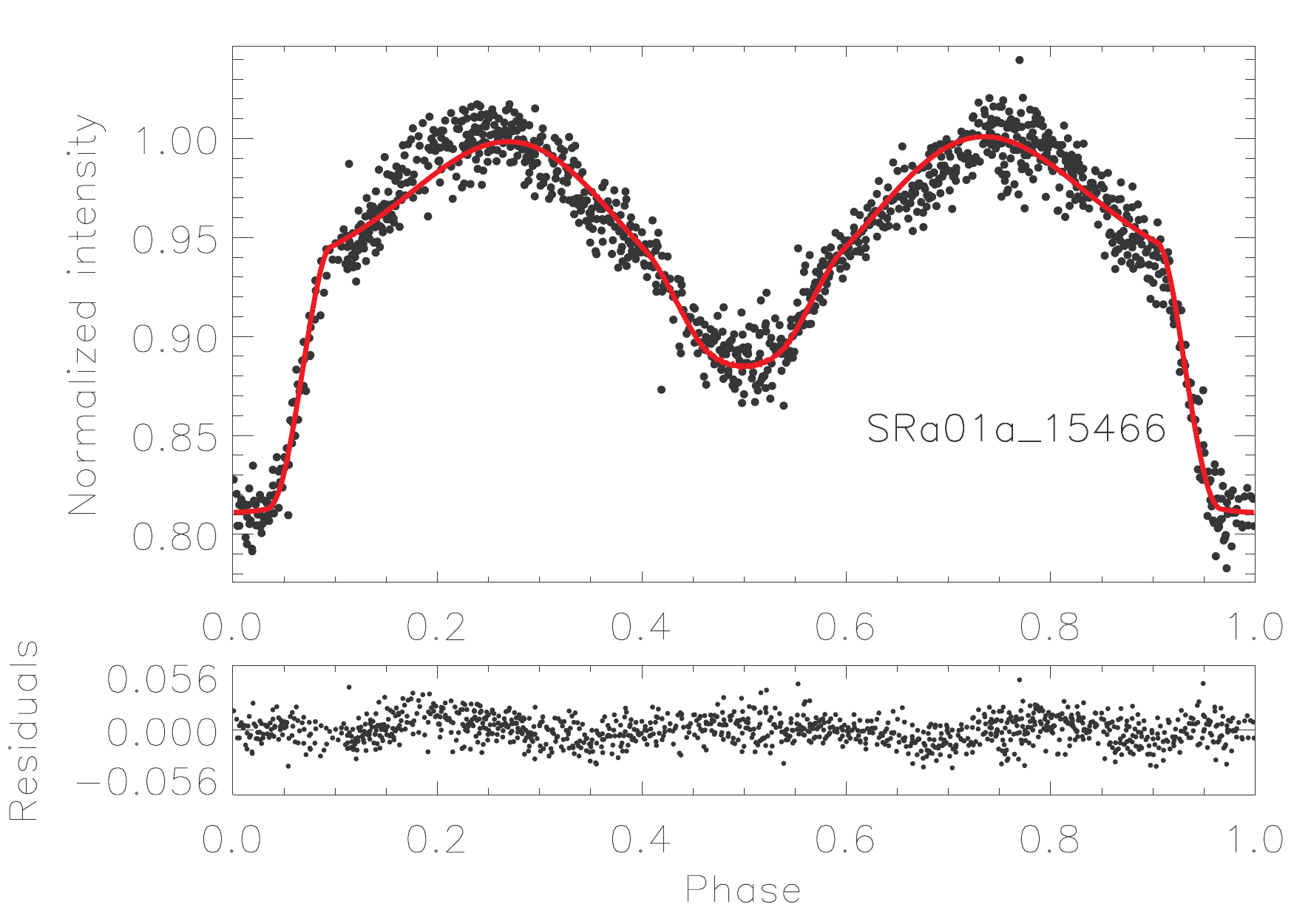}
\includegraphics[width=0.32\textwidth]{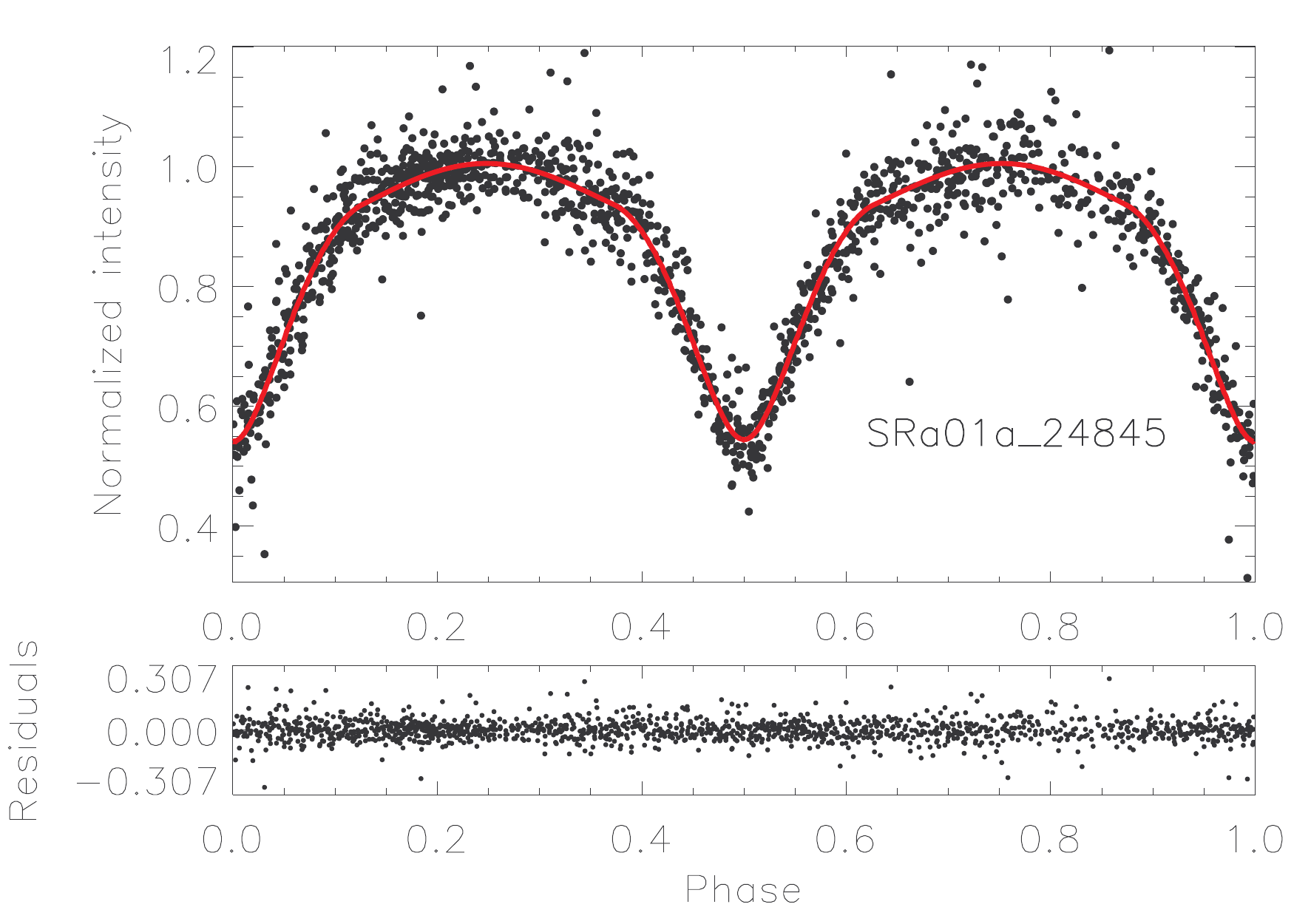}
\includegraphics[width=0.32\textwidth]{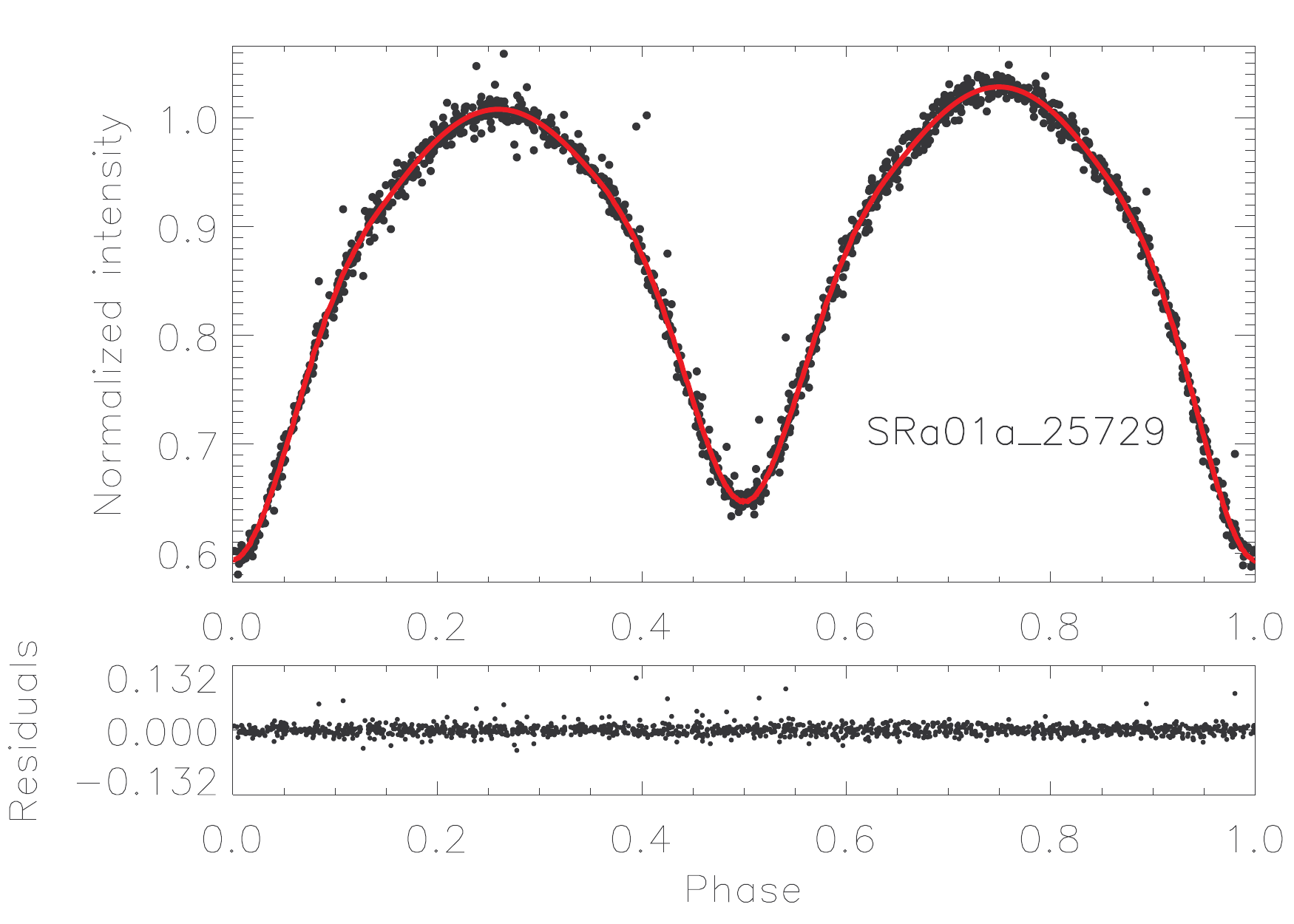}
\includegraphics[width=0.32\textwidth]{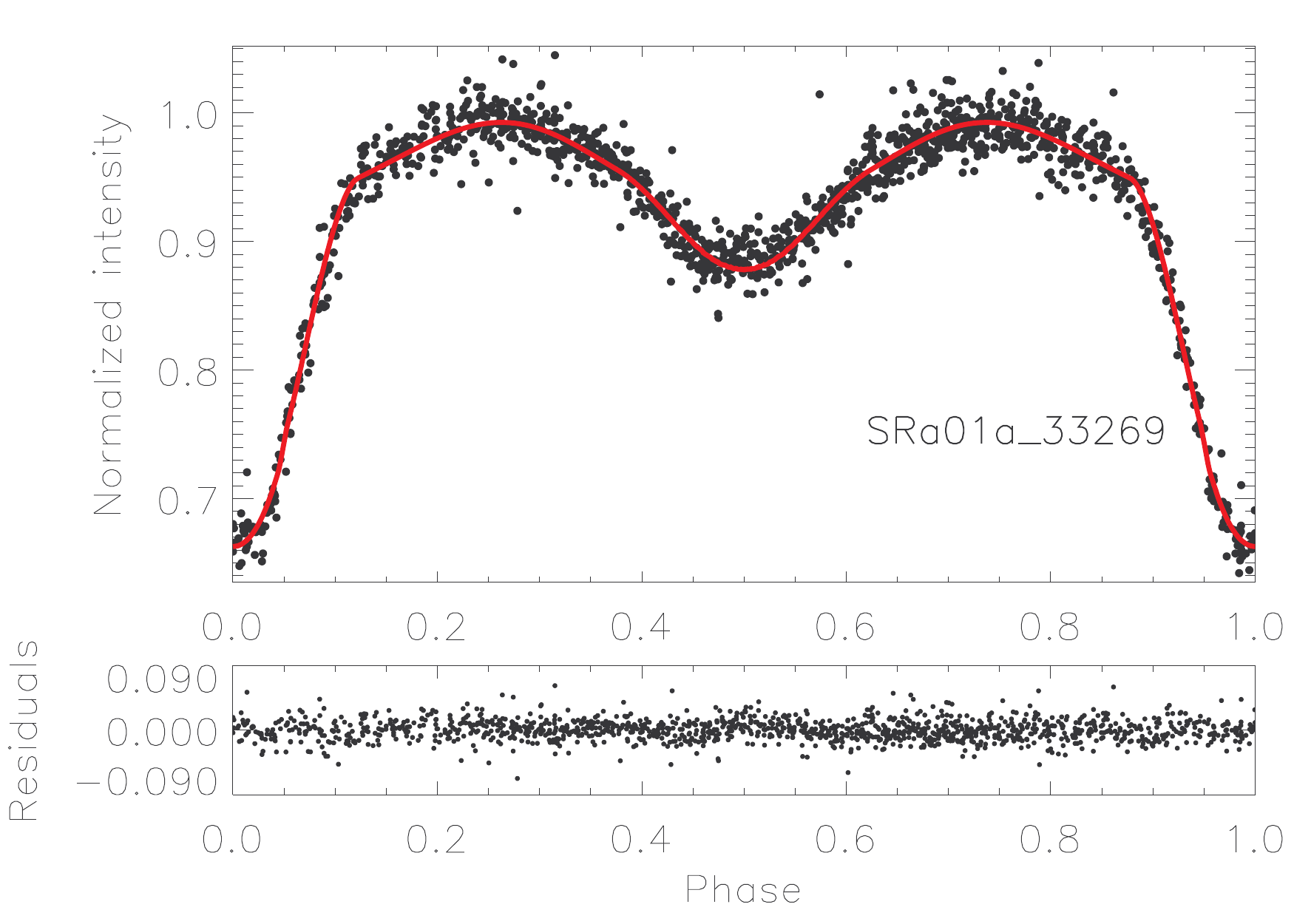}
\includegraphics[width=0.32\textwidth]{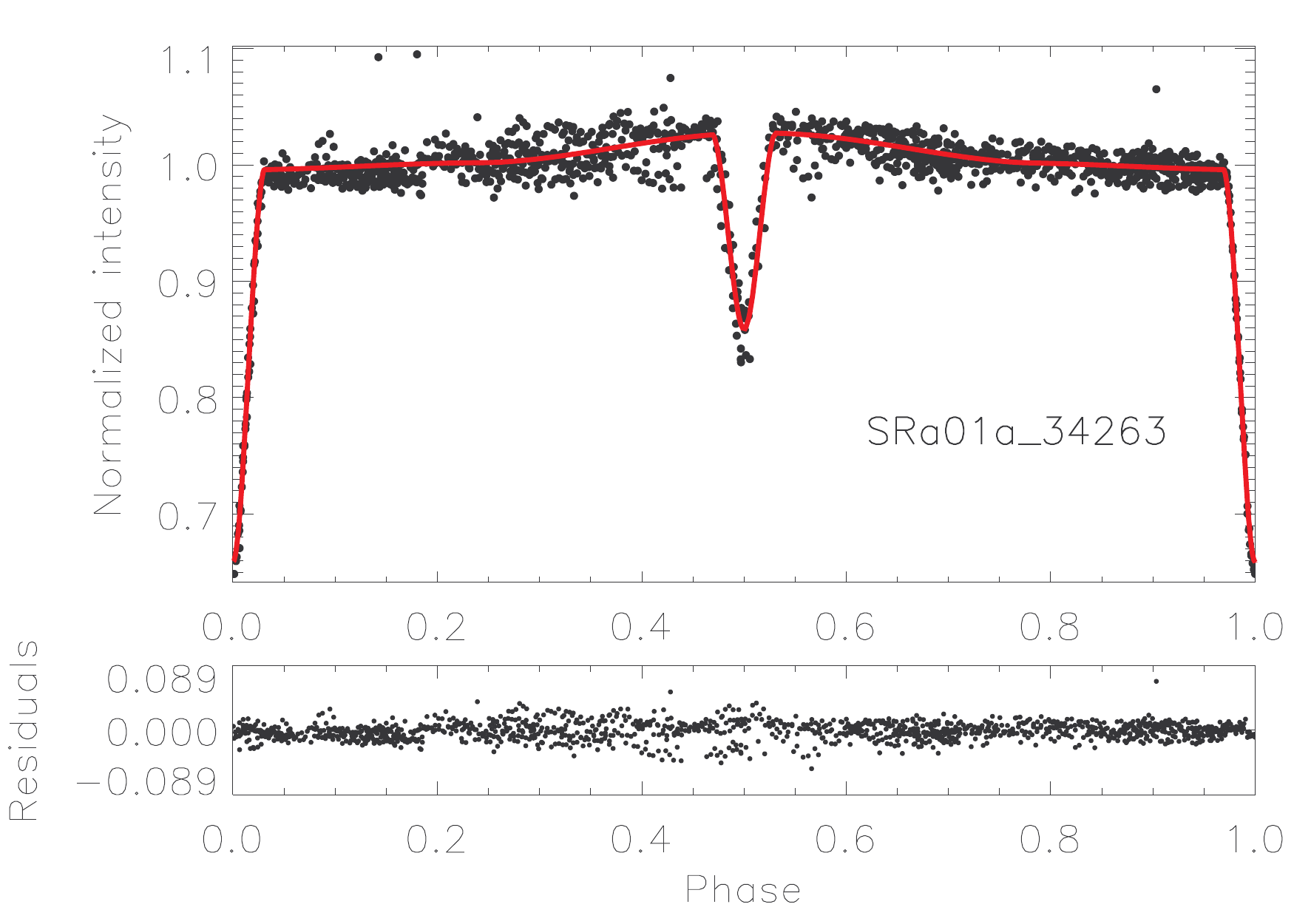}
\includegraphics[width=0.32\textwidth]{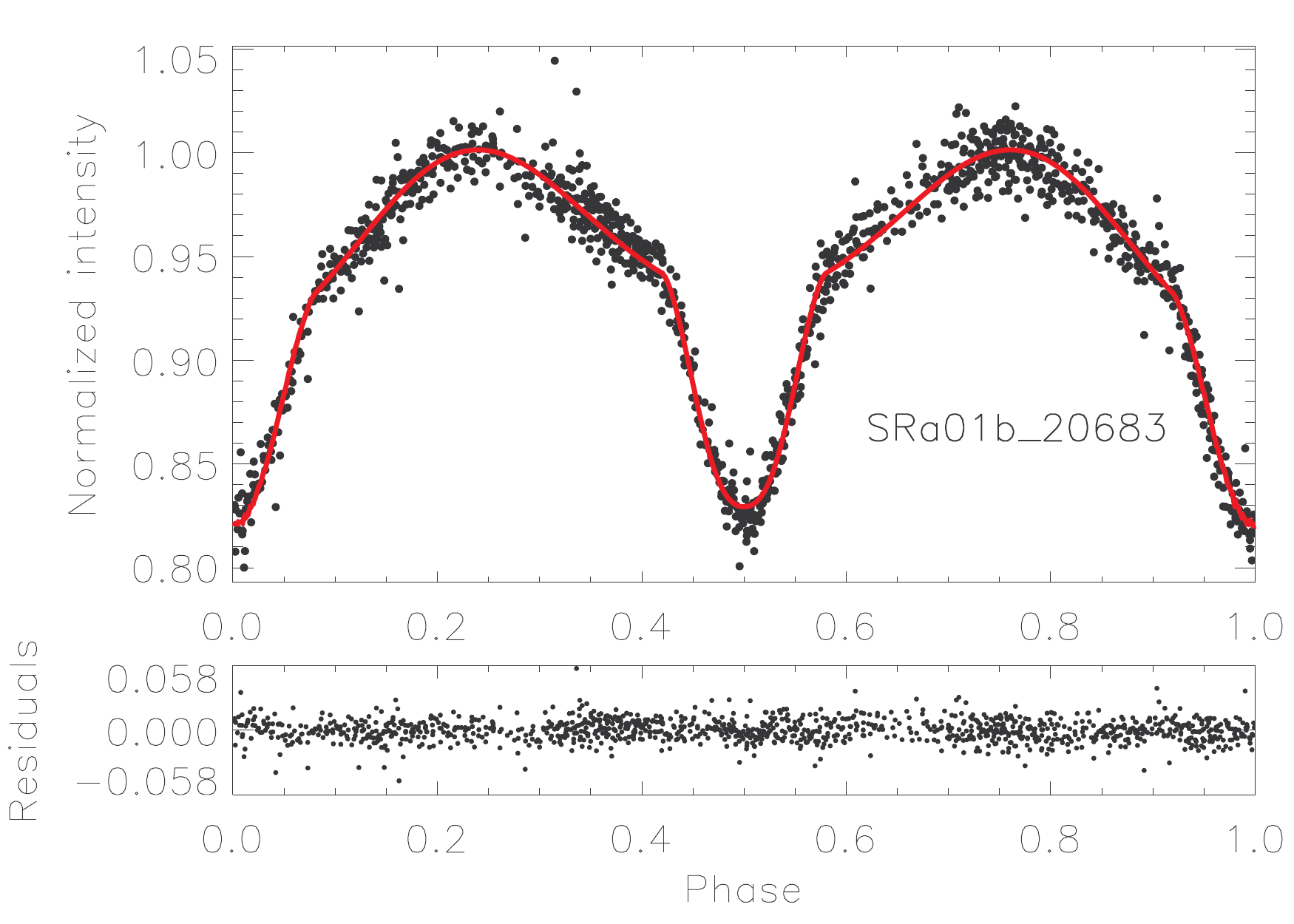}
\includegraphics[width=0.32\textwidth]{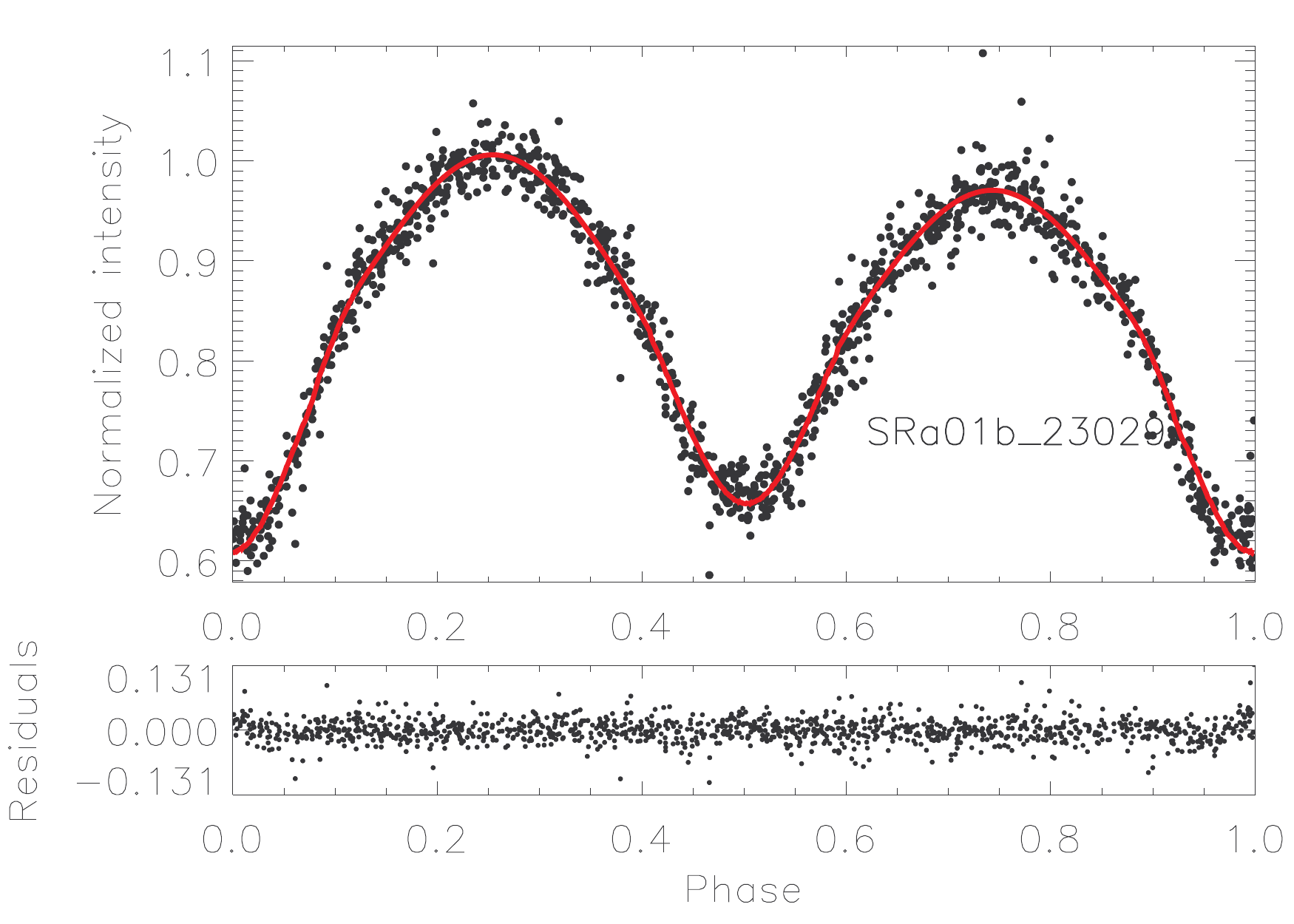}
\includegraphics[width=0.32\textwidth]{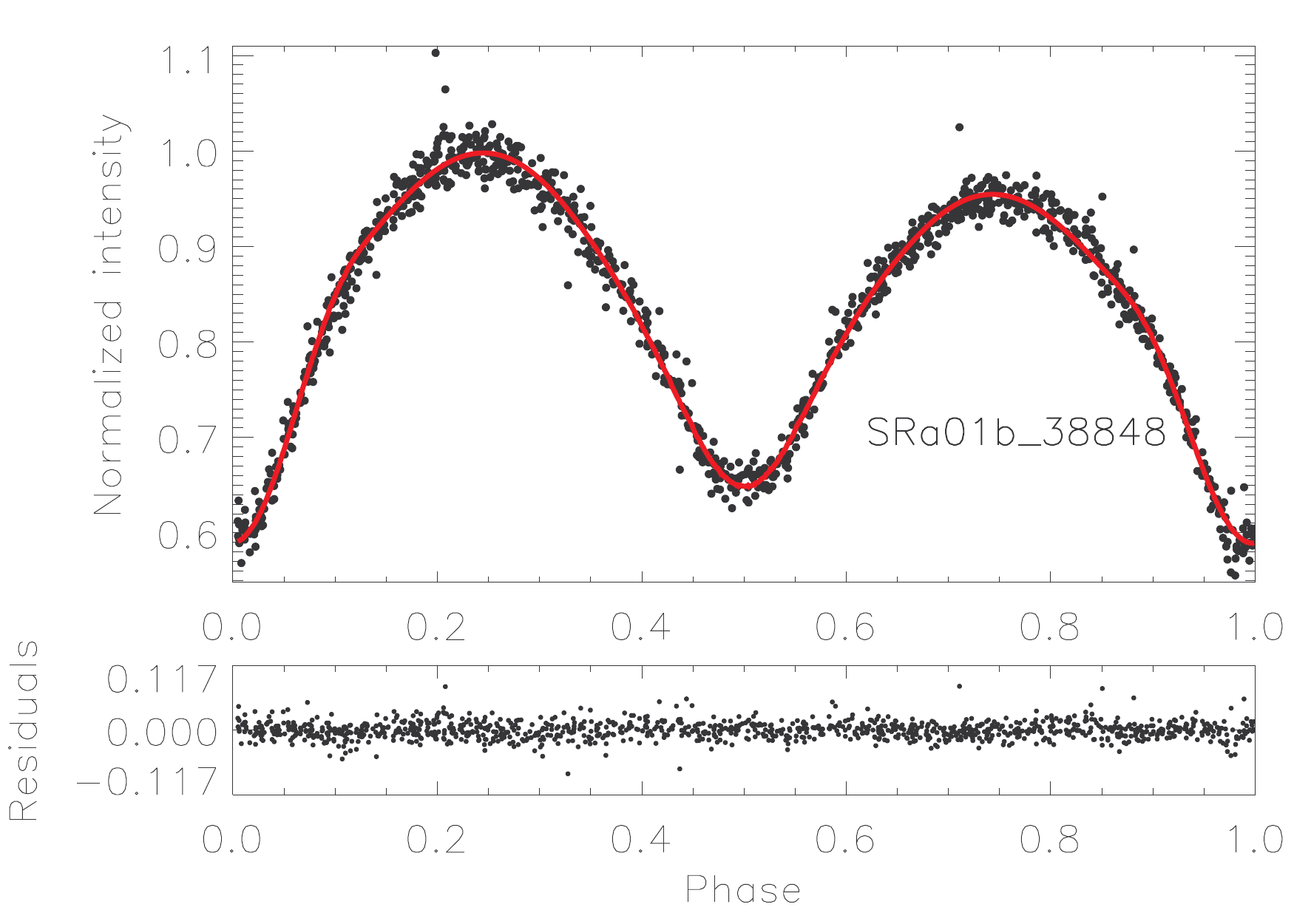}
\includegraphics[width=0.32\textwidth]{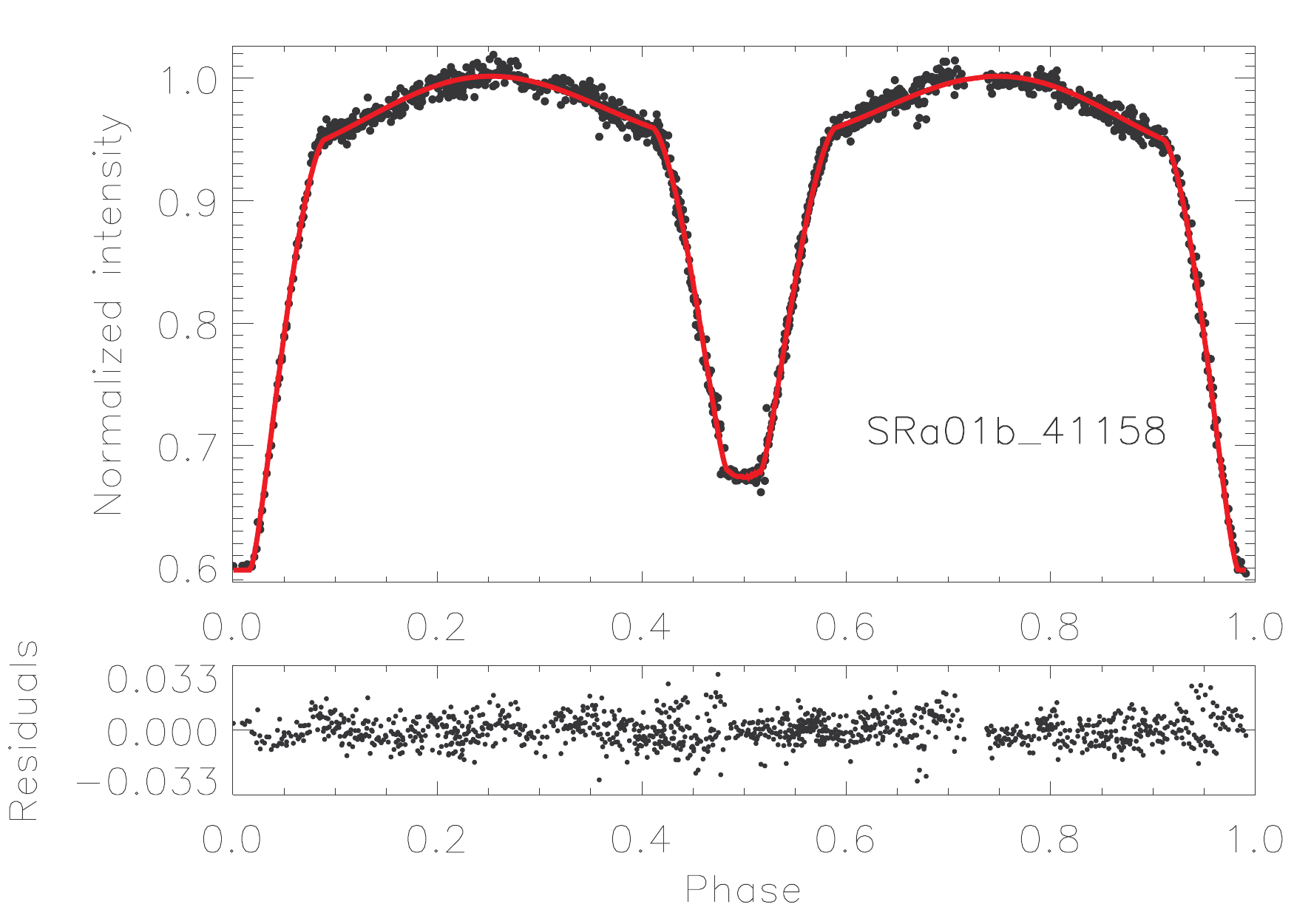}
\includegraphics[width=0.32\textwidth]{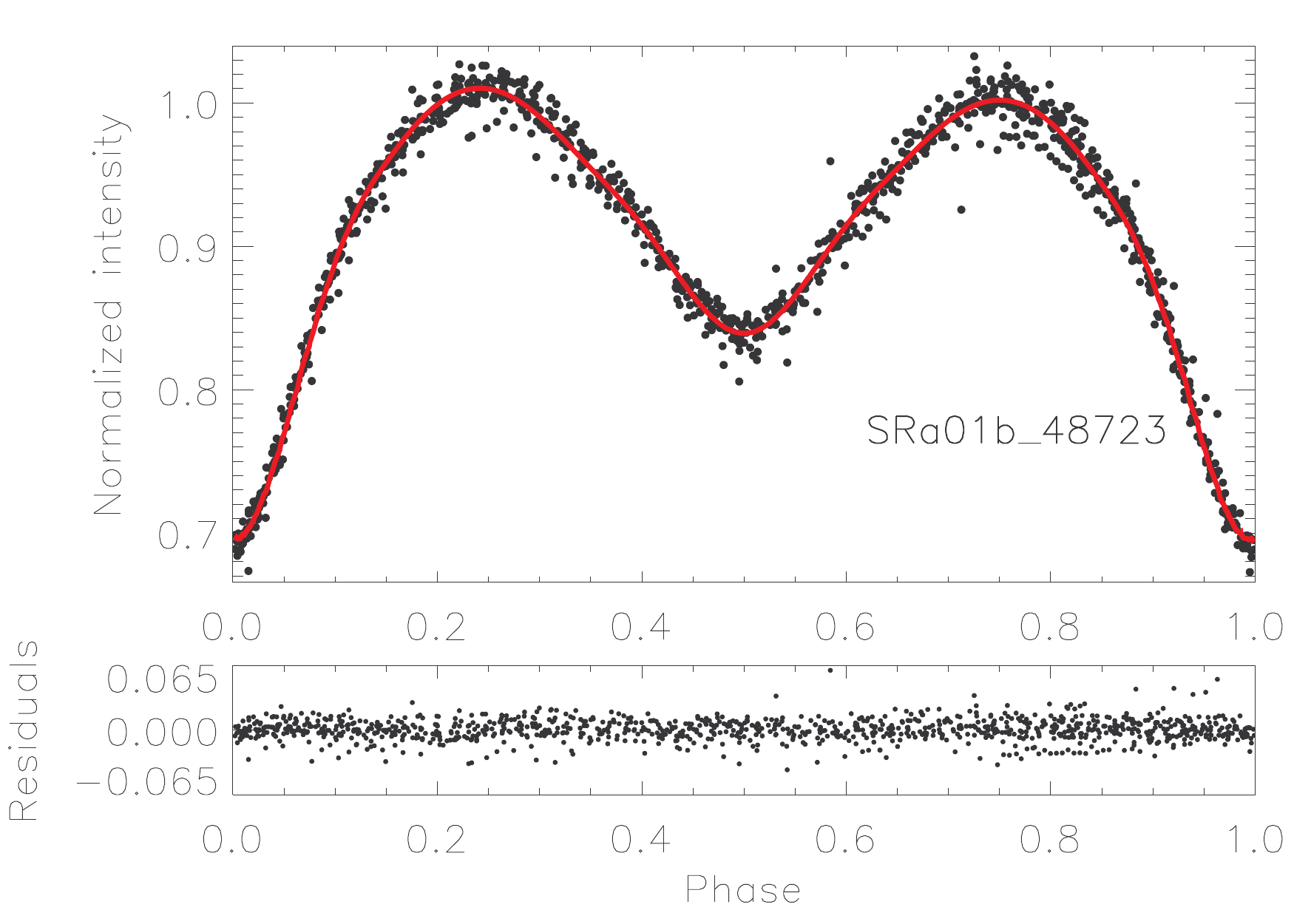}
\caption{Results of the light curve modeling of the selected eclipsing binaries.
Diamonds represent the raw observations, while the solid lines (colored red in the online version)
are the fits. The lower panels show the residuals.}
\label{fig:lcs_modelled}
\end{figure*}

Since some individual eclipsing binaries show interesting features,
we studied them in more detail by modeling their light curve
with our own code as described by \citet{csizmadia09} and Csizmadia~et~al.~(in prep.).
This code is based on Roche-geometry, and includes all proximity effects of binaries.

Since the variables presented here are either poorly studied
or are new discoveries, the literature has sometimes only limited information on them;
in particular there is no radial velocity curve.
Therefore we used the $J-K$ colour indices of the selected eclipsing binaries
from the 2MASS Point Source Catalogue \citep{skrutskie06} and estimated
the temperatures by assuming that both interstellar
extinction and the contribution of the secondary to the colour
index is negligible. We estimate that the derived temperatures have an error
of about $\pm250$K. We leave a detailed modeling for other studies, which are
able to utilize higher quality multi-colour photometry with bigger telescopes
and spectroscopic data.

\begin{table*}
\centering
\caption{Fitted parameters of the modeled binary systems. The temperature factor of the
spots is defined as $T_{spot}/T_{star}$.}
\label{tab:model_parameters}
\small
\begin{tabular*}{0.85\textwidth}{lccccccccccc}
\tableline
\tableline
BEST ID &  SRa01a\_10712  & SRa01a\_15466  & SRa01a\_24845  & SRa01a\_25729 \\
\tableline
Mass ratio $q$                       & $0.37\pm0.05$   & $6.37\pm0.22$   & $0.75\pm0.12$   & $0.98\pm0.05$ \\
Inclination $i$ ($^{\circ}$)         & $67.8\pm0.2   $ & $69.2\pm0.2   $ & $82.2\pm0.6   $ & $75.2\pm0.3   $ \\
Fill-out factor $f_1$                & $-1.43\pm0.10 $ & $-11.05\pm0.58$ & $-0.43\pm0.17 $ & $0.013\pm0.010$ \\
Fill-out factor $f_2$                & $-1.04\pm0.05 $ & $-0.47\pm0.03 $ & $-0.41\pm0.14 $ & $f_2 = f_1    $ \\
Temperature $T_1$ (K)                & $4150$ (fixed)  & $4400$ (fixed)  & $4060$ (fixed)  & $4940$ (fixed)  \\
Temperature $T_2$ (K)                & $4060\pm8     $ & $3294\pm16    $ & $4062\pm13    $ & $4730\pm20    $ \\
{\it Spot 1} &&&&\\
Colatitude $\phi_{1}$ ($^{\circ}$)   &  $119\pm14 $    & $134\pm38 $     & --              & $137\pm23 $ \\
Longitude $\lambda_{1}$ ($^{\circ}$) &  $226\pm102$    & $225\pm111$     & --              & $306\pm10 $ \\
Diameter $d_{1}$ ($^{\circ}$)        &  $1.9\pm2.9  $  & $22.4\pm8.4 $   & --              & $25\pm10$ \\
Temperature factor                   & $1.46\pm0.37  $ & $0.85\pm0.45  $ & --              & $0.73\pm0.19  $ \\
{\it Spot 2} &&&&\\
Colatitude $\phi$                    & --              &  $137\pm40 $    & --              & --              \\
Longitude $\lambda$                  & --              &  $113\pm122$    & --              & --              \\
Diameter $d$                         & --              &   $23\pm27$     & --              & --              \\
Temperature factor                   & --              & $0.09\pm0.39  $ & --              & --              \\
\tableline
$\chi^2$                             & $2.071        $ & $1.241        $ & $0.661        $ & $1.233        $ \\
\tableline
\tableline
BEST ID &  SRa01a\_33269  & SRa01a\_34263  & SRa01b\_20683  & SRa01b\_23029  \\
\tableline
Mass ratio $q$                       & $10.96\pm0.47$  & $2.10\pm0.77$   & $0.43\pm0.14$   & $0.65\pm0.09$   \\
Inclination $i$ ($^{\circ}$)         & $64.5\pm0.4   $ & $85.6\pm0.6   $ & $74.4\pm2.8   $ & $69.4\pm0.7   $ \\
Fill-out factor $f_1$                & $-1.54\pm0.09 $ & $-9.0\pm2.5 $   & $-1.03\pm0.03 $ & $0.30\pm0.05  $ \\
Fill-out factor $f_2$                & $-0.48\pm0.14 $ & $-30.1\pm5.3$   & $-6.6\pm2.5 $   & $f_2 = f_1$     \\
Temperature $T_1$ (K)                & $3570$ (fixed)  & $4270$ (fixed)  & $4050$ (fixed)  & $5000$ (fixed)  \\
Temperature $T_2$ (K)                & $2707\pm9     $ & $3828\pm44    $ & $4010\pm32    $ & $4731\pm101   $ \\
{\it Spot 1} &&&&\\
Colatitude $\phi$ ($^{\circ}$)       & $39\pm50  $     & $60\pm55      $ & --              & $36\pm34    $   \\
Longitude $\lambda$ ($^{\circ}$)     & $0\pm43   $     & $175\pm10 $     & --              & $81\pm20  $     \\
Diameter $d$ ($^{\circ}$)            & $2.9\pm1.7  $   & $5.3\pm5.7  $   & --              & $19.5\pm6.4 $   \\
Temperature factor                   & $1.6\pm0.6  $   & $1.4\pm0.3  $   & --              & $0.12\pm0.18  $ \\
{\it Spot 2} &&&&\\
Colatitude $\phi$ ($^{\circ}$)       & --              & --              & --              & $125\pm42 $ \\
Longitude $\lambda$ ($^{\circ}$)     & --              & --              & --              & $1\pm94   $ \\
Diameter $d$ ($^{\circ}$)            & --              & --              & --              & $102\pm11$\\
Temperature factor                   & --              & --              & --              & $1.81\pm0.66  $ \\
\tableline
$\chi^2$                             & $0.795        $ & $2.753        $ & $0.800        $ & $0.728        $ \\
\tableline
\tableline
BEST ID &  SRa01b\_38848  & SRa01b\_41158  & SRa01b\_48723 \\
\tableline
Mass ratio $q$                       & $3.71\pm0.55$   & $6.26\pm0.18$   & $0.456\pm0.009$ \\
Inclination $i$ ($^{\circ}$)         & $73.4\pm1.1   $ & $88.2\pm0.9   $ & $67.7\pm0.1   $ \\
Fill-out factor $f_1$                & $0.31\pm0.08  $ & $-0.60\pm0.04 $ & $0 (fixed)    $ \\
Fill-out factor $f_2$                & $f_2 = f_1$     & $-11.9\pm0.5$   & $-0.03\pm0.02 $ \\
Temperature $T_1$ (K)                & $5300$ (fixed)  & $5300$ (fixed)  & $5200$ (fixed)  \\
Temperature $T_2$ (K)                & $4901\pm76    $ & $4898\pm45    $ & $4059\pm22    $ \\
{\it Spot 1} &&&&\\
Colatitude $\phi_{1}$ ($^{\circ}$)   & $50\pm20  $     & --              & $133\pm3.0  $ \\
Longitude $\lambda_{1}$ ($^{\circ}$) & $332\pm10 $     & --              & $46.9\pm6.2   $ \\
Diameter $d_{1}$ ($^{\circ}$)        & $36\pm12$       & --              & $15.2\pm1.5 $ \\
Temperature factor                   & $1.15\pm0.07  $ & --              & $0.57\pm0.21  $ \\
\tableline
$\chi^2$                             & $0.875        $ & $1.709        $ & $2.817        $ \\
\tableline
\end{tabular*}
\end{table*}

For those systems where we could not find $J-K$ data we adopted an effective
temperature of 6000\,K for the primaries.
These cases require additional photometry or spectroscopy.
The linear bolometric and quadratic $R$-band limb darkening
coefficients were taken from \citet{vanhamme93} and \citet{claret11},
respectively. The $R$-band is the closest filter to our white light observations.
For $T_{\rm eff} < 6000$\,K we set the gravity darkening exponents and albedos to $g=0.32$ and $A=0.5$
and for $T_{\rm eff} < 6000$\,K we set $g=1.0$ and $A=1.0$ \citep[cf.][]{lucy67,rucinski69}.

The free parameters are: mass ratio, fill-out factors of the components,
inclination, effective surface temperature of the secondary,
epoch, and height of the maximum brightness. In some cases we added a
stellar spot to one or both of the components in order to achieve a better fit.
Then we adjusted the size and the temperature of the spot and its astrographic
position (stellar co-latitude and longitude). The temperature factor of the spots
is defined as the ratio of the temperature ratio of the spot and the star
(temperature factor = $T_{\rm spot}/T_{\rm star}$).

The results are summarized in Table \ref{tab:model_parameters} and Figure \ref{fig:lcs_modelled}.
The individual systems are discussed below.

\subsubsection{$\rm{SRa01a\_10712}$: a possible lower main sequence binary}

This system is a detached binary system with $P=1.43$ days orbital period
exhibiting a strong ellipsoidal variation. The
components are late types ($T_1 = 4150$\,K and $T_2 = 4060$\,K)
suggesting spectral types of K8-M1 for both components. Since the orbital
period is short, the objects are most likely dwarfs.
The brightness of the system ($R=13.4$ at maximum) will allow a more
detailed multicolour photometric and spectroscopic as well as radial velocity
study to obtain more precise and absolute dimensions of the system; this
improves our knowledge of the lower main sequence.

\subsubsection{$\rm{SRa01a\_15466}$}

This 14 magnitude system shows a remarkable scatter in its light curve
which varies from cycle to cycle. Spotted
solutions yield significantly worse fits than an unspotted light curve model.
However, if the star is spotted and the spots evolve fast, then a global spot
model describing the phase-folded light curve is probably not appropriate. The
system has a very high inclination and total eclipses which help to constrain
the mass ratio better \citep{rucinski73}.

\subsubsection{$\rm{SRa01a\_24845}$}

This system is similar to SRa01a\_10712
as concerns the shape of the light curve, the orbital period
(1.51 vs. 1.43 days), and the surface temperatures ($T_1 = 4060$\,K, $T_2 = 4061$\,K and $T_1 = 4150$\,K, $T_2 = 4060$\,K, respectively).
However, the fill-out factors are almost equal in
this system in contrary to SRa01a\_10712. Likewise, the mass ratio is much higher: the system
consists of almost two equal-sized stars.
Whether the components lie before the onset of a mass transfer or whether they will never fill out
their Roche-lobes can only be answered by a radial velocity curve.

This system, which is probably composed of low-mass objects ($J-K= \mathrm{+}0.85$)
will be an interesting target for the study of low-mass stars.

\subsubsection{$\rm{SRa01a\_25729}$}

This 14 mag system is a typical contact binary with an
inclination of $i\sim75^{\circ}$ and $P = 0.295$ days. What makes it interesting is its mass
ratio of $q=0.98$. Furthermore it exhibits the O'Connell-effect ($\sim0.02$ mag) which seems to
be very stable during the observing run. The O'Connell-effect is generally explained
by strong stellar activity \citep[for an overview see][]{kaszas98}.

The temperature of the larger star ($T_1 = 4930$\,K, estimated from $J-K$) is
higher than the temperature of the smaller secondary ($T_2 = 4730$\,K) which is
consistent with the H-subtype contact binary systems \citep{csiz04a}.
However, the error of mass ratio ($q=0.98\pm0.05$)
allows the reverse case, too.

\subsubsection{$\rm{SRa01a\_33269}$: a double M-dwarf system?}

This short period ($P\sim0.65$ days) system shows an extreme mass ratio as far as
one can estimate from photometry for such a low inclination (i$\sim64^{\circ}$).
The system is extremely red, the effective surface temperature of the primary component was
estimated to be $T_{\rm eff}=3570$\,K, while the secondary has 2700\,K. If these parameters
are realistic - which is the case if the system does not suffer from interstellar
reddening - then these stars form very likely an dM1 + dM8 binary system.
Not also are such systems very rare, but also binarity among dM stars. That is why
this system may serve as an important target for lower main sequence parameter
studies \citep{torres12}. Spectroscopy is needed to check its spectral type and
the reddening effect.

\subsubsection{$\rm{SRa01a\_34263}$}

This Algol-type binary has a strange light curve shape around the secondary minimum,
where the brightness of the system increases significantly. This increase however is
not stable; during our observations it changes, resulting in a larger scatter
at these phases. We modeled this feature successfully with a bright spot on the
back side of the primary component.

\subsubsection{$\rm{SRa01b\_20683}$}

Unfortunately, we did not find any $J-K$ color for this star and hence the
light curve modeling results are uncertain. The system seems to consist of two
equal stars close to each other, because the ellipsoidal effect is quite clear.
Although the two transit depths look similar (indicating nearly equal surface
temperatures), the ellipsoidal effect suggests a mass ratio of
$q = 0.43\pm0.14$. Therefore one of the stars may be an evolved object, because of its radius-excess.

\subsubsection{$\rm{SRa01b\_23029}$}

This faint ($R_{max}\sim15.4$ mag), common orbital period ($P=0.357$ days) contact
binary star belongs to the A-subtype of contact binaries exhibiting a
distinct O'Connell-effect ($\sim0.04$ mag). To fit its light curve,
two spots were required, one very dark ($TF=0.11$) and one very bright ($TF=1.81$), although
this bright spot has a large uncertainty, see Table \ref{tab:model_parameters}).
We propose to obtain a full phase-coverage multicolour light curve, maybe extending it with H$\alpha$
measurements to test the validity of our light curve solution and see how the
stellar surface inhomogeneities are distributed on this star. However, the
presence of strong spot activity is indisputable owing to the presence of a strong
O'Connell-effect.

\subsubsection{$\rm{SRa01b\_38848}$}

Although this contact binary system is relatively faint ($R_{max} \sim 14.9$),
we obtained a good signal-to-noise ratio light curve. The system belongs to the so-called
A-subtype contact binaries in which the bigger star is also the hotter one;
in W-subtype systems the smaller component is hotter \citep{binn65}. It is
expected that W-subtype systems will show higher spot-activity \citep[e.g.][]{hendry92}
and W-type systems have generally later spectral types while
A-type systems have earlier ones. In the compilation of \citet{maceroni96}
the coolest A-subtype system is VZ Psc (4500\,K + 4352\,K), although it is
classified as a H-subtype system in \citet{csiz04a}. We conclude that SRa01b\_38848 is a good
candidate for an exceptionally low temperature A-type system. This can be
proven by multicolour photometry because then one can establish the exact
temperature difference between the components from the colour-curves.

The system also shows a remarkably large ($\sim0.04$ mag) O'Connell-effect.
The spot is halfway between the equator and the pole (co-latitude of
the spot's center is $41$ degrees) and its longitude is close to $90^\circ$
(see Table~\ref{tab:model_parameters}).

\subsubsection{$\rm{SRa01b\_41158 = NSVS~9838884}$}

Here, to obtain a satisfactory fit, we must fit the limb darkening coefficients instead of using the table
values. The system shows total eclipses which are deep,
with an inclination of $88^{\circ}$. Since the components are well separated
its brightness of $R\sim13.2$ mag makes it an easy target for further
studies. Its variability was discovered by the Northern Sky Variability Survey
\citep{wozniak04}.

\subsubsection{$\rm{SRa01b\_48723}$: near contact binary?}

The light curve modeling yields near contact binary with $P = 0.5$ days.
For the importance of near contact binaries see \citet{eggleton12}.
No signs of spot activity or O'Connell-effect can be seen, which agrees well
with the expectation of \citet{hendry92}. It is expected that A-subtype
systems have lower spot-activity.

\section{Other interesting variables}
\label{others}

\begin{figure*}
\centering
\includegraphics[width=0.24\textwidth]{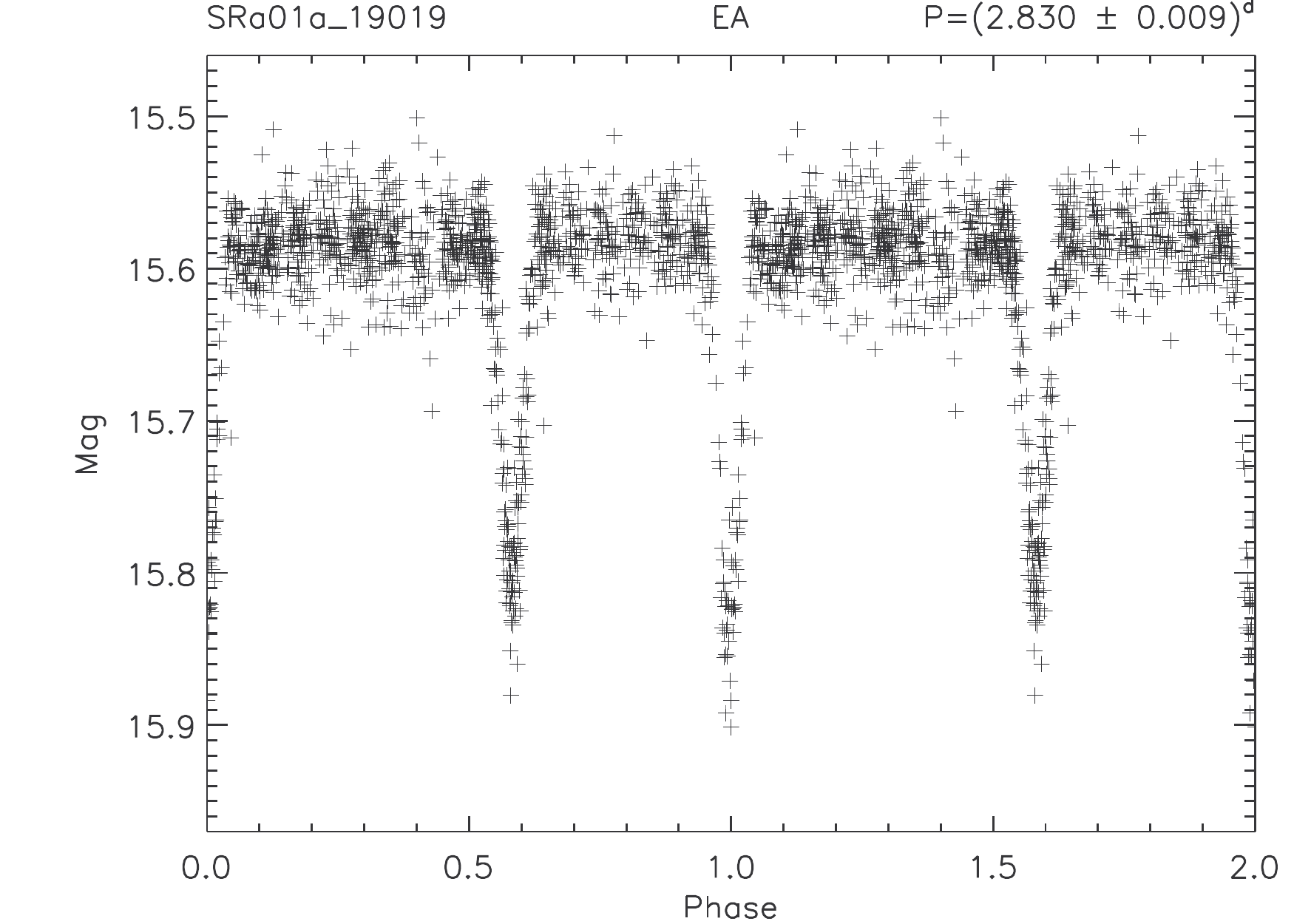}
\includegraphics[width=0.24\textwidth]{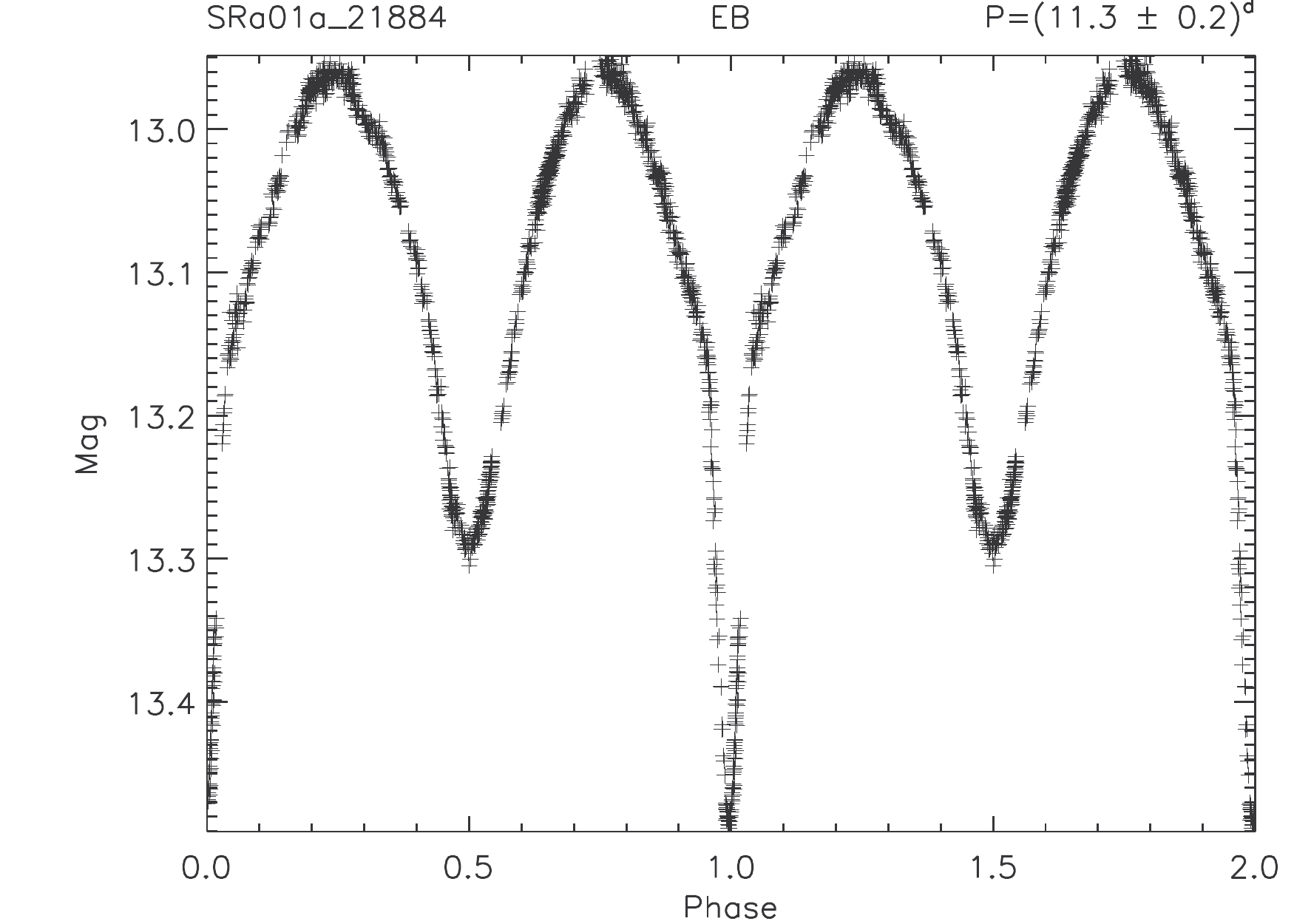}
\includegraphics[width=0.24\textwidth]{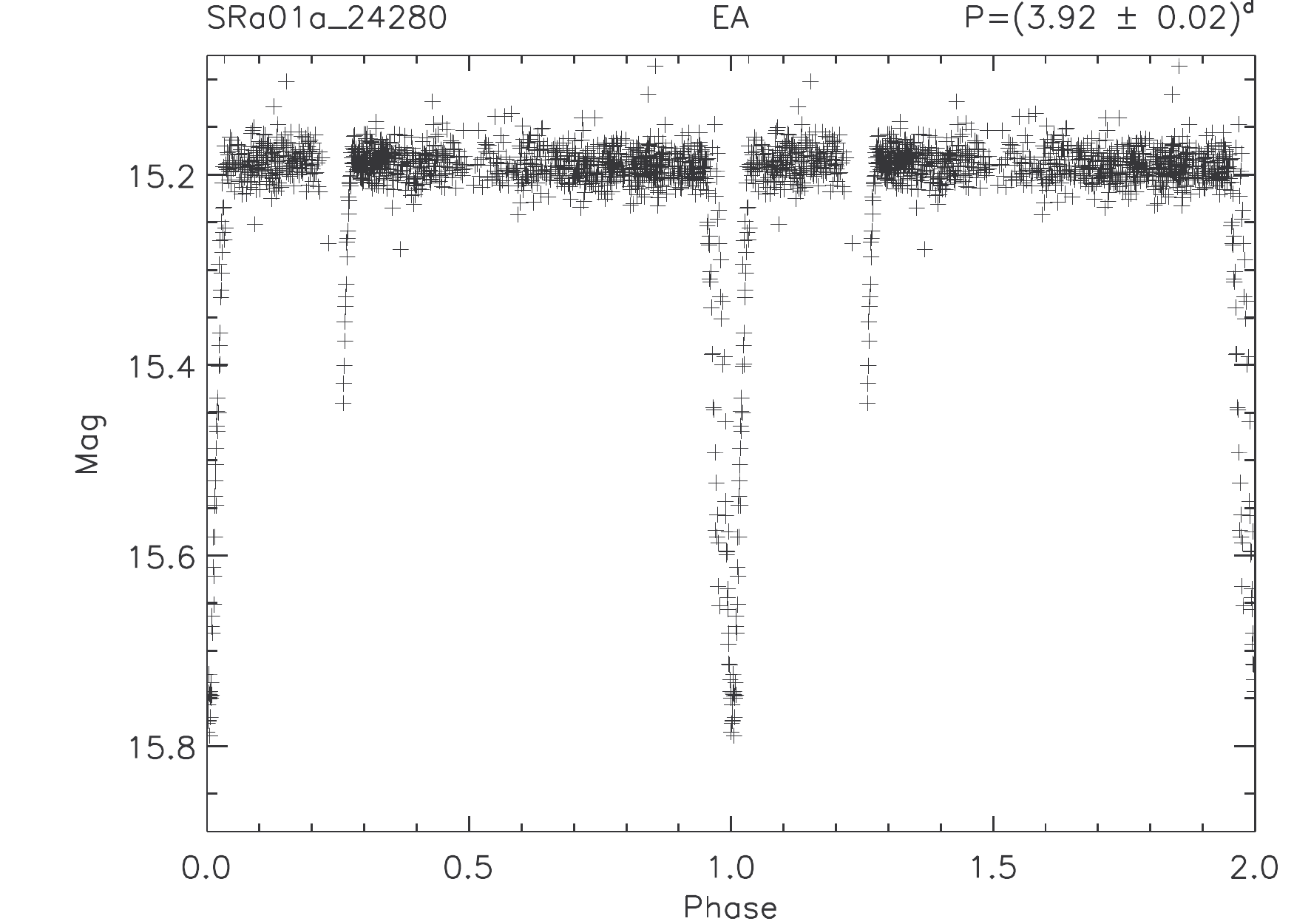}
\includegraphics[width=0.24\textwidth]{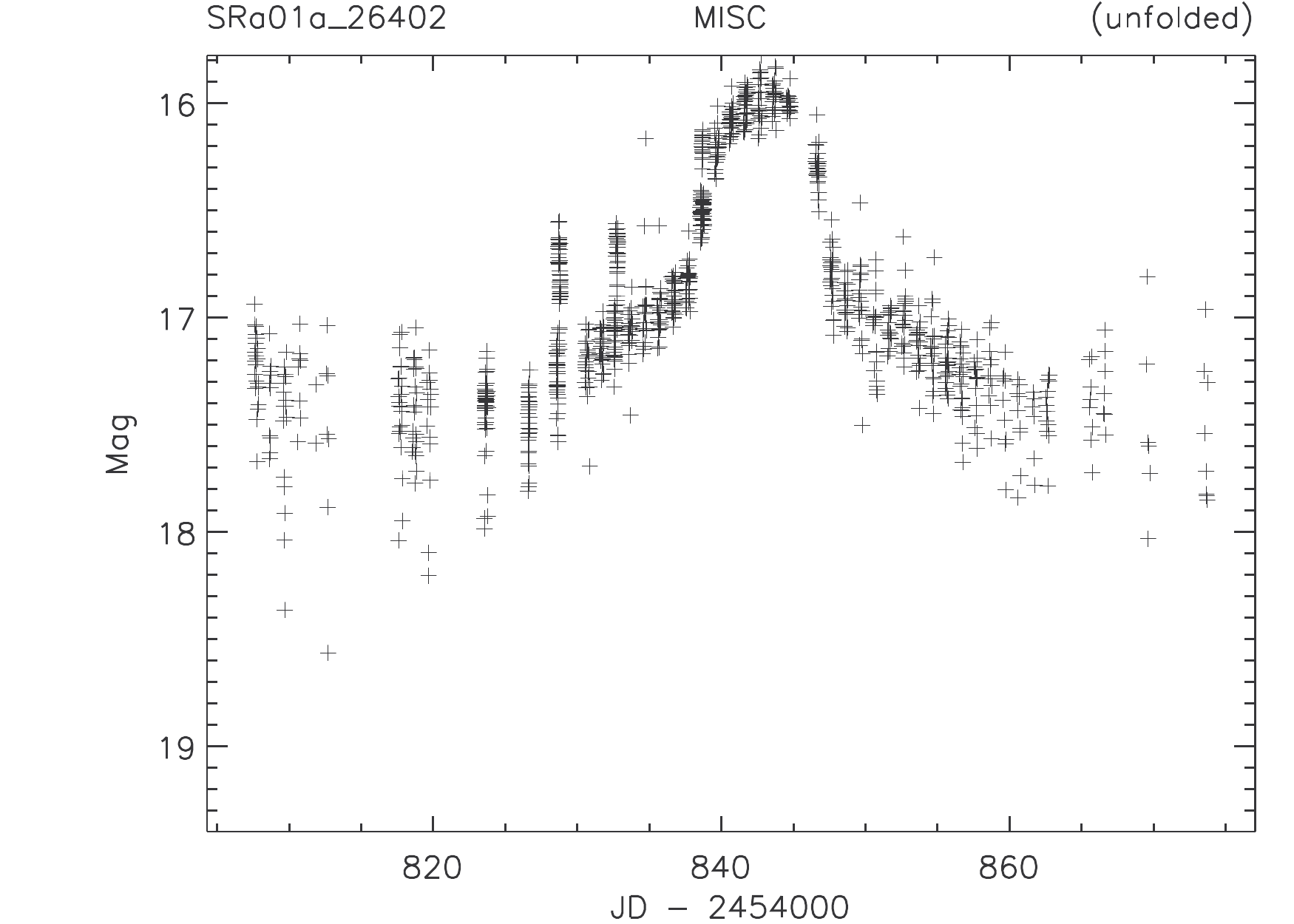}
\includegraphics[width=0.24\textwidth]{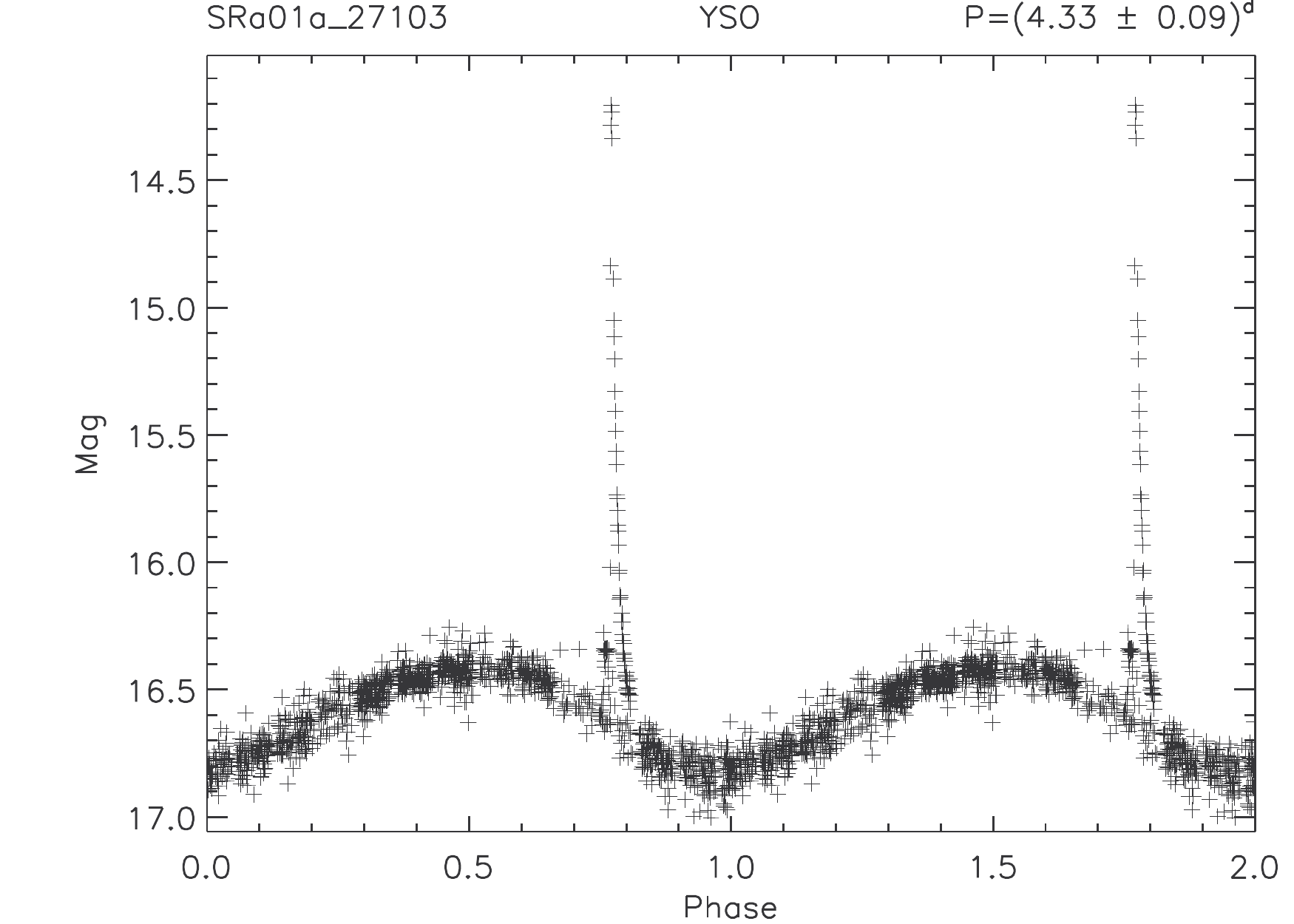}
\includegraphics[width=0.24\textwidth]{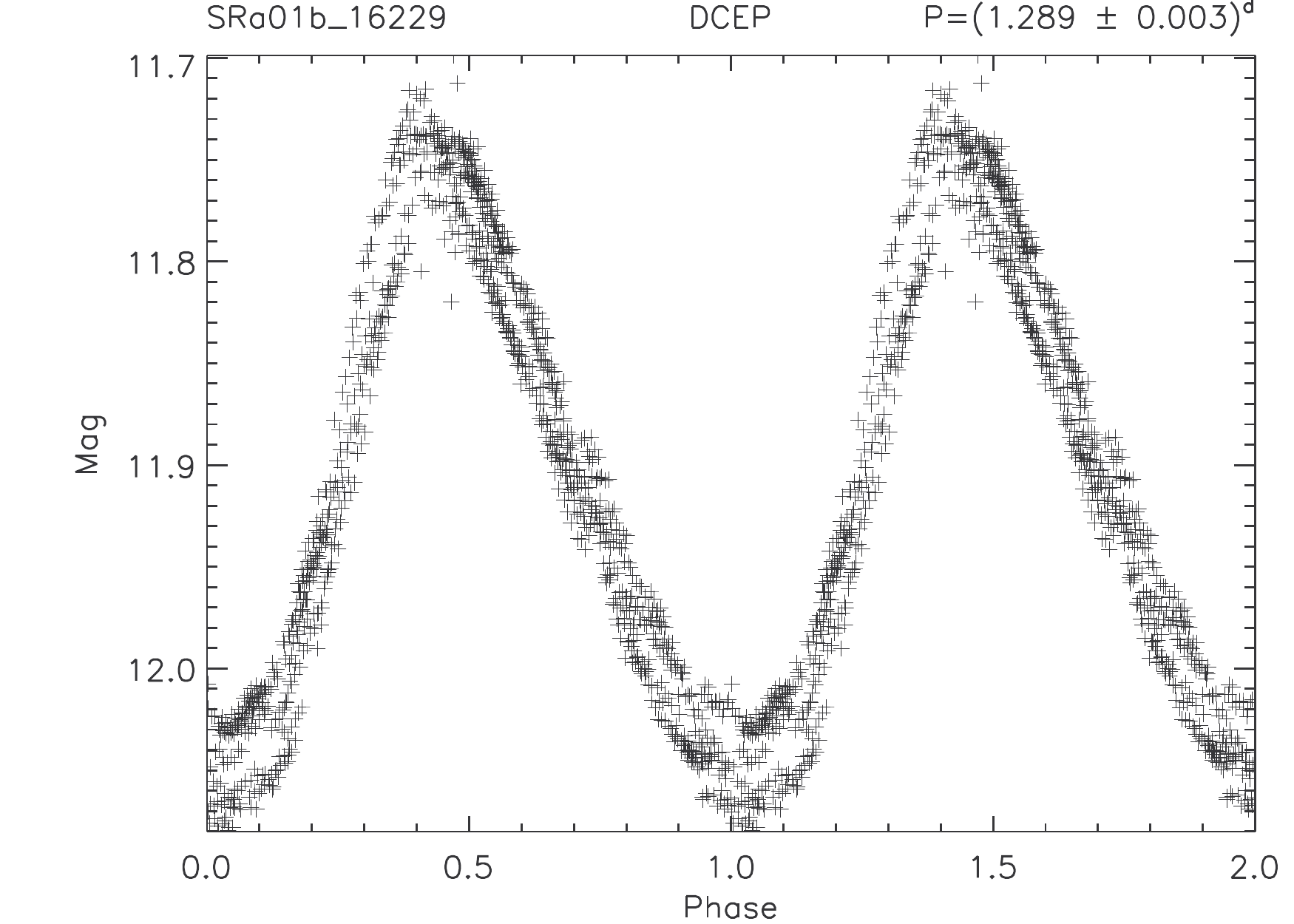}
\includegraphics[width=0.24\textwidth]{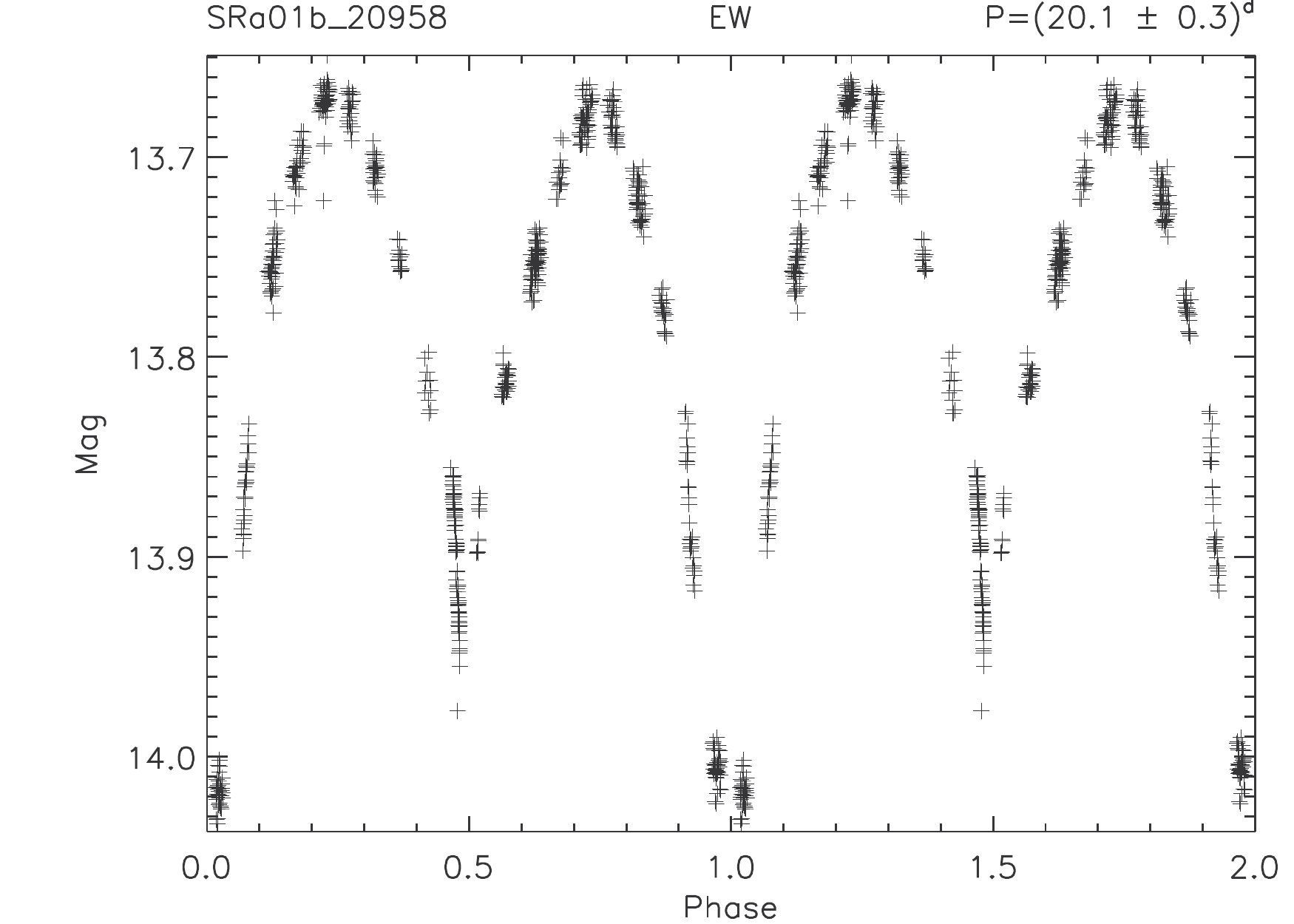}
\includegraphics[width=0.24\textwidth]{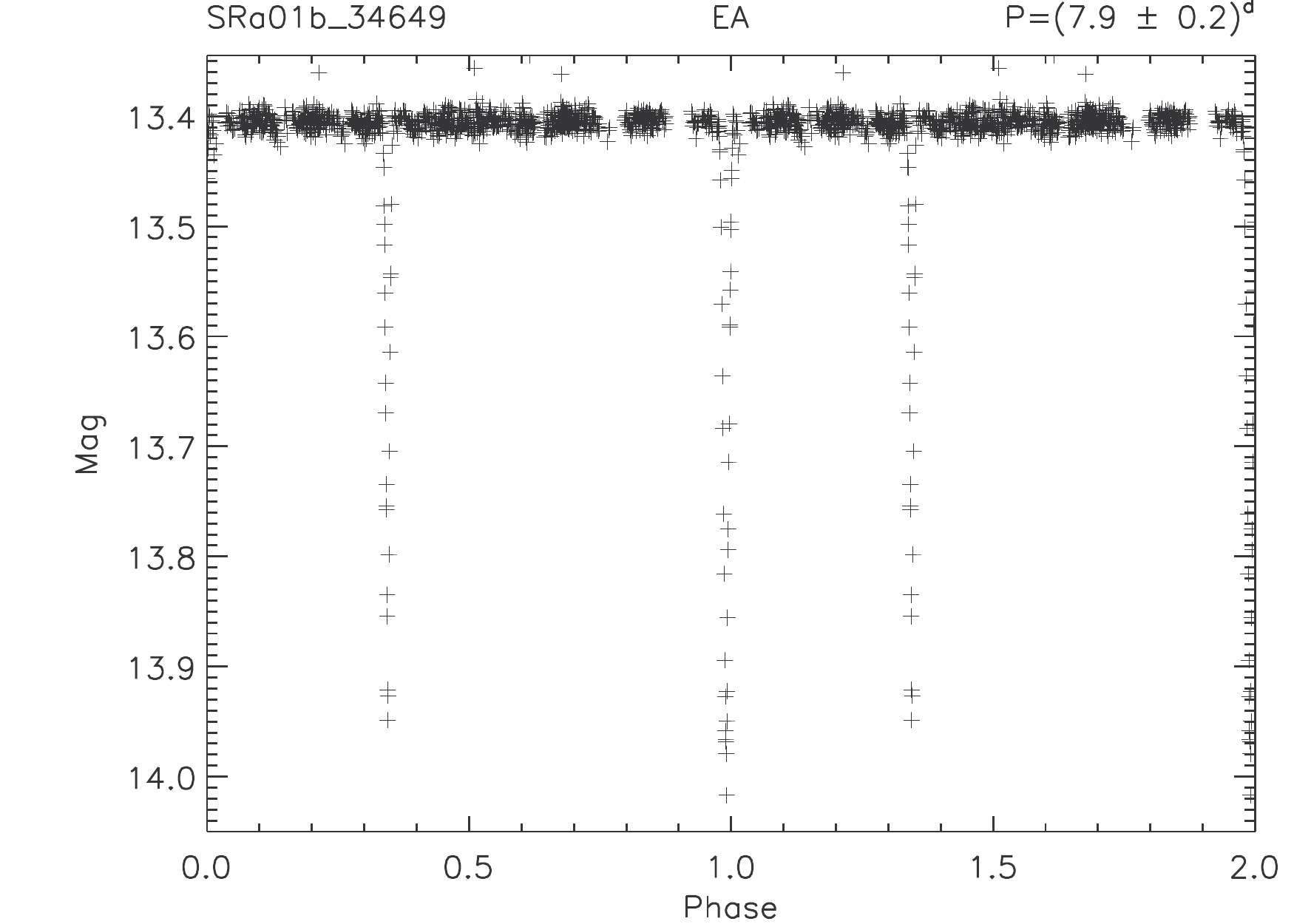}
\caption{Light curves of interesting variable stars. For a short description of these objects
see Section~\ref{others}.}
\label{fig:special}
\end{figure*}

There are many additional interesting variable stars in our dataset.
Here we present a selection of these objects. They are shown in Fig. \ref{fig:special}.

\subsection{$\rm{SRa01a\_19019}$}

The object SRa01a\_19019 is an eccentric Algol-type
eclipsing binary with an orbiting period $P=2.830$
days. The secondary eclipse is at $\phi=0.59$ and the widths of the
eclipses are almost the same. This yields an eccentricity of $e\sim0.14$
and $\omega\sim76^{\circ}$. $e$ and $\omega$ were estimated from
the widths and the phases of the eclipses.

\subsection{$\rm{SRa01a\_21884 = NSVS~9839705}$}

This object seems to be another near contact or semi-detached system
with a long period of 11.3 days. However, widths of the eclipses are not
the same (0.1 and 0.2 in phase) and the maxima are located not at 0.25 and 0.75 in phase, but
at 0.23 and 0.77. Therefore this system should have an eccentricity of $e\sim0.33$ and
$\omega\sim0^{\circ}$ meaning that we are viewing the system from the direction of periastron.

\subsection{$\rm{SRa01a\_24280}$}

SRa01a\_24280 is also an eccentric Algol-type eclipsing
binary system with $P = 3.92$ days. The secondary eclipse is located at $\phi = 0.24$
which corresponds to an eccentricity of $\sim0.44$ and
$\omega\sim315^{\circ}$.

\subsection{$\rm{SRa01a\_26402 = V0582 Mon = KH15D}$}

KH15D is one of the most intriguing objects in NGC 2264. Its variability is
caused by a precessing, warped circumbinary disc \citep{herbst10}. It has showed variable flux
since at least 1965 \citep{johnson05}, and started to show
eclipses around 1994-1995. The duration as well as the amplitude of such eclipses were continuously
increasing \citep{hamilton05}. In 2005, it featured 45 days long and 4 magnitude
deep eclipses every 90 days. After 2005, the system started to fade, its
maximum brightness was $I\sim15$ mag in 2008/09, but its amplitude was still four
mags \citep{herbst10}. Our light curve shape (Fig. \ref{fig:special}) is fully consistent with the one presented
in Fig. 3 of \citet{herbst10}.
Our scatter in the minimum is due to the faintness of the star and the fact
that we used a smaller telescope.

\subsection{$\rm{SRa01a\_27103}$}

The star shows a clear rotation modulation with a period of 4.33 days
and an amplitude of 0.5 mag. During one night we observed a
large flare with an amplitude of 2.5 mag.

\subsection{$\rm{SRa01b\_16229 = ASAS J064135+0756.6}$}

This variable star was discovered by \citet{khrus09}.
It is a double mode Cepheid. The previously published pulsation periods are
$P_1 = 1.28861$ and $P_2 = 1.03153$ days, so the period ratio is $P_2/P_1 = 0.8005$.
This ratio is typical for beat Cepheids pulsating
in the first and second overtone modes.
Our observations confirm the double mode behaviour
of the variable star with the periods $P_1 = 1.28848\pm0.00004$ days
and $P_2 = 1.03274\pm0.00021$ days and the period ratio of $P_2/P_1 = 0.8015\pm0.0002$.

In the Magellanic Clouds, dozens of double-mode Cepheids
pulsating in the first and second overtone modes are known, yet
in our Galaxy only three are known \citep{khrus09}.
Therefore, this star requires permanent multicolour observations.

\subsection{$\rm{SRa01b\_20958}$}

This object is a contact or near contact eclipsing binary
system with an extremely long orbital period of $P = 20.1$ days.
For comparison, in the catalogue of contact binaries
collected by \citet{csiz04a},
the longest period object is V729 Cyg with $P = 6.598$ days.
Further observations are needed with high priority at
the minima to be sufficient for modeling.

\subsection{$\rm{SRa01b\_34649}$}

SRa01b\_34649 is another Algol-type eclipsing binary with large
eccentricity and with a large amplitude of $0.6$ mag and an orbital
period of 7.9~days. The secondary minimum is at $\phi = 0.34$
which corresponds to an eccentricity of $e\sim0.42$ and $\omega\sim250^{\circ}$.
Due to the low number of measurements during
the eclipses, the system needs further observations.
It also requires spectroscopic and radial velocity observations
as well as precise eclipse timings.

\section{Special spatial distribution of eclipsing binaries}
\label{eclipsing}

We found 175 eclipsing variables in our dataset. Only 10 of them were previously known.
The large number of eclipsing binaries among the variable stars allows
us to investigate their distribution in the young open cluster.
This is also a very good opportunity to study the early phase of the evolution of these systems.

\begin{figure}[tb]
\centering
\includegraphics[bb=0 0 408 571, width=6cm]{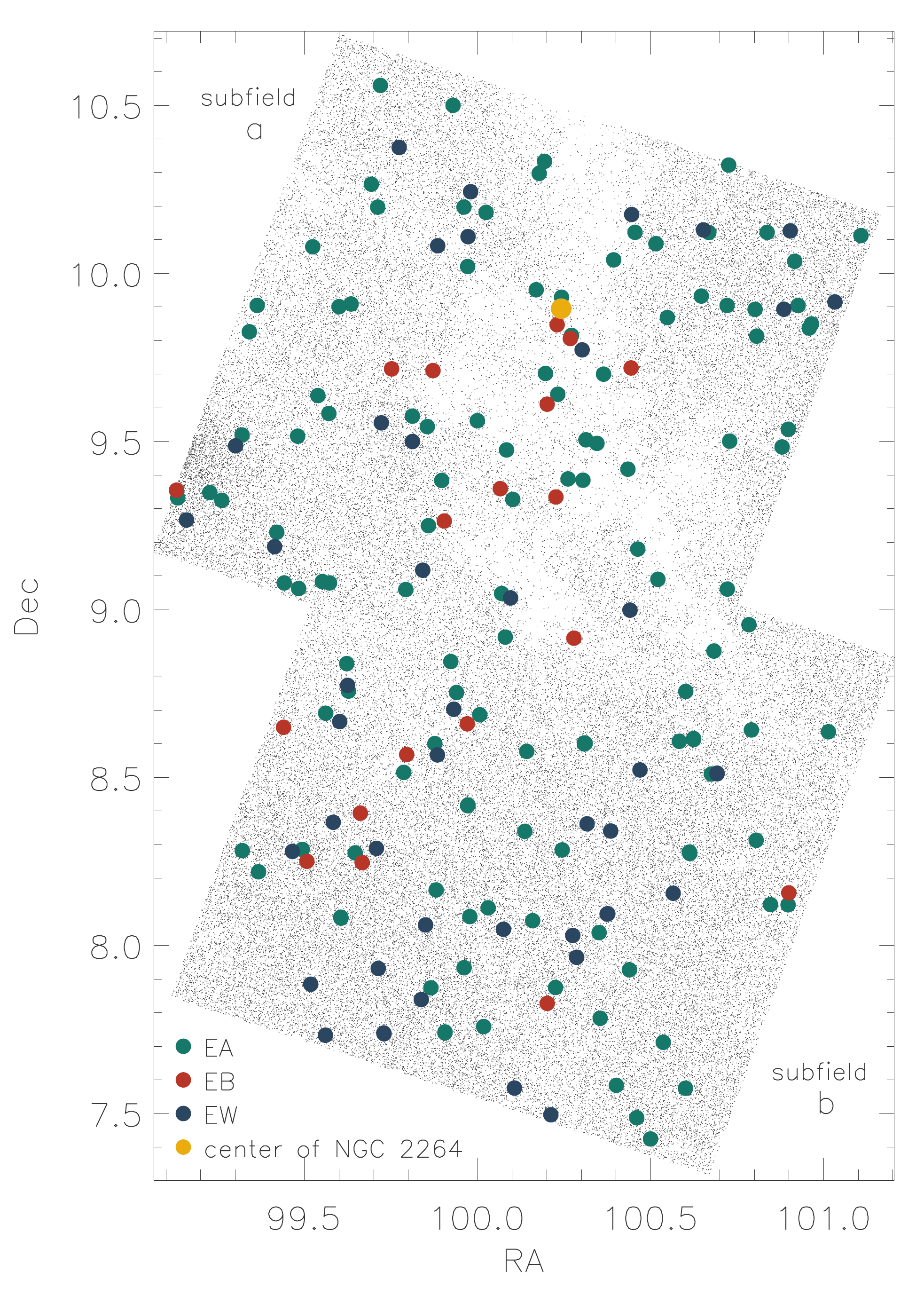}
\caption{The map of the eclipsing binary systems in our two subfields.
All observed stars are plotted with black dots. The circles (colored green in the online version)
show the positions of EAs, stars (red circles in the online version)
represent EBs and the triangles (blue circles in the online version)
are EWs. The square (yellow circle in the online version) is
the center of NGC 2264.
}
\label{binary_map}
\end{figure}

\subsection{EA excess in the cluster}
\label{EA_excess}

To create a spatial distribution statistics of binary stars in
NGC 2264, we calculated the fraction of different types (EA, EB and EW) of eclipsing binaries
among all detected stars at a given distance from the center of the cluster.
The surface density of cluster members decreases from the cluster center outwards.
Therefore, if the fraction of a variable star type
does not change with distance from the cluster center, this variable type
is not connected to the cluster.
The center of NGC 2264 (RA = 06:40:58, Dec = +09:53.7) was taken from the SIMBAD database.
The distances from the center were binned into $0.5$ degree
wide intervals. The stars classified as EW/DSCT were not involved.

\begin{figure}[tb]
\centering
\includegraphics[width=0.46\textwidth]{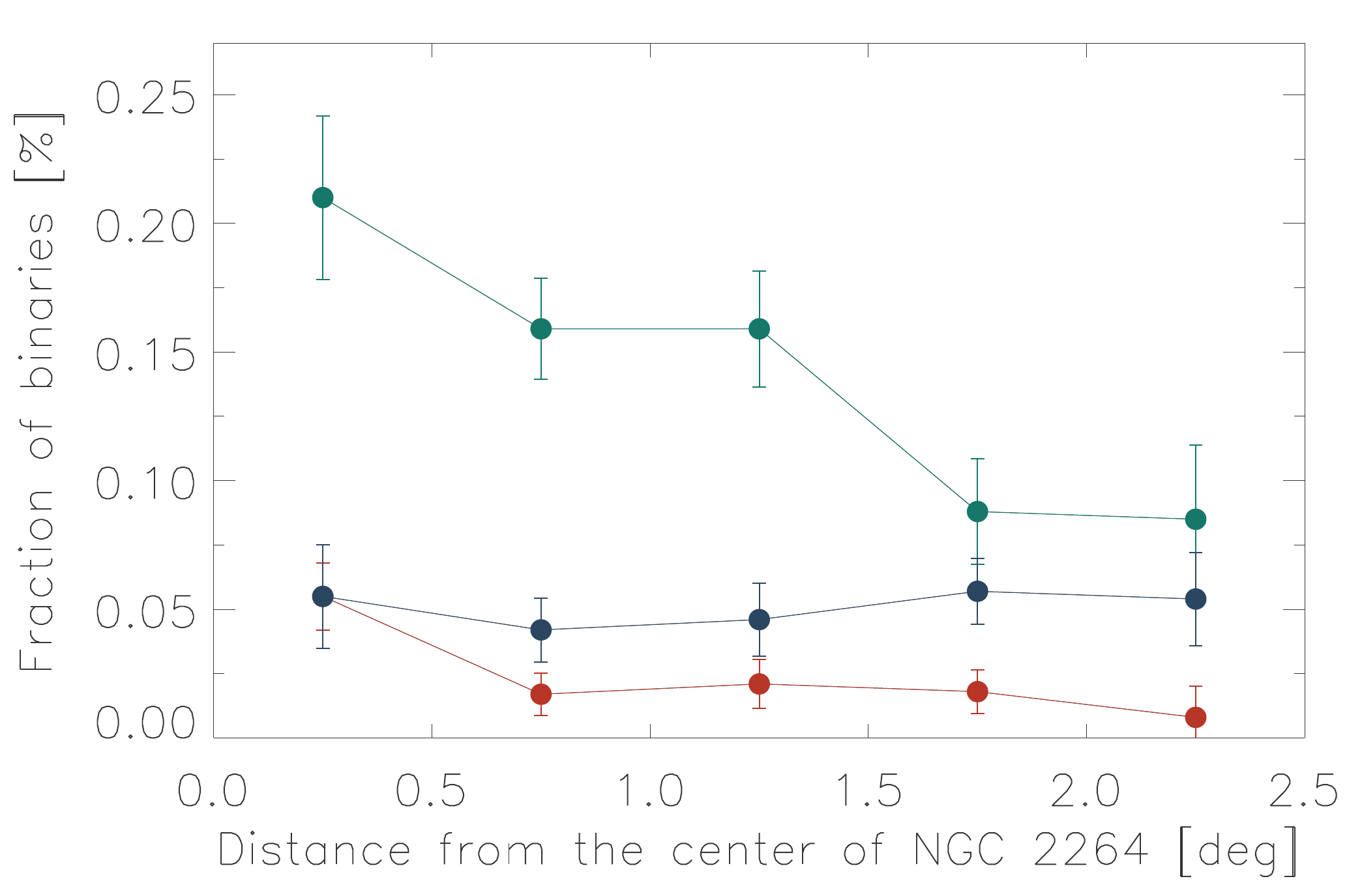}
\caption{Fraction of different type of eclipsing binaries among all stars in $\%$.
The color code is the same as in Fig. \ref{binary_map}.}
\label{binary_fraction}
\end{figure}

\begin{table*}
\centering
\caption{Number of all stars and different type of eclipsing binaries
in $0.5$ degree distance bins from the center of NGC 2264 outwards
and the fraction ($f_{binaries} = 100 \times N_{binaries}/N_{all~stars}$)
of binaries among all stars in $\%$.}
\label{tab:EA_excess}
\begin{tabular*}{0.85\textwidth}{cccccccc}
\tableline
Distance & $N_{all~stars}$ & $N_{EA}$ & $f_{EA}$ & $N_{EB}$ & $f_{EB}$ & $N_{EW}$ & $f_{EW}$\\
\tableline
$<$ 0.5     & 10937 & 23 &  0.210$\pm$0.032 &  6 &  0.055$\pm$0.013 &  6 &  0.055$\pm$0.020\\
0.5 $-$ 1.0 & 23869 & 38 &  0.159$\pm$0.020 &  4 &  0.017$\pm$0.008 & 10 &  0.042$\pm$0.012\\
1.0 $-$ 1.5 & 19468 & 31 &  0.159$\pm$0.023 &  4 &  0.021$\pm$0.009 &  9 &  0.046$\pm$0.014\\
1.5 $-$ 2.0 & 22779 & 20 &  0.088$\pm$0.020 &  4 &  0.018$\pm$0.009 & 13 &  0.057$\pm$0.013\\
2.0 $<$     & 12940 & 11 &  0.085$\pm$0.029 &  1 &  0.008$\pm$0.012 &  7 &  0.054$\pm$0.018\\
\tableline
\end{tabular*}
\end{table*}

The results are shown in Figure \ref{binary_fraction} and are summarized
numerically in Table \ref{tab:EA_excess}.
Only the Algol-type binaries show a significant decrease on moving outward from the center
of the cluster. The fraction of contact systems (EWs) is constant which
indicates that these type of eclipsing binaries are not members
of the cluster, in contrast to the EA type ones. The ratio of EB type
binaries also shows a slight decrease, but there are only 19 of these systems in the whole field.
First of all, these results may confirm theoretical calculations
suggesting that contact and semi-detached binaries are outcomes of
the dynamical evolution of close binary systems after several
tens or hundreds of Myrs \citep{eggleton06}.

Using a hypergeometric distribution of the number of EAs it is possible to test
the probability that the EA excess is not real but due to a random fluctuation.
The chance of detecting 92 EAs among 54,274 stars that are closer than 1.5 degrees
to the center of NGC 2264 is only $1.6\times10^{-6}$. The estimated error of the number of
EAs is the square root of the variance of the statistical distribution, which can be calculated using the total number of all stars,
the total number of EAs and the total number of stars in each bin. In this case $\sigma = 5.22$.
The expected value of the number of EAs in this area is 68.7 based on the fraction of all EAs
($f_{observed}$, see Sect. \ref{sec:EA_freq}) and 47.1 if we use
the fraction of EAs in the field ($f_{field}$). The difference between the number of detected and expected EAs
is $4.6 \sigma$ and $8.5 \sigma$, respectively, indicating a real excess.
An investigation on the origin of this excess is in Section \ref{sec:membership}.

Note that we included also EAs without known periods; their inclusion
does not change the results, because we are presently only interested
in their relative frequency.

\subsubsection{EA frequency inside the cluster}
\label{sec:EA_freq}

The Algol-type systems are detached eclipsing binaries with a typical
period of a few days. The lower limit is around 1 day \citep{paczynski06},
while the longest known period is 27 years ($\epsilon$ Aur, GCVS).

If we define the relative frequency of Algols as $f_{\rm EA}=N_{\rm EA}/N_{\rm total}$.
This frequency is twice of the field's value in the region the cluster.

We separate the stars into field and cluster member stars.
Actually, we do not know which is a field object and which is a cluster member, but
we are only interested in the fraction of these two groups and
this quantity is known: $x=N_{\rm cluster}/N_{\rm total}$. The observed fraction of EAs
is then the weighted average of the fractions of the two groups, i.e.,

\begin{equation}
 N_{\rm EA} = N_{\rm field} \times f_{\rm field} + N_{\rm cluster} \times f_{\rm cluster},
\label{eq_N_EA}
\end{equation}
dividing by $N_{\rm tot}$ and taking $N_{\rm field} = N_{\rm tot} - N_{\rm cluster}$
we get:

\begin{equation}
 f_{\rm observed} = (1-x) \times f_{\rm field} + x \times f_{\rm cluster},
\label{eq_observed_fraction}
\end{equation}
where $\rm{f_{\rm observed}}$, $\rm{f_{\rm field}}$ and $\rm{f_{\rm cluster}}$ are the fractions of EAs
among all stars, among the field stars and in the cluster, respectively.

There are $\sim 2000$ known cluster members \citep[][and references therein]{cody13},
therefore we set the number of observable cluster members to 2000. This results in $x=0.0222$.
The fraction of EAs among all stars is $f_{\rm observed}=0.00137\pm0.00013$.
The errors of the observed frequencies are calculated as $\sigma=1/\sqrt{N}$.
$f_{\rm field}$ was set to 0.00087 which is the
ratio of EAs beyond 1.5 degrees from the center of the cluster. All these values result in
$f_{\rm cluster}=0.0219\pm0.0033$. This means that more than $2\%$ of the cluster member stars are Algol-type eclipsing binaries,
which is 25 times higher than for field stars.
In order to check our field frequency we calculated the same ratio for the CoRoT-field
LRa02 observed also by BEST II \citep{fruth12}. This field contains 115 EAs among a total of 98,219
stars, which yields $f_{\rm obs,LRa02}=0.00117\pm0.00011$.
Given that we did not find any difference in the magnitude distribution
for stars within and outside of the cluster, we regard this value as real.

In the field we can assume that the fraction of binary systems
among all stars is around $50\%$ \citep{yamasaki13}.
If all the stars in NGC 2264 were located in binary systems we could expect
a doubled number of EAs in the cluster. This is, however, still
12 times smaller than what we observe.

Another possibility is that during the stellar and binary formation
the systems preserve the original angular momentum vector of their parental
cloud resulting in a more or less parallel orientation of the orbital planes of the binaries.

\subsubsection{Cluster membership of EAs}
\label{sec:membership}

We tried to separate the Algols that belong to the cluster from the eclipsing
binaries in the field to confirm
that the observed excess is in the cluster and not in the field.

We checked the proper motions of our stars using the PPMXL proper motion catalog
\citep{roeser10}. To calculate the probability of the EAs we used the method described by \citet{sand71}.
The probability that a given star belongs to the cluster can be calculated as

\begin{equation}
 \Phi_c = e^{-\frac{1}{2} \left( \frac{d_{\alpha}^2}{\sigma_{\alpha_{cl}}^2} + \frac{d_{\delta}^2}{\sigma_{\delta_{cl}}^2} \right)},
\end{equation}
where $d_{\alpha}$ and $d_{\delta}$ are the differences between the proper motion of a star
and the average proper motion of the cluster member stars and $\sigma_{\alpha_{cl}}$ and $\sigma_{\delta_{cl}}$
the standard deviations of the proper motions of the cluster member stars. The same equation can
be used for the field membership ($\Phi_f$) but with the average proper motion of the
field stars and their standard deviation.

Since we do not know which ones are cluster members among our 90,065 stars,
during the determination of the proper motion of the cluster we used the
cluster members identified by \citet{furesz06}. The average proper motion of
the cluster is $\mu_{\alpha_{cl}} = 0.35 \pm 3.83$ mas/year and 
$\mu_{\delta_{cl}} = -5.09 \pm 4.12$ mas/year. The average proper motion
of the field is $\mu_{\alpha_{field}} = -0.03 \pm 5.07$ mas/year and 
$\mu_{\delta_{field}} = -3.24 \pm 5.17$ mas/year based on all 90,065 stars.

The normalized probability that a given star belongs to the cluster \citep{sand71} is:

\begin{equation}
 P_c = \frac{\Phi_c}{\Phi_c + \Phi_f}.
\end{equation}
The sum of these probabilities for all of the EAs in the cluster area is 30.1, which is the expected
number of the cluster member EAs. We were able to identify in the PPMXL catalogue only 87 out of the 92 EAs,
therefore this value is underestimated.
Using the EA frequency of $f_{field} = 0.00087$ the expected value of field EAs is 76.6.
Since the total number of Algols is 114, it follows that the expected number
of cluster member EAs is $114 - 76.6 = 37.4$. This is in good agreement
with the value we expect from the proper motion analysis. Since the distance of the cluster is
900 pc, the proper motion values are small and hence it is difficult to select
the cluster members individually.

Another way of selecting cluster members is via the use of a color-magnitude diagram.
The problem with this method however is that EAs are binary stars and both the brightness
and the color are affected by the companion. Moreover, the NGC 2264 is a very young
open cluster embedded in a dense interstellar cloud and this cloud produces a non-negligible
absorption and reddening even in near-infrared $J$, $H$ and $K$ bands. Therefore, the
de-reddened position of the EAs in a color-magnitude diagram is doubtful. We compared the
location of the cluster member stars \citep{furesz06} and our EAs in the $(J-K)$ - $K$
color-magnitude diagram. We shifted the binaries with the maximum value of 0.75 mag
to correct for the additional intensity of the companion in case of two equally
bright stars. 46 EAs are located in the same area
as the cluster members, which is 25\% higher than the number of expected EAs in the cluster.

The most accurate test would be to check the parallaxes of the stars,
but, due to the large distance of the cluster, for sufficiently accurate
parallaxes we have to wait for the results of the Gaia satellite.

In conclusion, based on both the proper motions and on the color-magnitude diagram
there are a sufficient number of cluster member candidates in our EA sample to confirm
the EA excess as an extreme excess of eclipsing binaries in NGC 2264.
The expected value of Algols among the 2000 cluster members based on the Algol frequency in the field
is only 1.74 contrary to the 30.1 and 46 using the proper motions and the color-magnitude
diagram, respectively.

\subsubsection{Model of the fraction of EAs}

It is reasonable to assume that the increased frequency os Algols
in the cluster area is due to the cluster members, because the chance
of being only a random fluctuation in the field is extremely low.
In order to explain the excess we modeled the fraction of observed
Algol type binaries. We assume an initial inclination ($i_0$) and
a scatter ($\sigma$) with a normal distribution.

It is possible to calculate the theoretical fraction of EAs in the field
and in the cluster using only basic physical equations and empirical relations.
We assume the period distribution as follows:
\begin{equation}
 P (\rm{days}) = 5 \times 10^4 \left(\frac{X}{1-X}\right)^{3.3}
\label{eq:period_distr}
\end{equation}
where $X$ is a random variable distributed uniformly over the range [0,1].
Since there are only a few EAs with an orbital period less than 1 day
\citep{paczynski06} we remove the binary stars with $P < 1$ day, and
taken into account only stars with longer periods. In our sample
there are only 6 EAs with a period shorter than 1 day. For longer periods
this distribution is appropriate.

The mass distribution is:
\begin{equation}
 M_1 = 0.3 \left(\frac{Y}{1-Y}\right)^{0.55},
\label{eq:mass_distr}
\end{equation}
where $Y$ is another random number uniformly distributed between 0 and 1.
Finally the mass ratio is:
\begin{equation}
 q = \frac{M_2}{M_1} = 1 - Z^3,
\label{eq:mass_ratio}
\end{equation}
where $Z$ is a third random number also uniformly distributed in [0,1].
We use only circular orbits ($e=0$). Relations \ref{eq:period_distr} - \ref{eq:mass_ratio}
were taken from \citet{eggleton06}.

The mass-radius relation was \citep{tingley05}:
\begin{equation}
 R_{1,2} = M_{1,2}^{0.8}.
\end{equation}

We see an eclipse, if
\begin{equation}
 i > i_{cr},
\end{equation}
where $i$ is the inclination of the orbital plane and $i_{cr}$ is the critical inclination.
$i_{cr}$ can be expressed as:
\begin{equation}
 i_{cr} = \arccos{\left(\frac{R_1+R_2}{a}\right)},
\end{equation}
where $a$ is the semi-major axis of the orbit. This can be calculated from
Kepler's 3rd law:
\begin{equation}
 a (R_{\odot}) = [M_1 (1+q) (P/365.25)^2]^{1/3} \times 214.8339.
\end{equation}
$M_{1,2}$ and $R_{1,2}$ are in solar units, while $P$ is in days.

First, we simulate the Algol-type variables with a
uniformly distributed angular momentum unit vector in space.
Using a Monte Carlo simulation, we generated 50,000
binary systems with the parameter distribution described above
and a distribution of the inclinations varying as:
\begin{equation}
 i = \arccos{G}
\end{equation}
where $G$ is another random number uniformly distributed in [0,1].

Assuming $50\%$ of binary systems, this model
yields an $0.260\pm0.008\%$ fraction of eclipsing binaries.

There are two more factors that influence the observability
of these systems. One is the depth of the eclipse. There are many
grazing eclipsing systems with very small amplitudes. In our sample
all the eclipsing systems have at least 0.04 mag deep eclipses, so
we set the lower limit to 0.04 mag. The depth of the eclipse is calculated
without taking into account the limb darkening.

\begin{figure}[t]
\centering
\includegraphics[width=0.46\textwidth]{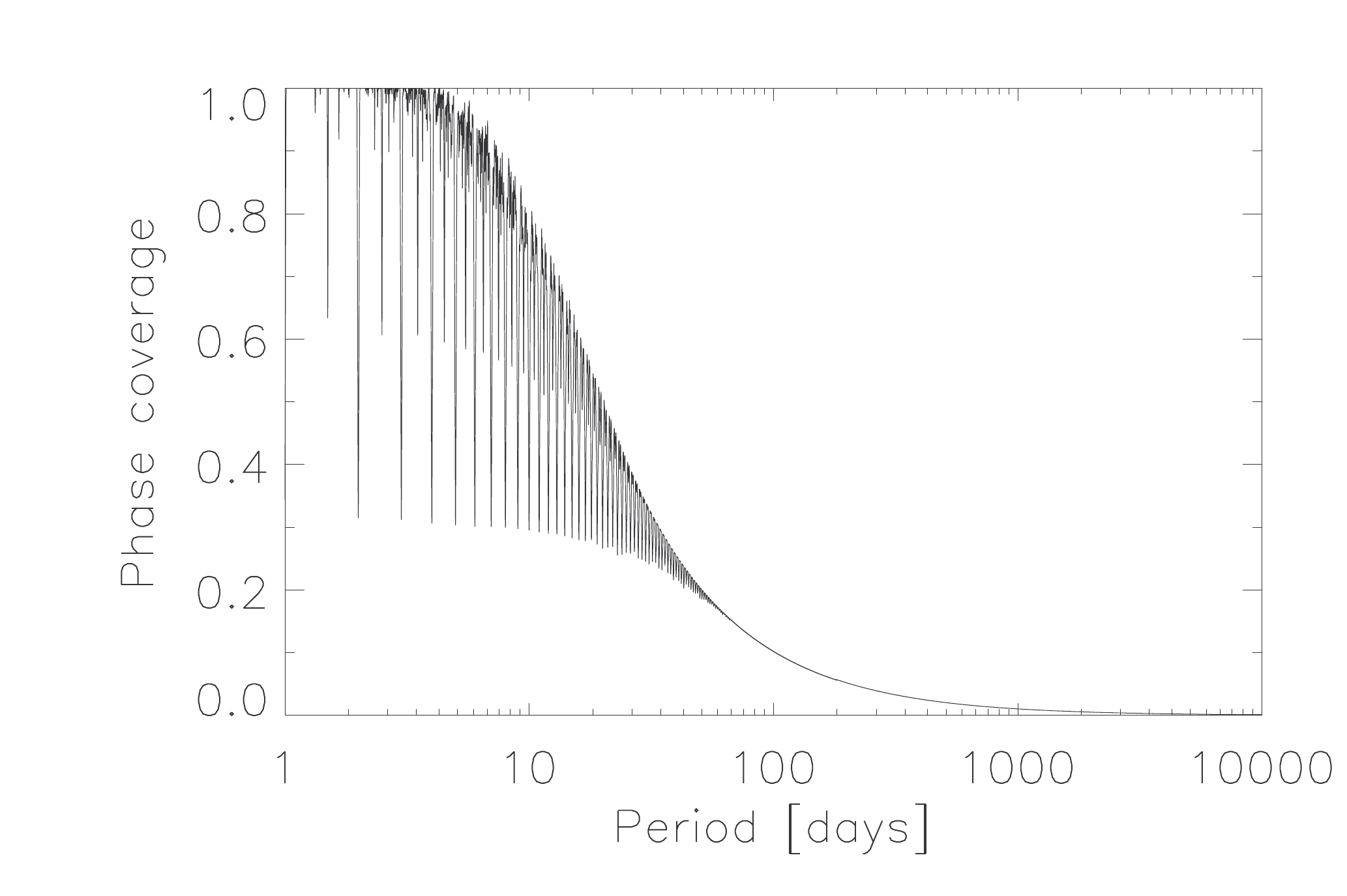}
\caption{Observed phase coverage for at least one eclipse of an eclipsing
binary based on the epochs of our observations.}
\label{observability}
\end{figure}

The other factor comes from the epochs of the observations itself.
If an Algol-type system has an orbital period of e.g. 2.0 days but the
eclipses appear in the daytime we have no chance to observe them.
Therefore, we calculated the fractional phase coverage as a function of orbital period.
This function gives the probability that we observe at least one eclipse
in our dataset. This is shown in Fig. \ref{observability}. The calculation takes
periods until $P=10000$ days into account.
Then instead of counting the EAs with at least 0.04 mag amplitude
and $P > 1$ day, we summarize these probabilities for the detectable binaries.

Finally the relative frequency of the observable Algol type systems in our dataset
is $0.104\pm0.004\%$. This is acceptably close to the observed value
of $0.087\pm0.016\%$. For fainter stars the limiting eclipse depth of 0.04 mag is
somewhat underestimated. If we set the limiting amplitude to 0.1 mag
the observable frequency is $0.102\pm0.004\%$, and thus not significantly different.

The observed fraction of Algols in NGC 2264, however, is $\sim2.2\%$, more
than 20 times higher. If all stars were binaries, this could
increase the field's frequency to $0.2\%$ inside the cluster but not to $2.2\%$.
This leads us to suggest that the inclinations cannot be isotropically distributed inside the cluster.
Using our model we can check whether there is a combination
of the initial inclination and a scatter around it that can reproduce
this high fraction. Therefore we run a Monte Carlo simulation. The inclinations are
taken as follows:
\begin{equation}
 i_{initial} = i_0 + \sigma_i
\label{eq:i_0}
\end{equation}
where $i_0$ is a random number uniformly distributed in [0$^{\circ}$,90$^{\circ}$] and
$\sigma_i$ is another random number normally distributed around 0$^{\circ}$
with $\sigma_i$ standard deviation in [0$^{\circ}$,45$^{\circ}$].

\begin{figure}[t]
\centering
\includegraphics[width=0.46\textwidth]{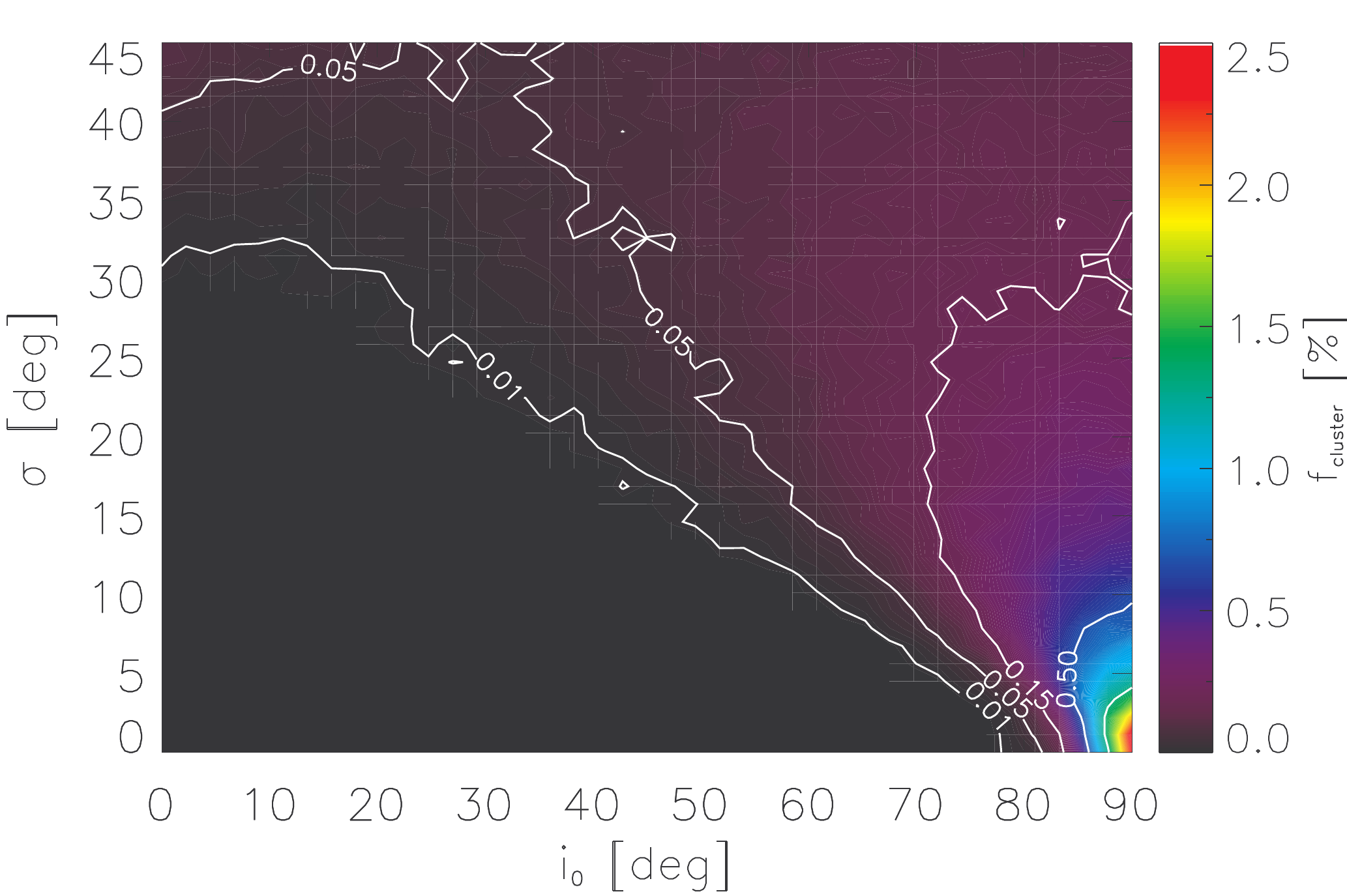}
\includegraphics[width=0.46\textwidth]{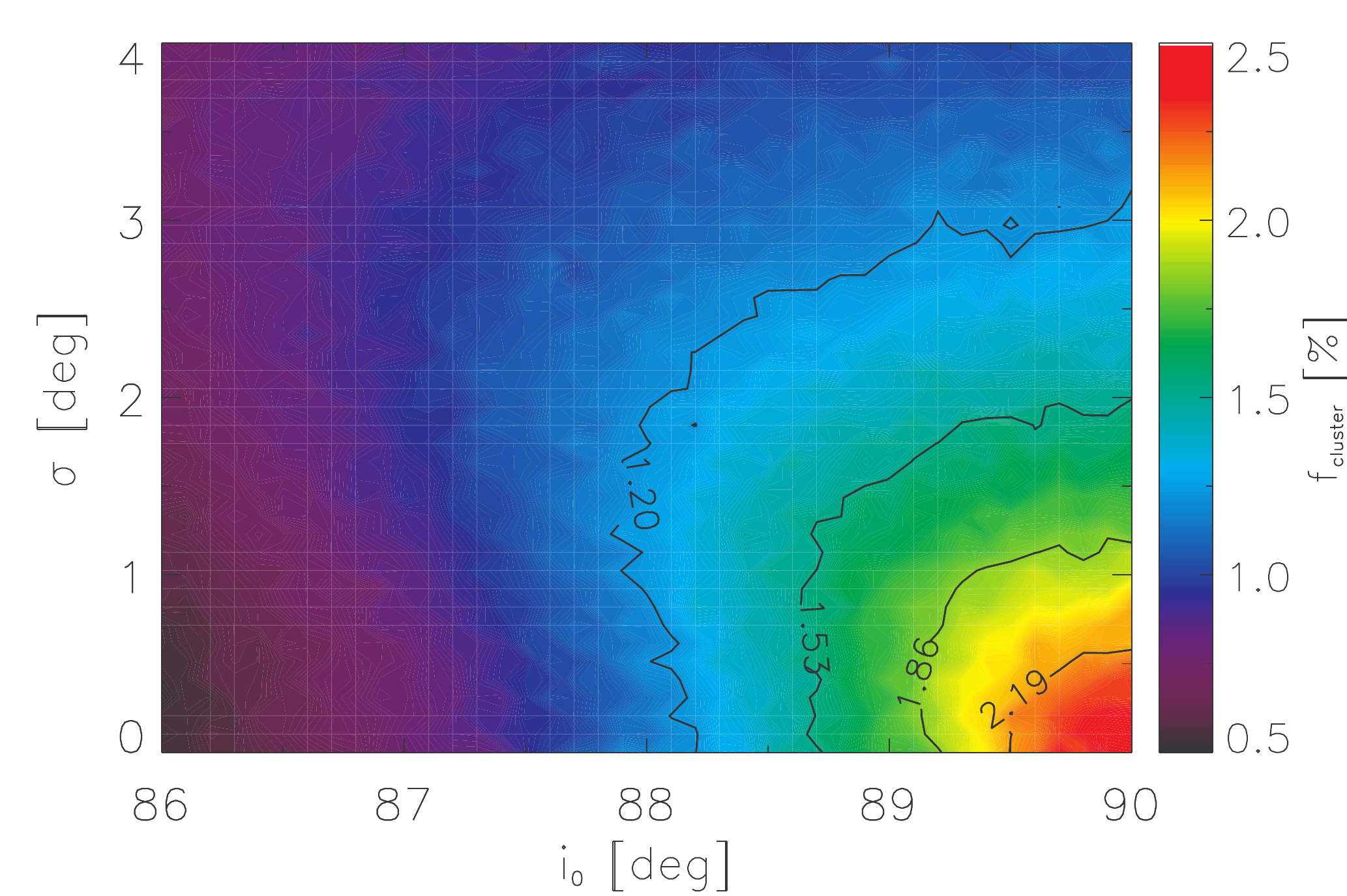}
\caption{The modeled value of the fraction of observable eclipsing binaries in the cluster
($f_{\rm cluster}=N_{\rm EA}/N_{\rm cluster}$) with an initial inclination $i_0$ and a standard deviation $\sigma$ around $i_0$.
The top panel shows the whole simulated parameter space, while the bottom panel
enlarges on high inclinations and small scatter.
In the bottom panel our observed value of $f_{\rm cluster}=2.19\pm0.33\%$ and the $1\sigma$, $2\sigma$
and $3\sigma$ error areas are marked by a white line (black line in the online version).}
\label{binary_fraction_map}
\end{figure}

The results are shown in Fig. \ref{binary_fraction_map}.
Both panels of this figure are based on a Monte Carlo simulation of 10,000
inclination distributions described in Eq. \ref{eq:i_0}.
In the top panel of Fig. \ref{binary_fraction_map} we plot the whole
modeled range, while in the bottom panel we zoom into the right bottom
corner region. As one can see, there is only a very
small region where the fraction of detectable EAs is around $2.19\%$,
our measured value. So the average inclination of the NGC 2264
should be above $89^{\circ}$ and the standard deviation is less than
$1^{\circ}$. Practically we see the cluster's eclipsing binaries edge on.

\subsubsection{Possible explanation of the excess}

There are two possible solutions for the origin of this distribution.
First, there may be a physical mechanism that uniformises the
angular momentum vector distribution inside the cluster. Second, and more likely,
the original direction of the angular momentum vector of the cloud from
which the cluster was formed was conserved even if the total amount of
the angular momentum was decreased during the star formation process.
The parallel orientation of the outflows in star forming regions
also supports this scenario \citep[e.g.][]{froebrich03, trinidad04}.

Although the suggested common inclination distribution has some implications on
the star formation processes, it is out the scope of this paper to investigate
which star formation theory includes the conservation
of the direction of the angular momentum vector.

The proposed strict inclination distribution of the binaries may explain
also the lack of eclipsing binaries in the Pleiades \citep{bukowiecki12}.
In the lower left part of the top panel of Fig. \ref{binary_fraction_map}
the fraction of observable eclipsing binaries is practically 0, although
about 2/3 of the stars in the Pleiades are binaries \citep{kahler99}.

These results also have an important consequence for transiting exoplanet
searching surveys. If we can assume that the same process works for planet formation
as for multiple star formation including the Kozai-mechanism \citep{fabrycky07,mazeh79}
and the planets remain in the same orbital plane for longer timescales,
one is unlikely to find any transiting exoplanet in
open clusters that do not contain eclipsing binaries. Thus the more favourable targets
for transit searches are eclipsing binary-rich clusters.

\section{Summary}
\label{summary}

We presented a study of the variable stars in and around the young open
cluster NGC 2264 observed by BEST II in 2008/2009. We detected 1,161
variable stars out of 90,065 stars in our field of view. Only 241 of
these stars were previously reported as variables, 920 are newly detected.

We collected a set of young stellar objects with unusual
light curves. A similar light curve was already found by \citet{rodrigez-ledesma12},
but our amplitudes are smaller and the periods are shorter.
The large number of such objects is maybe due to the
extreme distribution of the direction of angular momentum vectors (see Sect.~\ref{eclipsing}).

We studied the light curves of 11 eclipsing binaries.
SRa01a\_10712 seems to be a lower main sequence binary,
while SRa01a\_33269 is a possible double M-dwarf system.

Investigating the spatial distribution of the Algol-type
eclipsing binaries in the field we found that the fraction of these binaries
is extremely high in the cluster. The incidence of these systems is
25 times more frequent than outside the cluster among the field stars.
We checked whether this excess is real and we found that the probability of
it being a statistical fluctuation is extremely low. Therefore, it is reasonable
to assume that the excess is caused by the Algols inside the cluster. We also searched for
the origin of the excess and we concluded that the number of cluster member
candidate EAs is in very good agreement with our assumption.
This feature can be explained if we assume that the angular momentum
vectors of the orbital motions are directed anisotropically.
This direction originates most likely from the original angular
vector of the parental cloud. In the case of NGC 2264, the average inclination
of the orbital planes should be more than $89^{\circ}$ with a standard
deviation less than $1^{\circ}$. This model can explain also the lack of eclipsing
binaries in the Pleiades. If the inclination of a cluster is low enough,
we cannot see any eclipsing binaries even if the scatter of the inclinations
is higher. It is the task of star formation theories to explain the extremely high ratio
of Algol type variables. Our finding is in agreement with the outflows
in star forming regions \citep{froebrich03,trinidad04}.

\acknowledgments

Peter Klagyivik acknowledges support from the Hungarian State E\"otv\"os Fellowship.
Petr Kabath acknowledges the co-funding under the Marie Curie Actions of the European
Commission (FP7-COFUND). This research has made use of the SIMBAD database, operated at CDS, Strasbourg, France.
We also made use of 2MASS, GCVS catalogs, and AAVSO variable star search index.
The authors thank Drs. M\'aria Kun, M\'aria S\"uveges for discussion, Dr. Lee Grenfell
for correcting the English text and the referee for the remarks leading to a better presentation of the results.

\end{document}